\pdfoutput=1
\documentclass[cernpreprint,atlasdraft=true,texlive=2016,UKenglish,texmf]{atlasdoc}
 
\usepackage[subfigure]{atlaspackage}
\usepackage{pdflscape}
\usepackage{multirow}
\usepackage{atlasbiblatex}
 
\usepackage{atlasphysics}
 
\graphicspath{{logos/}{figures/}}
 
\usepackage{ANA-STDM-2018-30-PAPER-defs}

\AtlasTitle{Measurements of differential
cross-sections in four-lepton events in 13~TeV proton--proton
collisions with the ATLAS detector}
 
\AtlasAbstract{
Measurements of four-lepton differential and integrated fiducial cross-sections
in events with two same-flavour, opposite-charge
electron or muon pairs are presented.
The data correspond to 139~fb$^{-1}$ of $\sqrt{s}=13$~TeV proton--proton collisions,
collected by the ATLAS detector during Run~2 of the Large Hadron
Collider (2015--2018).
The final state has contributions from a number of interesting
Standard Model processes that dominate in different  four-lepton
invariant mass regions, including single $Z$ boson production,
Higgs boson
production and on-shell $ZZ$ production, with a complex mix of
interference terms, and possible contributions from
physics beyond the Standard Model.
The
differential cross-sections include  the four-lepton invariant mass
inclusively, in slices of other
kinematic variables, and in different lepton flavour categories. Also
measured are dilepton
invariant masses, transverse momenta, and angular correlation
variables, in four regions of four-lepton invariant mass, each
dominated by different processes.
The measurements are corrected for detector effects and are compared
with state-of-the-art Standard Model calculations, which are found to
be consistent with the data. The $Z\to 4\ell$ branching fraction is
extracted, giving a value of $\left(4.41 \pm 0.30\right) \times 10^{-6}$.
Constraints on effective field theory parameters and
a model based on
a spontaneously broken $B-L$ gauge
symmetry are also evaluated.
Further reinterpretations can be performed with the provided information.
 
}
 
\author{The ATLAS Collaboration}
 
\AtlasRefCode{STDM-2018-30}
\AtlasJournalRef{JHEP 07 (2021) 005}
\AtlasDOI{10.1007/JHEP07(2021)005}
\AtlasJournal{JHEP}
\PreprintIdNumber{CERN-EP-2021-019}

\hypersetup{pdftitle={ATLAS document},pdfauthor={The ATLAS Collaboration}}
 
\begin{document}
 
\maketitle
\tableofcontents
 
\section{Introduction}
\label{sec:intro}
 
This paper
presents measurements of differential and
integrated fiducial
cross-sections of four-lepton events, containing two same-flavour,
opposite-charge electron or muon pairs.
The data used correspond to 139~\ifb{} of $\sqrt{s}=13$~\TeV\ proton--proton collisions,
collected by the ATLAS detector~\cite{PERF-2007-01, PIX-2018-001}  during Run~2 of the Large Hadron
Collider (LHC)~\cite{Evans:2008zzb} between 2015 and 2018.
 
Several interesting Standard Model (SM) processes
contribute to this final state, with the possibility of additional contributions from beyond-the-SM
(BSM) physics. The dominant SM
contribution  is from the
quark-induced $t$-channel \qqFourL{} process,
shown in
Figure~\ref{fig:feyn-qq}. Gluon-induced \ggFourL{} production
contributes at next-to-next-to-leading order (NNLO)  in
quantum chromodynamics (QCD), via a quark loop, as shown in Figure~\ref{fig:feyn-gg}.
The \ZFourL{} process, shown in Figure~\ref{fig:feyn-Z},  dominates in
the four-lepton invariant mass, \mFourL, region close to the \Z{} boson
mass, \mZ{}~\cite{Zyla:2020zbs}.
The \HZZFourL{} process, shown in Figure~\ref{fig:feyn-H} for the
gluon--gluon production mode,  dominates in
the \mFourL{} region close to the Higgs boson
mass, \mH{}~\cite{Zyla:2020zbs}. Here the superscript $(*)$ refers to a particle that can be either
on-shell or off-shell.

BSM contributions can arise from modifications to the SM couplings of the
Higgs boson, the gauge bosons and from possible four-fermion
interactions.
Contributions are also possible from models producing four leptons either via the decay
of \Z{} bosons or of new BSM particles.
For example, cascade decays of new particles  introduced by the Minimal Supersymmetric SM,
with parameters set such that
searches based on missing transverse momentum~\cite{Athron:2018vxy}
are insensitive, can nevertheless
contribute to four-lepton final states~\cite{Brooijmans:2020yij}.
Other examples include generic models with additional gauge boson(s), \Zp, which may be pair-produced and decay into lepton
pairs, or models with additional Higgs bosons which may decay into a
pair of \Z{} bosons.

\begin{figure}[htb]
\centering
\subfigure[]{\includegraphics[width = 0.21 \textwidth]{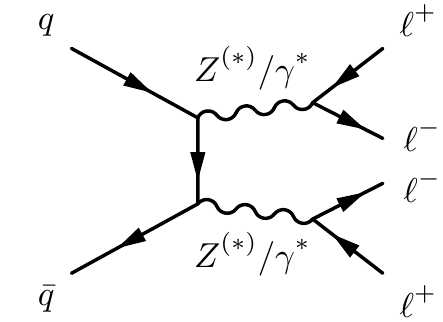}\label{fig:feyn-qq}}
\subfigure[]{\includegraphics[width = 0.27 \textwidth]{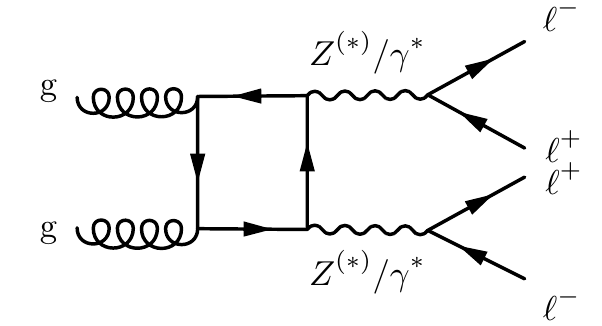}\label{fig:feyn-gg}}
\subfigure[]{\includegraphics[width = 0.21 \textwidth]{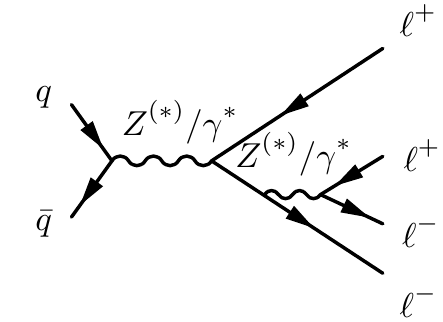}\label{fig:feyn-Z}}
\subfigure[]{\includegraphics[width = 0.23 \textwidth]{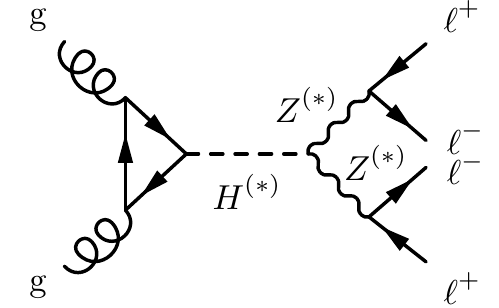}\label{fig:feyn-H}}
 
\caption{Main contributions to the $pp \to 4\ell (\ell =
e,\mu)$ process: (a) $t$-channel \qqFourL production, (b)
gluon-induced \ggFourL production via a quark loop, (c)
internal conversion in \Z{} boson decays and (d) Higgs-boson-mediated
$s$-channel production (here: gluon--gluon fusion).
The superscript $(*)$ refers to a particle that can be either
on-shell or off-shell, whereas $*$ indicates that it is
always off-shell. \label{fig:ZZproduction} }
\end{figure}

The measurements are corrected for the effects of the detector. Observables  are
defined in terms of final-state particles, rather than in terms of a particular process.
The  definition is inclusive of particles in addition to the two lepton
pairs,
including neutrinos, other leptons, hadrons, photons  and any possible BSM
particles, although the leptons are required to be isolated from other particles.
There are therefore small SM contributions from top-quark pair production in association with a
dilepton pair, from triboson processes, where at least two bosons
decay leptonically, and from events where $\tau$-leptons decay to
muons or electrons. Using this definition results in minimal
dependence of the measurement on the modelling of these other SM
processes. The dependence on SM modelling is transferred to the
theoretical predictions that are used in comparisons with the data.
 
Cross-sections are measured differentially as a
function of
various kinematic variables, and integrated fiducial cross-sections are
also provided. The primary observable, \mFourL, is
measured inclusively,  sliced
in other kinematic variables, and in different lepton flavour
categories.
Additional differential cross-sections are measured, including dilepton
invariant masses and transverse momenta, and angular correlation
variables between leptons, in four regions of \mFourL, each
dominated by different processes.
The \mFourL{} distributions and a subset of the other variables,
measured in a region where \mFourL{} is above the on-shell \Z\Z{} production threshold,
have been presented previously
at this centre-of-mass energy, using a smaller
dataset~\cite{STDM-2017-09,STDM-2016-15,CMS-SMP-16-017} and by the CMS
collaboration using a similar dataset~\cite{Sirunyan:2020pub}.
The current result is more inclusive than previous measurements, in
particular the previous requirement that the invariant mass of at least one of the
dilepton pairs be  close to \mZ{} is removed. This gives sensitivity to BSM
processes where there are sources of dilepton pairs other than \Z{}
boson decays.
The cross-section measurement in the region close to \mZ{} is used to extract the \ZFourL{} branching fraction.

BSM contributions to this final state may lead to discrepancies between the
measured cross-sections and the SM predictions. The measurements can
therefore be used to set limits on a wide range of BSM models. The
methods used to correct the data for the effects of the detector are shown to be robust against the addition of BSM contributions, as
is discussed in Section~\ref{sec:unfolding}.
Since the
cross-sections are defined at particle-level, they can be compared with
BSM theory predictions or improved SM predictions without the need
to simulate the ATLAS detector.
The data and the SM predictions are
available in HEPData~\cite{Maguire:2017ypu}, with the analysis included in the Rivet~\cite{Bierlich:2019rhm} library, allowing straightforward comparison with other predictions.

\section{ATLAS detector}
\label{sec:detector}
\newcommand{\AtlasCoordFootnote}{
ATLAS uses a right-handed coordinate system with its origin at the
nominal interaction point (IP) in the centre of the detector and the
\(z\)-axis along the beam pipe. The \(x\)-axis points from the IP to
the centre of the LHC ring, and the \(y\)-axis points
upwards. Cylindrical coordinates \((r,\phi)\) are used in the
transverse plane, \(\phi\) being the azimuthal angle around the
\(z\)-axis. The rapidity is defined as $y = (1/2)\ln\left[(E + p_z)/(E - p_z)\right]$,
where $E$ is the energy of a particle and $p_z$ is the momentum component in the beam direction.
The pseudorapidity is defined in terms of the polar angle \(\theta\) as \(\eta = -\ln \tan(\theta/2)\).
Angular distance is measured in units of \(\Delta R \equiv \sqrt{(\Delta\eta)^{2} + (\Delta\phi)^{2}}\).}
The ATLAS detector at the LHC covers nearly the
entire solid angle around the collision
point, as detailed below for each sub-detector.\footnote{\AtlasCoordFootnote} It consists of an inner tracking
detector (ID) surrounded by a thin superconducting solenoid,
electromagnetic and hadronic calorimeters, and a muon spectrometer
incorporating three large superconducting toroidal magnets.
 
The ID, immersed in a \SI{2}{\tesla} axial magnetic field,  provides charged-particle tracking for \(|\eta| < 2.5\).
It consists of a high-granularity silicon pixel detector covering the vertex region and typically provides four measurements per track.
This is followed by the silicon microstrip tracker, which usually provides eight measurements per track. These silicon detectors are complemented by the transition radiation tracker (TRT), which enables radially extended track reconstruction up to \(|\eta| = 2.0\). The TRT also provides electron identification information based on the fraction of hits (typically 30 in total) above a higher energy-deposit threshold corresponding to transition radiation.
 
The calorimeter system covers the pseudorapidity range \(|\eta| <
4.9\). Within \(|\eta|< 3.2\), electromagnetic calorimetry is provided
by barrel and endcap high-granularity lead/liquid-argon (LAr)
calorimeters, with an additional thin LAr presampler covering \(|\eta|
< 1.8\) to correct for energy loss in material upstream of the
calorimeters. Hadronic calorimetry is provided by the
steel/scintillator-tile calorimeter, segmented into three barrel
structures within \(|\eta| < 1.7\), and two copper/LAr hadronic endcap
calorimeters covering $1.5 < |\eta| < 3.2$. The solid angle coverage is completed with forward copper/LAr and tungsten/LAr calorimeter modules optimised for electromagnetic and hadronic measurements respectively.
 
The muon spectrometer (MS) comprises separate trigger and high-precision tracking chambers measuring the deflection of muons in a magnetic field generated by the superconducting air-core toroids.
A set of precision chambers covers the region \(|\eta| < 2.7\) with three layers of monitored drift tubes, complemented by cathode-strip chambers in the forward region, where the background is highest. The muon trigger system covers the region \(|\eta| < 2.4\) with resistive-plate chambers in the barrel, and thin-gap chambers in the endcap regions.
 
Interesting events are selected by the first-level trigger system implemented in custom hardware, followed by algorithms implemented in software in the high-level trigger~\cite{TRIG-2016-01}.

\section{Fiducial region definition and measured variables}
\label{sec:fidxs}
\subsection{Fiducial definition}
\label{sec:fiddef}
The fiducial phase space is defined according to the kinematic acceptance of the
detector, with kinematics that ensure a high efficiency to trigger on the events, and is designed to be as inclusive as possible while keeping
backgrounds from non-prompt\footnote{`Prompt' refers to leptons and photons that do not
originate from hadron decays. This definition of prompt excludes leptons originating from
hadronic resonances such as  $J/\psi \to \ell^{+}\ell^{-}$ and
$\Upsilon{} \to \ell^{+}\ell^{-}$~\cite{ATL-PHYS-PUB-2015-013}.} leptons relatively small.
This phase space is defined by a kinematic selection applied at
particle-level\footnote{`Particle-level' refers to a definition based on
final-state particles equivalent to the particles produced
from a Monte Carlo event-generator simulation, without simulating
the effects of the detector. The data are corrected to this level
such that they can be compared directly with theoretical predictions.} using  final-state, prompt
leptons (including those from $\tau$-lepton decays).
In order to align the particle-level definition of lepton kinematics
as closely as possible to their measurements in the detector, prompt electrons are
`dressed' by adding to their four-momenta the four-momenta of prompt
photons within a cone of size $\Delta R =  0.1$ around the electron. This is done because the experimental
measurement of electron energies in the calorimeter includes the
energy of nearby photons. If the photon is near more than one
electron, it is assigned to the nearest.
This dressing is not applied to prompt muons because their  four-momenta are
measured using the ID and the MS, and the
four-momenta of nearby photons are not included in the measurement.
Prompt leptons are required to be isolated from other particles.
Their isolation is tested by
forming a scalar sum of the transverse
momenta, \pt{}, of all charged particles
within a cone of $\Delta R  = 0.3$ around the lepton. The ratio
of this sum to the \pt{} of the lepton is required to be less
than 0.16. If another selected lepton is within the cone, the momentum of this lepton is not included in the sum.
 
Electrons (muons) are required to have $\pt > 7~(5) ~\GeV{}$ and $|\eta| <
2.47~(2.7)$.
Events are required to contain at least four such leptons that can be
grouped into at least  two
same-flavour, opposite-charge pairs. This results in three
possible flavour configurations: $e^{+}e^{-}e^{+}e^{-}$ ($4e$),
$e^{+}e^{-}\mu^{+}\mu^{-}$ ($2e2\mu$) and  $\mu^{+}\mu^{-}\mu^{+}\mu^{-}$ ($4\mu$).
Additional particles (leptons, neutrinos, photons, hadrons and
possible BSM particles) are allowed to be present in the event.
Events are also required to satisfy the following:
\begin{itemize}
\item $\pt > 20~\GeV{}$ for the leading lepton (in $\pt$).
\item $\pt > 10~\GeV{}$ for the sub-leading lepton (in $\pt$).
\item The invariant mass of any same-flavour, opposite-charge lepton
pair that can be formed in the event is required to satisfy $\mll{} > 5~\GeV$.
\item The angular separation between any two leptons in the event
is required to satisfy $\Delta R > 0.05$.
\end{itemize}
These requirements are made to minimise experimental uncertainties
and are justified in Section~\ref{sec:eventsel}.
 
\subsection{Lepton pairing}
Leptons are paired in order to define some of the measured
variables. The same-flavour, opposite-charge pair with an invariant mass
closest to \mZ{} is selected as the primary pair in the event. Of
the remaining leptons, the same-flavour, opposite-charge pair with an invariant mass
closest to \mZ{} is selected as the secondary pair, completing a
quadruplet of leptons.
Therefore, only one quadruplet is defined even in events containing more than four leptons.
This selection strategy ensures that pairs corresponding to on-shell
\Z{} bosons for the dominant \Z\Z{} pair production process
are formed preferentially.
 
\subsection{Measured variables}
\label{sec:vars}
The integrated fiducial cross-section is measured over the full
fiducial phase space  and in four \mFourL{}
regions dominated by: single \Z{} boson production ($60 <
\mFourL < 100$~\GeV), Higgs boson production ($120 <
\mFourL < 130$~\GeV), on-shell \Z\Z{} production ($180 <
\mFourL < 2000$~\GeV) and off-shell \Z\Z{} production ($20 <
\mFourL < 60$~\GeV\ or $100 <
\mFourL < 120$~\GeV\ or $130 <
\mFourL < 180$~\GeV).
 
A number of differential fiducial cross-sections are also measured,
providing  kinematic information about the events, which are each varyingly sensitive to the modelling
of the SM processes, including QCD and electroweak corrections, and to
possible BSM contributions.
The  \mFourL{} distribution is measured single-differentially as well as
double-differentially with the four-lepton transverse momentum,
\ptFourL, and the absolute four-lepton rapidity, |\yFourL|.
The single-differential \mFourL{} distribution is also measured in $4e$, $4\mu$
and $2e2\mu$ events separately.
 
The following variables are measured in the four regions of \mFourL{}
defined above:
\begin{itemize}
 
\item The invariant mass of the primary (secondary) lepton pair: $\mZOne$ ($\mZTwo$).
\item The transverse momentum of the primary (secondary) lepton pair:
$\ptZOne$ ($\ptZTwo$).
\item  A variable sensitive to the polarisation of the decaying
particle:
$\costhetastar_{12}$ $(\costhetastar_{34})$ is the cosine of the angle between the negative lepton in the  primary (secondary) dilepton
rest frame, and the primary (secondary)  lepton pair in the laboratory frame.
\item  The absolute rapidity difference between the primary and secondary lepton pairs, $\dYPairs$.
\item  The difference
in the azimuthal angle between the primary and secondary lepton pairs, $\dPhiPairs$.
\item  The difference
in the azimuthal angle between the leading lepton and sub-leading
lepton (in \pt) of
the quadruplet, $\dPhill$.

\end{itemize}

\section{Theoretical predictions and simulation}
\label{sec:theory}
 
Simulated events are used to correct the observed events for detector
effects, as well as to provide the particle-level predictions with
systematic uncertainties for comparisons with the measured data. The various
simulated samples and their uncertainties are described in this
section.

The \qqFourL{} process, including \ZFourL{}, was simulated with the
\SHERPAV{v2.2.2} event generator~\cite{Bothmann:2019yzt}.
Matrix elements were calculated at next-to-leading-order (NLO) accuracy in QCD for up to one additional parton
and at leading-order (LO) accuracy for up to three additional parton
emissions. The higher-order corrections include initial states with a gluon, but the \qqFourL{}  notation is kept for simplicity.
The calculations were matched and merged  with the \SHERPA{} parton shower based on Catani--Seymour
dipole factorisation~\cite{Gleisberg:2008fv,Schumann:2007mg}, using the MEPS@NLO
prescription~\cite{Hoeche:2011fd,Hoeche:2012yf,Catani:2001cc,Hoeche:2009rj}.
The virtual QCD corrections were provided by the
\openloops{} library~\cite{Cascioli:2011va,Denner:2016kdg}.
The
\nnpdfnnlo{} set of PDFs~\cite{Ball:2014uwa} was used, along with the
dedicated set of tuned parton-shower parameters (tune) developed by the
\SHERPA{} authors. All \SHERPAV{v2.2.2} samples discussed below use
this same PDF set and tune.
 
An alternative \qqFourL{} sample was generated at NLO accuracy in QCD
using \POWHEGBOXV{v2}~\cite{Alioli:2010xd,Melia:2011tj,Nason:2013ydw}.
Events were interfaced to \PYTHIAV{8.186}~\cite{Sjostrand:2007gs}
for the modelling of the parton shower, hadronisation, and underlying
event, with parameters set according to the AZNLO
tune~\cite{STDM-2012-23}. The \texttt{CT10} PDF set~\cite{Lai:2010vv} was used
for the hard-scattering processes, whereas the \texttt{CTEQ6L1} PDF
set~\cite{Pumplin:2002vw} was used for the parton shower.
A correction to higher-order precision, defined for this
process as the ratio of the cross-section at NNLO QCD accuracy to the
one at NLO QCD accuracy, was obtained using a \textsc{Matrix} NNLO QCD
prediction~\cite{Cascioli:2014yka,Grazzini:2015hta,Grazzini:2017mhc,Kallweit:2018nyv}, and applied  as a function of $\mFourL$.

A reweighting for virtual NLO electroweak
effects~\cite{Biedermann:2016yvs,Biedermann:2016lvg} was applied as a
function of $\mFourL$ to both \qqFourL{} samples.
A 100\%{} uncertainty was assigned to this reweighting function to account for
non-factorising effects in events with
high QCD activity~\cite{Gieseke:2014gka}.
The real higher-order electroweak contribution to $4\ell$ production
in association with two jets (which includes vector-boson scattering,
but excludes processes involving the Higgs boson)
was not included in the sample discussed above but was simulated
separately with the \SHERPAV{v2.2.2}
generator. The LO-accurate matrix elements were matched to a parton
shower based on Catani--Seymour dipole factorisation using the MEPS@NLO
prescription.

Uncertainties due to missing higher-order QCD corrections are
evaluated for both  \qqFourL{} samples~\cite{Bothmann:2016nao} using seven variations of the QCD
factorisation and renormalisation scales in the matrix elements by
factors of one half and two, avoiding variations in opposite directions.
The envelope of the effects of these variations is taken as the
uncertainty.
For the \SHERPA{} sample,
uncertainties from the choice of nominal PDF set are evaluated using 100 replica
variations, as well as by reweighting to the
alternative \ctfourteennnlo~\cite{Dulat:2015mca} and
\mmhtnnlo~\cite{Harland-Lang:2014zoa} PDF sets, and
taking the envelope of these contributions as a combined PDF uncertainty.
The effect of the uncertainty in the
strong coupling constant $\alphas$ is assessed by variations of $\pm 0.001$.
For the \POWHEG{} + \pythia{} sample, the PDF uncertainty is evaluated using the
26 pairs of upwards and downwards internal PDF variations within
\texttt{CT10 NLO}, as well as reweighting to the \nnpdfnnlo{} and MSTW2008~\cite{Martin:2009iq} PDF sets, and
taking the envelope of the variations.

The gluon-initiated $4\ell$ production process was simulated using
\SHERPAV{v2.2.2}~\cite{Cascioli:2013gfa} at LO precision for up to one additional parton emission, with the
parton-shower modelling being the same as for the \qqFourL{} sample
described above. The generator requires $\mll > 10~\GeV$ for
any same-flavour and
opposite-charge lepton pair, which removes a small
amount of the phase space included in the measurement. This is
recovered by the correction to the NLO QCD calculation described below, for all but
the \mZOne{} and \mZTwo{} distributions in the region  below $10~\GeV$, where
the prediction from this sample is missing. This is a few percent of
the total prediction in the fiducial phase space in this region.
The sample includes the  \ggFourL{} box diagram, the $s$-channel
process proceeding via a Higgs boson, \ggS{}, and the interference
between the two.
In the
region $\mFourL > 130$~\GeV\ this sample is used to simulate all
three contributions.
In the region $\mFourL < 130$~\GeV, where on-shell
Higgs boson production dominates, this sample is only
used to simulate the \ggFourL{} box diagram, with the dedicated
samples described below used to simulate Higgs boson production. In this
region the interference between the two is negligible and is not simulated.
A NLO QCD calculation~\cite{Caola2015,Caola2016}  allowing $\mFourL$ differential $K$-factors to be
calculated is used to correct
each of these contributions  separately, together with an
associated uncertainty. The
details are the same as those described in Ref.~\cite{STDM-2017-09}.
An additional correction factor of 1.2, taken from the ratio of a NNLO QCD calculation~\cite{Passarino:2013bha,deFlorian:2016spz}
to the NLO prediction for off-shell Higgs production, is assumed to be
the same for all three
components.
Scale and PDF uncertainties are obtained in the same way as for the
\SHERPA{} \qqFourL{} sample described above.

In the region $\mFourL < 130~\GeV$, where on-shell Higgs boson production is
important, and the effect of interference is negligible, dedicated
samples are used to model the Higgs boson production processes as
accurately as possible.
Higgs boson production via gluon--gluon
fusion~\cite{Campbell:2012am}, which dominates, was simulated at NNLO accuracy in QCD using the \powheg{} NNLOPS program~\cite{Hamilton:2013fea,Hamilton:2015nsa,Alioli:2010xd,Nason:2004rx,Frixione:2007vw}. The simulation achieves NNLO accuracy for arbitrary inclusive $gg\to H$ observables by reweighting the Higgs boson rapidity spectrum in \texttt{Hj-MiNLO}~\cite{Hamilton:2012np,Campbell:2012am,Hamilton:2012rf} to that of HNNLO~\cite{Catani:2007vq}.
\pythia~\cite{Sjostrand:2014zea} was used with parameters set
according to the AZNLO tune to simulate the parton shower and non-perturbative effects.
The \powheg{} prediction used the \texttt{PDF4LHC15nnlo} PDF set~\cite{Butterworth:2015oua}.
The prediction from the Monte Carlo samples is normalised to the
next-to-next-to-next-to-leading-order QCD cross-section plus electroweak corrections
at
NLO~\cite{deFlorian:2016spz,Anastasiou:2016cez,Anastasiou:2015ema,Dulat:2018rbf,Harlander:2009mq,Harlander:2009bw,Harlander:2009my,Pak:2009dg,Actis:2008ug,Actis:2008ts,Bonetti:2018ukf}.
Higgs boson production via vector-boson
fusion (VBF)~\cite{Nason:2009ai}, in association with a vector boson
(\VH)~\cite{Luisoni:2013kna}, and in association with a top-quark pair
were all simulated with \powheg~\cite{Nason:2009ai,Alioli:2010xd,Nason:2004rx,Frixione:2007vw} and interfaced with \pythia{} for
the parton shower and non-perturbative effects, with parameters set
according to the AZNLO tune. The \powheg{} prediction used the
\texttt{PDF4LHC15nlo} PDF
set~\cite{Butterworth:2015oua}.
For VBF production, the \powheg{} prediction
is accurate to NLO in QCD,
and
is normalised to an approximate-NNLO QCD cross-section with NLO electroweak
corrections~\cite{Ciccolini:2007jr,Ciccolini:2007ec,Bolzoni:2010xr}.
For \VH{} production, the \powheg{} prediction
is accurate to NLO in QCD  for up to one additional jet, and is normalised to a cross-section calculated at NNLO in QCD with NLO electroweak corrections~\cite{Ciccolini:2003jy,Brein:2003wg,Brein:2011vx,Denner:2014cla,Brein:2012ne}.
The uncertainties in on-shell Higgs boson production are the same
as reported in Ref.~\cite{HIGG-2016-22}. The largest components are from
the QCD scale and PDF uncertainties affecting the gluon--gluon fusion component.

Other SM processes making smaller contributions to the final state used in the
analysis include triboson production (\W\W\Z, \W\Z\Z{} and \Z\Z\Z{}),
collectively referred to as \V\V\V{}, and $t\bar{t}$ pairs produced in association with vector bosons ($t\bar{t}$\Z,
$t\bar{t}$\W\W), collectively referred to as $t\bar{t}$\V(\V).
The production of triboson events was simulated with
\SHERPAV{v2.2.2} using factorised gauge-boson decays.
Matrix elements, accurate to NLO for the inclusive process and to LO for up to
two additional parton emissions, were matched and merged with the \SHERPA{} parton
shower based on Catani--Seymour dipole factorisation
using the MEPS@NLO prescription.
The virtual QCD corrections for matrix elements at NLO accuracy were
provided by the \openloops{} library.
Uncertainties are evaluated in the same way as discussed above for the
\SHERPA{} \qqFourL{} sample.
Two sets of  $t\bar{t}$\V(\V)  samples are used.
A sample produced with the \madgraph~2.3.3~\cite{Alwall:2014hca} generator at NLO with the
\nnpdfnlo{}~\cite{Ball:2014uwa} PDF set is used to provide the particle-level predictions to
compare with the measurements.
The events were interfaced to \PYTHIAV{8.210}~\cite{Sjostrand:2014zea} using the A14
tune~\cite{ATL-PHYS-PUB-2014-021} and the \nnpdftwo{} PDF set~\cite{Ball:2012cx}. Uncertainties due to missing
higher-order corrections are evaluated as in the \SHERPA{} samples
discussed above. Uncertainties in the PDFs are evaluated using the 100
replicas of the \nnpdfnlo{} PDF set.
A second sample  produced with  \SHERPAV{v2.2.0} at LO accuracy, using
the MEPS@LO set-up with up to one
additional parton, is used to perform the detector corrections.
The default
\SHERPAV{v2.2.0} parton shower was used along with the
\nnpdfnnlo{} PDF set. A flat uncertainty of $\pm15\%$ is assigned to these samples to cover any differences between them and the
\madgraph{} samples, and also their uncertainties.

A small contribution from double-parton scattering, with the dilepton
pairs produced in different parton--parton interactions, is expected to
contribute at the level of 0.1\%. This is included in the definition of
the final state but neglected in the predictions due to the negligible
contribution.

To correct for detector effects, generated events were passed through a $\GEANT4$-based simulation of the
ATLAS detector and trigger~\cite{Agostinelli:2002hh, SOFT-2010-01}, and then through the same reconstruction and
analysis software as applied to the data. Corrections are applied to
simulated leptons to account for differences seen between simulation and data. These include
differences in lepton reconstruction, identification, isolation and vertex-matching efficiencies, as well as in the momentum scale and resolution, with associated uncertainties~\cite{EGAM-2018-01, PERF-2015-10}.
The effect of multiple proton--proton interactions in the same bunch
crossing, known as
pile-up, was emulated by overlaying inelastic proton--proton
collisions, simulated with \PYTHIAV{8.186}
using the \nnpdftwo{} set of PDFs and the
A3 tune~\cite{ATL-PHYS-PUB-2016-017}.
The events are then reweighted to reproduce the distribution of the
number of collisions per bunch-crossing observed in the data, which
have an average of 33.7 for the total dataset, with an
associated uncertainty coming from the uncertainty in the inelastic
cross-section.
 
\section{Data analysis}
\subsection{Event selection}
\label{sec:eventsel}
Events are selected by requiring  at
least one of a set of triggers to accept the event. Each trigger requires the
presence of one, two or three electrons
or muons satisfying a variety of \pt{} thresholds~\cite{TRIG-2018-05,Aad:2020uyd}. The trigger efficiency
increases from around 80\%{} for \mFourL{} below 80~\GeV, to nearly 100\%{} at $\mFourL{}
\sim 200$~\GeV\ and above.
Events are required to contain at least one reconstructed $pp$ collision vertex candidate with at least two
associated ID tracks. The vertex with the
largest sum of $\pt^{2}$ of tracks is considered to be the primary
interaction
vertex.

Electron identification utilises a likelihood-based method,
combining information from the shower shapes of
clusters in the electromagnetic calorimeters, properties of tracks
in the ID, and the quality of the track--cluster matching, the latter
being based on spatial separation as well the ratio of the cluster
energy to the track momentum.
The shower shape variables include variables sensitive to the lateral
and longitudinal
development of the electromagnetic shower, and variables designed to
reject clusters from multiple incident particles.
A `loose'
identification working point is used~\cite{EGAM-2018-01}, with the
additional requirement of a hit in the innermost layer of the pixel detector.
Muons are
identified using information from various combinations of the MS, the
ID and the calorimeters.
As with electrons, a `loose'
identification working point is used~\cite{PERF-2015-10}, which
focuses on
recovering efficiency in poorly instrumented detector regions. In
particular,
ID tracks identified as muons on the basis of their calorimetric energy deposits or
the presence of individual muon segments are included in the region
$|\eta|<0.1$, where the muon spectrometer is only partially
instrumented. In addition, stand-alone MS tracks, supplemented using
short tracks
in the forward pixel detector where possible, are added in the region
$2.5<|\eta|<2.7$, where ID coverage does not permit full independent
ID track reconstruction.
 
The kinematic requirements described in Section~\ref{sec:fidxs} for the
particle-level selection are also applied to these
reconstruction-level\footnote{`Reconstruction-level' refers to the
identification and kinematic measurements of
final-state object candidates,
as defined by measurements with the
detector, or full detector simulation, and a subsequent reconstruction
software step.} electrons and muons, ensuring very little
extrapolation into unmeasured regions when correcting for detector
effects. The leading and sub-leading leptons must have $\pt > 20$~\GeV\
and $> 10$~\GeV, respectively, to ensure efficient triggering of events. The requirements of
$\mll{} > 5~\GeV$ and  $\Delta R > 0.05$ between the leptons suppress
contributions from $J/\psi$ decays and conversion electrons
respectively.
To select leptons originating from the primary proton--proton
interaction, their tracks are required to have a longitudinal impact
parameter, $z_{0}$, satisfying $|z_{0} \sin(\theta)| < 0.5$~mm from the primary
interaction vertex.  For MS-only muons, no such requirement is made.
To avoid the double-counting of particles, if leptons share an ID
track, only one survives the selection. Preference is given to higher-\pt{} leptons
and to muons over electrons, unless the muon has no
associated MS track, in which case the electron survives.
The leptons satisfying the above criteria are referred to as
\emph{baseline} leptons and are used to form the quadruplet, as detailed in Section~\ref{sec:fidxs}.
Once the quadruplet is formed the following further selection
requirements are made.
The electrons and muons are required to be isolated from other
particles using information from the ID and the calorimeters.
The isolation
variables form a ratio of the scalar sum of the transverse
momenta of all charged particle tracks within a lepton-\pt{}
dependent cone size, with an additional contribution from calorimeter
measurements, to the \pt{} of the lepton. The variables
have a correction for pile-up, and another correction for
tracks or energy deposits originating from other leptons in the event,
in order to retain events with closely spaced prompt leptons.
Additionally, the transverse impact parameter significance,
$|d_{0}|/\sigma_{d_{0}} $, calculated relative to the measured beam-line position
must be less than 5 (3) for electrons
(muons).
The leptons satisfying the above criteria are referred to as
\emph{signal} leptons, and they are a subset of the \emph{baseline} leptons.
 
The overall efficiency to reconstruct, identify, isolate and
vertex-match electrons varies from 30\%{} at low \pt{} and high $|\eta|$ to 98\%{} at high \pt{}. For muons it varies  from
30\%{} at low \pt{} and  high $|\eta|$ to more than 99\%{} at
high \pt{}, and is higher on average than for electrons~\cite{Aad:2020gmm}.

\subsection{Background estimation}
\label{sec:bkg}
 
Backgrounds where one or more of the reconstructed leptons
entering a quadruplet did not originate from a prompt lepton
(non-prompt leptons) are
estimated using data-driven methods rather than simulation.
The backgrounds are subtracted from
the data prior to correcting for detector effects.
Simulations suggest that the main source of  background
events is where the quadruplet is formed by a combination of prompt leptons from $\Z /
\gamma^*$ or $t\bar{t}$ production processes, with additional leptons
originating from the decay of hadrons.
 
The background is estimated with
a \emph{fake factor} method.
The method uses three classes of leptons, the  \emph{signal} and
\emph{baseline} leptons defined in Section~\ref{sec:eventsel}, as well as
\emph{baseline-not-signal} leptons, defined as those which pass the baseline selection but fail the signal selection.
A quantity called the \emph{fake factor} is defined as the ratio of
the number of signal leptons to the number of baseline-not-signal
leptons and measured in a dedicated control region in data, which is enriched in
non-prompt leptons.
The background
yield is estimated by applying the fake factor to each
baseline-not-signal lepton in events passing an event selection
requiring only baseline instead of signal leptons. Four-lepton events
containing prompt baseline-not-signal leptons are removed from the
estimation by using simulation predictions.
 
The control region in which the fake factor is evaluated is defined by
selecting events containing a same-flavour and
opposite-charge lepton pair, with an invariant mass within $15~\gev$ of
\mZ, and at least one other baseline lepton.  The leptons from this
pair are required to have triggered the event.
Events with $\Z\rightarrow\ell\ell$ decays compose 90\%{} of this sample.
The remaining leptons in an event that do not form the candidate \Z{} pair are
likely to be non-prompt and to have originated from hadron decays or,
to a lesser extent, from jets misidentified as leptons.
There is a small contribution of prompt leptons, primarily from \W\Z{}
decays, and an even smaller contribution from four-prompt-lepton events. These contributions are subtracted using \SHERPAV{v2.2.2} simulations.
After this subtraction, the fake factor is measured in bins of lepton \pt{} and the number of jets in the event.
 
An important assumption of the fake-factor method is that the
probability of any given lepton to be prompt is uncorrelated with the
equivalent probabilities for other leptons in the event. Since this
analysis accepts leptons which are separated by as little as $\Delta R
= 0.05$, this assumption can break down. For example, cascade decays
of $b$-hadrons
can lead to two or more leptons being close to each other. In such cases, the leptons' prompt/non-prompt probabilities are highly correlated.
To account for this, if two leptons are found to be within $\Delta R
= 0.4$ of each other, one of them is omitted when determining the fake contribution expected from that event.
The choice of which lepton to omit depends on the lepton flavour and \pt.
If the leptons are different flavours, the electron is omitted, and if
the leptons are of the same flavour, the one with the lower
transverse momentum is omitted.
This approach was validated in  closure tests of the method
using simulated events.
 
In order to limit the impact of statistical fluctuations in the
non-prompt background, a smoothing
procedure
based on  a variable-span smoothing technique~\cite{Friedman:155846} is
used to obtain a smoother background shape and reduce the impact of
single outlier events. The overall predicted background  is 4\%{} of
the selected data in the signal region,
and falls quickly from 38\%{} for
\mFourL{} between 20 and 60~\GeV\ to around 2.5\%{} for \mFourL{} above 200~\GeV.
 
Five sources of uncertainty in the background yield are
considered.
First, the dominant uncertainty for $\mFourL < 150~\GeV$ and $\mFourL >  350~\GeV$
comes from the statistical uncertainty of the number
of events with four baseline leptons, used as input to the fake-factor method.
Second, dominant in the region  $150 < \mFourL < 350~\GeV$, are the theory uncertainties of the prompt-lepton subtraction in the control
region, predominantly from \W\Z{} events, which are dominated by QCD
scale variations.
Third, the uncertainties in the subtracted contribution of genuine
four-prompt-lepton events containing baseline-not-signal leptons
(which is estimated from simulation) are propagated to the background
estimate, which is a subdominant contribution.
Fourth is the statistical uncertainty of the data in the control
region used to measure the fake factors,
which is also subdominant.
Finally, a very small uncertainty in the smoothing technique is included.
The total size of the uncertainty in the predicted number of
background events varies between $15\%$ for
\mFourL{} between 20 and 60~\GeV\ and
$150\%$ for \mFourL{} of 1000~\GeV.
 
The background estimation was validated by comparing the prediction
of the method
to data in dedicated background-enriched phase-space regions.
The first validation region is the  `different-flavour validation
region', which is defined to be like the signal region,
but requires the leptons of
one of the pairs to have different flavours.
The second is the `same-charge validation region', which instead requires the leptons in one of the pairs to have the same charge.
After application of the estimated background yield, the
data and simulation in both validation regions are in agreement within
the statistical uncertainties of the data, as shown in
Figure~\ref{fig:validation} for the \mFourL{} observable.

\begin{figure}[h]
\centering
\subfigure[]{\includegraphics[width = 0.49 \textwidth]{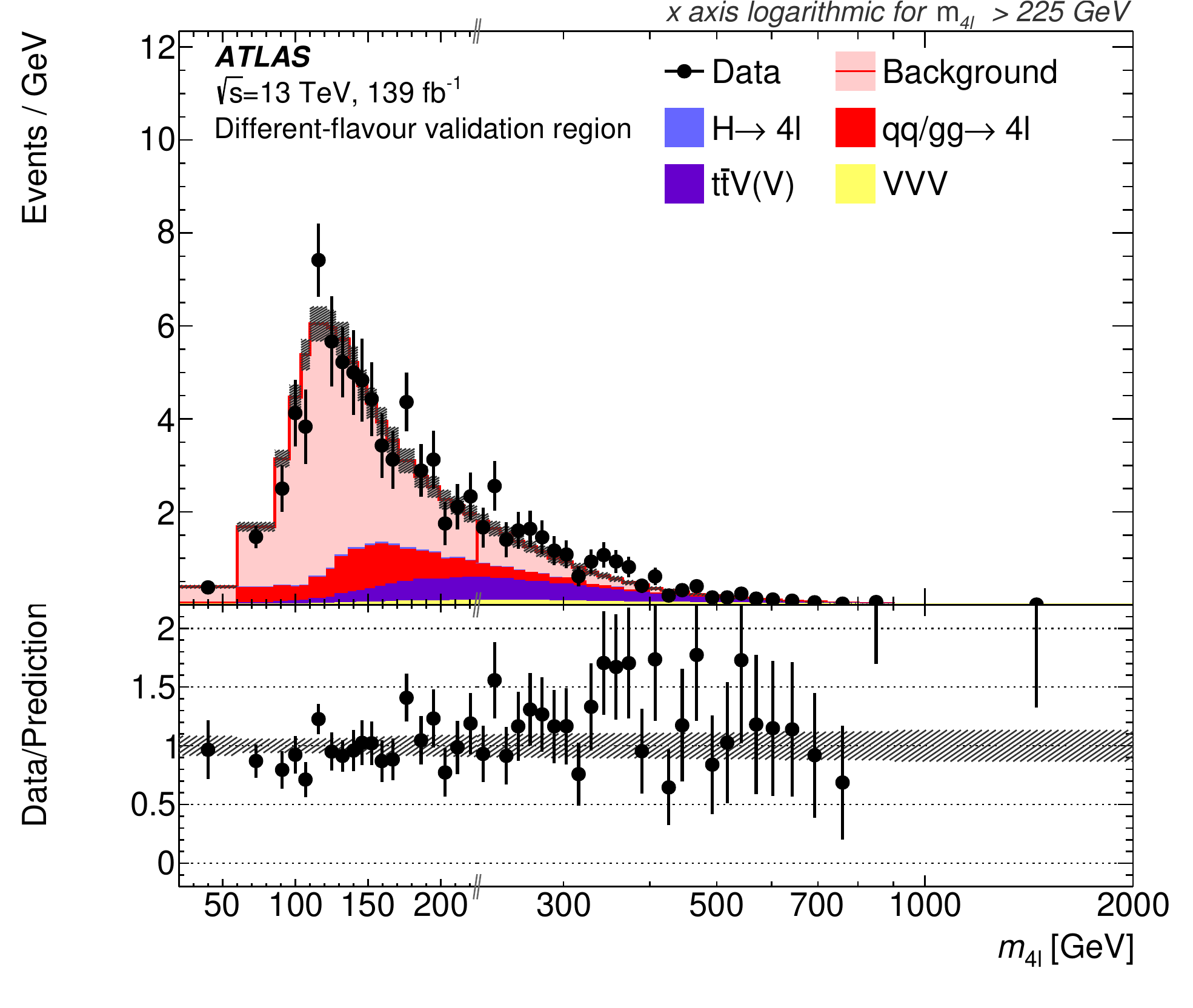}}
\subfigure[]{\includegraphics[width = 0.49 \textwidth]{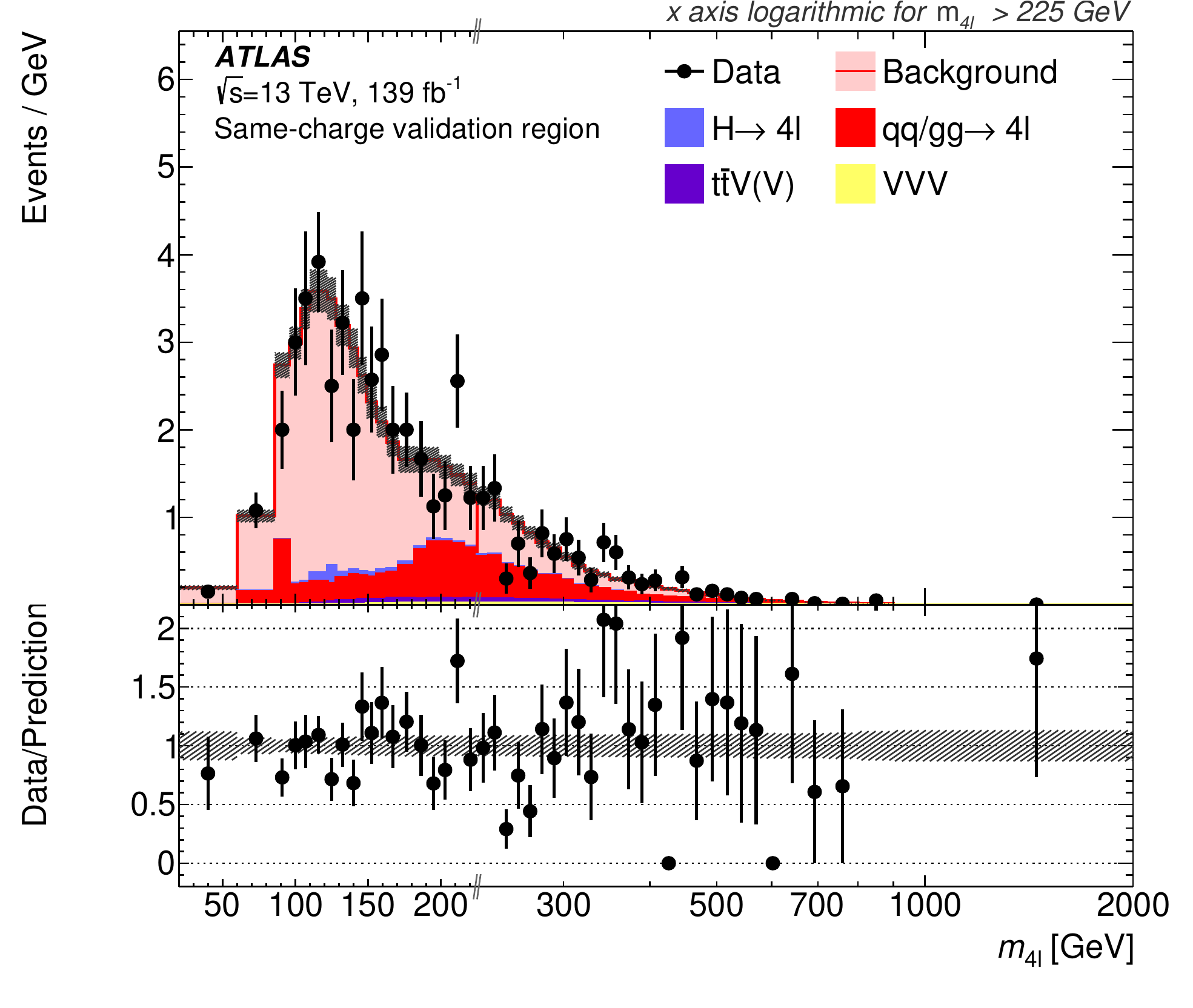}} \\
 
\caption{Observed \mFourL{}  distribution compared to the
data-driven estimation of the
background from non-prompt leptons
in the (a) different-flavour and (b) same-charge validation
regions. Contributions from events with four prompt leptons,
obtained from simulation, are also shown. The hashed band
represents the total uncertainty on the background estimation.
The $x$-axis is on a linear scale until $\mFourL = 225$~\GeV,
where it switches to a logarithmic scale.
\label{fig:validation}}
\end{figure}

The fake factor method does not account for a very small background contribution from
$\Z{} + \Upsilon{}$ events. These events amount to 0.2\%{} of the selected
data, and 0.8\%{} in the \ZFourL{} region, and are removed prior to
correcting for detector effects, using an
estimate from \pythia{}
simulation.

\subsection{Selected events}
\label{sec:events}
Table~\ref{tab:RecoYieldTablePerProcess} shows the number of selected
events over the full fiducial phase space and in each \mFourL{} region. Also shown
is the predicted contribution from each SM process contributing to the
final state, as well as the
predicted background contribution from non-prompt leptons.
The \SHERPA{} simulation is used for the \qqFourL{} process.
Uncertainties in the predictions arise from the
sources discussed in Sections~\ref{sec:theory},~\ref{sec:bkg}
and~\ref{sec:unc}.
 
\begin{table}
\centering
\caption{Predicted reconstruction-level yields per process and in total,
compared with observed data counts, over the full fiducial phase space and in the
following regions of
$\mFourL$: \ZFourL{}  ($60 < \mFourL < 100$~\GeV), \HFourL{}  ($120 <
\mFourL < 130$~\GeV), off-shell $\Z\Z$  ($20 <
\mFourL < 60$~\GeV\ or $100 <
\mFourL < 120$~\GeV\ or $130 <
\mFourL < 180$~\GeV) and  on-shell \Z\Z{} ($180 <
\mFourL < 2000$~\GeV).
Uncertainties in the predictions include both the
statistical and systematic sources. The uncertainty in the total
prediction takes into account correlations between
processes, and therefore contributions in a given column do not
trivially add up in quadrature to give the total.
The background row is events with non-prompt leptons,
including those from $\Z{} + \Upsilon{}$ events.
The \HFourL{} row includes only the
on-shell Higgs boson contribution, with off-shell contributions included in
\ggFourL{}. \label{tab:RecoYieldTablePerProcess} }
\begin{tabular} {l  c  c  c  c  c }
\hline
& \multicolumn{5}{c}{Region} \\
& Full   & \ZFourL{}  & \HFourL{}  & Off-shell $\Z\Z$  & On-shell $\Z\Z$   \\
 
\hline
\qqFourL{} & $6100 \pm 500$  & $1490 \pm 120$  & $128 \pm 10$  & $800 \pm 60$  & $3640 \pm 280$ \\
\ggFourL{} & $680 \pm 90$  & $10.8 \pm 2.9$  & ~~$3.9 \pm 0.7$  & $49 \pm 6$  & $620 \pm 80$ \\
\HFourL{}  & $245 \pm 20$  & ~~$2.16 \pm 0.18$  & $207 \pm 17$  & $33.5 \pm 3.1$  & $1.98 \pm 0.20$ \\
\V\V\V{}  & $35 \pm 4$  & ~~$0.018 \pm 0.005$  & ~~$0.127 \pm 0.018$  & ~~$2.05 \pm 0.22$  & $32.9 \pm 3.4$ \\
$t\bar{t}$\V(\V) & $123 \pm 19$  & ~~$1.37 \pm 0.22$  & ~~$1.2 \pm 0.2$  & $15.5 \pm 2.4$  & $105 \pm 16$ \\
Background & $330 \pm 50$  & $44 \pm 8$  & $26 \pm 5$  & $129 \pm 19$  & $139 \pm 30$ \\
\hline
Total Pred. & $7500 \pm 500$  & $1540 \pm 110$  & $367 \pm 19$  & $1030 \pm 60$~~  & $4530 \pm 290$ \\
\hline
Data & $7755 $  & $1452 $  & $379 $  & $1095 $  & $4828 $ \\
\hline
\end{tabular}
\end{table}

Figure~\ref{fig:recoresults1} shows the \mFourL{} distribution,
comparing data with these predictions at
reconstruction-level, together with the uncertainties.
The distribution shows a number of interesting features. There is a
peak in the region $\mFourL \sim \mZ$, dominated by the \ZFourL{}
process, and a peak in the region  $\mFourL \sim \mH$, dominated by the \HFourL{}
processes. At the threshold for producing two on-shell \Z{} bosons,
$\mFourL \sim
180$~GeV, there is an increase in the cross-section. The cross-section then falls steeply.
\begin{figure}[h]
\centering
\includegraphics[width = 0.7\textwidth]{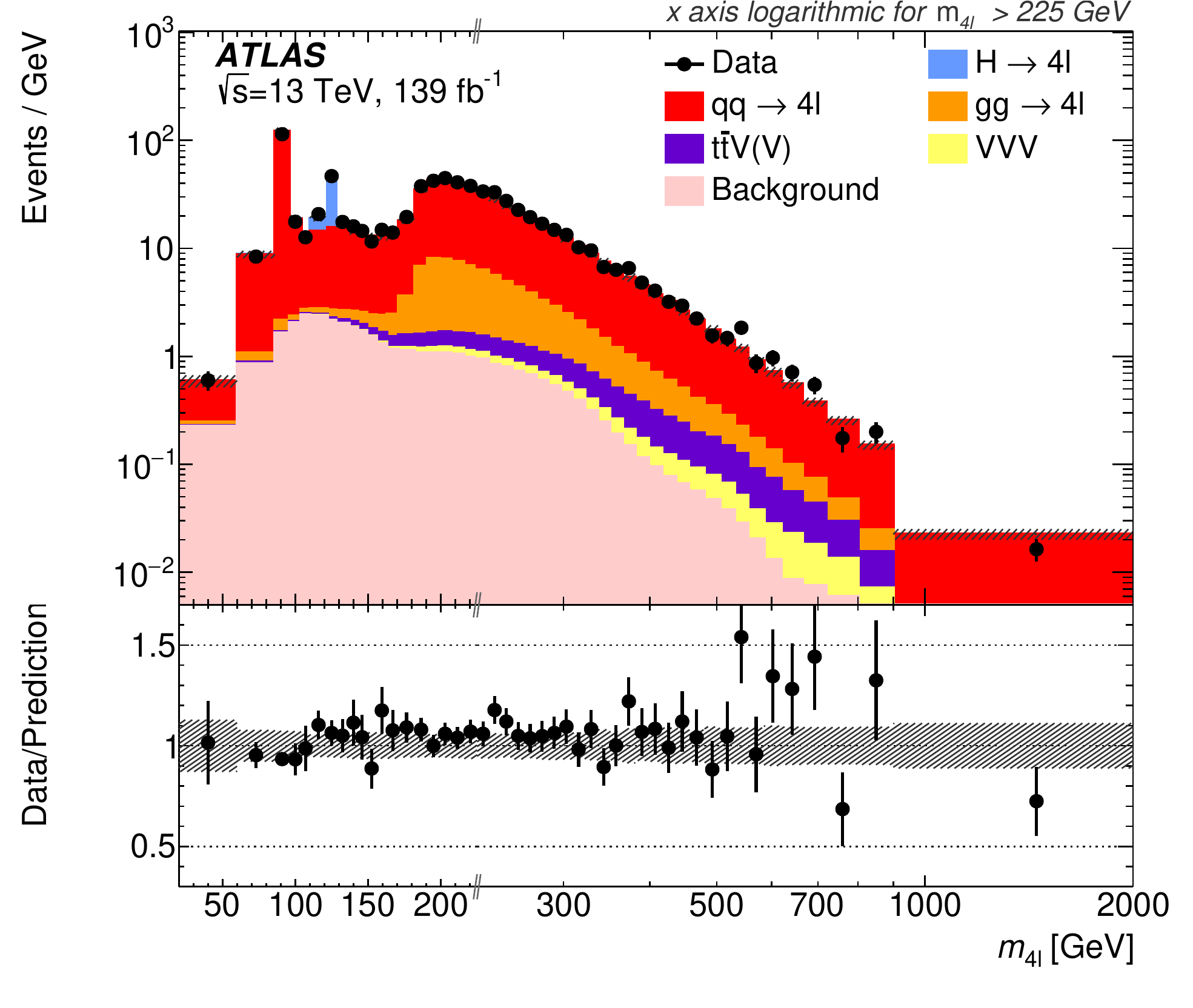}
\caption{Observed reconstruction-level \mFourL{} distribution compared with the SM prediction, using
\SHERPA{} for the \qqFourL{} simulation.
The statistical uncertainty of the
data is displayed as error bars and systematic uncertainties
in the prediction are shown as a grey hashed band.  The
ratio of the data to the prediction is shown in the lower
panel.  The $x$-axis is on a linear scale until $\mFourL = 225$~\GeV,
where it switches to a logarithmic scale, as indicated by the double
dashes on the axis.
There is one additional data event reconstructed with
$\mFourL = 2.14$~\TeV, while 0.4 events are expected from simulation for
$\mFourL > 2$~\TeV. \label{fig:recoresults1}}
\end{figure}
\subsection{Detector corrections}
\label{sec:unfolding}
 
After subtracting the background due to non-prompt leptons, the
measured event yields are corrected for detector effects
using a combination of a per-lepton efficiency correction and an
iterative Bayesian unfolding technique~\cite{DAgostini:1994fjx}. The
detector effects include the resolution of the measured kinematic variables and the inefficiencies of reconstructing leptons and triggering on the events.
The sum of the SM  simulations described
in Section~\ref{sec:theory} are used to provide the relationship
between the
particle-level observables defined in Section~\ref{sec:fidxs} and the
reconstruction-level observables defined in Section~\ref{sec:eventsel}.
 
The first step is a correction for the reconstruction,
identification, isolation and vertex-matching efficiency of each lepton in
the quadruplet, which is referred to as a
pre-unfolding efficiency correction. The efficiency is measured in the
simulation as a function of $\eta$
and \pt{} for electrons and muons, treating those from $\tau$ decays
separately, due to a lower $|d_{0}|/\sigma_{d_{0}} $ efficiency.
It is defined as the
ratio of the number of reconstruction-level leptons (that are matched
with $\Delta R < 0.05$ to particle-level leptons) to the number of
particle-level  leptons.
A per-event weight is given
by: $\prod\limits_{i=1}^4 \left[\epsilon_i \left(p_{\mathrm{T}i},\eta_i\right)\right]^{-1}$, where
$\epsilon_i\left(p_{\mathrm{T}i},\eta_i\right) $ is the efficiency for the $i^{th}$ lepton in the
quadruplet,
treated as uncorrelated between leptons.
The efficiency for a given lepton-flavour is obtained from
a weighted-average of the efficiency from leptons that originate from $\tau$
decays and those that do not, using the \mFourL{}-dependent admixture
expected from the simulation.
The per-event weight defined above  is applied to events in both
data and simulation, taking the $\eta$ and \pt{} values from the
leptons in data and
simulation respectively. This ensures there is minimal dependence on the SM description of the
lepton kinematics when correcting the data.

The data are then corrected for events that pass the
reconstruction-level selection but fail the particle-level selection.
This primarily occurs due to resolution effects, and is corrected by a multiplicative
factor, known as a fiducial correction, in each bin of each distribution.
Then, the iterative Bayesian
procedure is applied using the SM particle-level distribution as the
initial prior. The data are unfolded using a
migration matrix containing probabilities that an event in a given
particle-level bin of a distribution is found in a particular
reconstruction-level bin of that distribution. This matrix is formed
from events that pass both the particle-level and reconstruction-level
event selections.
This process is
iterated with the prior being replaced by the unfolded data from the previous
iteration: three iterations are used for all distributions, with the exception of
$\mZOne$, $\dYPairs$ and $\dPhill$ where only two iterations are
applied.
The iterations chosen represent the optimal compromise between the increase of the statistical uncertainty with too many iterations, and the increase of the residual bias due to over-regularisation with too few iterations.
 
The final step is to divide the resulting unfolded
distribution by the ratio of the number of events passing
both particle- and reconstruction-level selections to the number passing
the particle-level selections. This is known as an
efficiency correction. Since the reconstruction-level
selections have already had the pre-unfolding weights applied, this
correction is fairly close to one, but it accounts for any residual effects
such as resolution and trigger efficiencies.
 
For sliced variables the distributions are unfolded
simultaneously, properly taking into account the  migration between regions.
The integrated cross-section is obtained by correcting the total number
of observed events, after background subtraction and application of pre-unfolding
weights, with the fiducial and efficiency corrections calculated for
the inclusive phase space. When unfolding the per-region
cross-sections the inter-region migrations are accounted for.

The binning of each distribution is driven by the requirement that
the fraction of events in a reconstruction-level bin that originate from the same bin at particle-level
is at least 60\%{} (70\%{}, 80\%{}) if 25 (20, 14) or more
events are predicted for the reconstruction-level yield of the bin.
At least 14 events must be expected in each bin.
These requirements ensure that the number of events is approximately
Gaussian distributed.
There is an additional
constraint that bins should be centred on the various resonant peaks in the distributions.

The unfolding method is designed to minimise the dependence on the
simulation of the underlying kinematics of the particles. To validate this, a data-driven closure test is
performed, where a simulated pseudo-data sample is formed by
reweighting the simulation such that the distribution agrees with the
data. This pseudo-data sample is then unfolded with the nominal simulation
and the resulting unfolded distribution is compared with the input
reweighted particle-level prediction. They are in good agreement with each other, and the
small differences, averaging much less than 1\%{} but reaching 3\%{}
in a few bins in some distributions, are taken as a systematic uncertainty of the final result.

In order to demonstrate that these measurements are robust against the
presence of BSM physics in the data and can be used to  constrain BSM models, a number
of BSM signal injection tests are performed.
Pseudo-data consisting
of SM+BSM simulations are unfolded with the nominal SM simulation
and the result is compared with the particle-level SM+BSM
simulation. Any differences are interpreted as a bias in the unfolded
SM+BSM result.
Overall, the results are
found to be extremely robust with only small differences seen, all of
them within the experimental uncertainties.
Models that predict a broad excess over the SM prediction, such
as wide resonances and modifications to SM couplings, lead to
very small biases in the  unfolded distributions.
As an example, the addition of a heavy Higgs boson, with
various Higgs masses ranging from 300~\GeV\ to 1400~\GeV, and a width of
15\%{} of the mass, is studied. The
cross-section of the process is scaled such that the change relative to the SM
prediction is equivalent to 2$\sigma$ of the data uncertainty. The
bias in the unfolded result is always less
than 20\%{} of the
total experimental uncertainty in any given bin, with most cases and bins being
considerably less affected.
Without the pre-unfolding corrections applied the
bias is up to a factor of two larger, indicating that the
pre-unfolding step improves the robustness of the unfolding.
The same tests are performed with narrow-width heavy Higgs bosons,
which lead to predicted enhancements in a single bin only. These tests
lead to slightly larger biases due to the more drastic changes in the predicted
shapes of distributions. In this case the bias observed is
still always within experimental uncertainties, with the  maximum being
50\%{} of the total experimental uncertainty (which is dominated by
the statistical component) in any given bin. This
bias  slightly reduces the cross-section in the bin of interest. This implies that limits placed on narrow resonances will be
slightly more stringent, or that claims of an excess would be slightly less
significant, than they would be without the bias.

\subsection{Uncertainties}
\label{sec:unc}
The statistical uncertainty of the data is dominant for the vast
majority of the differential cross-section bins. It is estimated with
two approaches: the first is a model-independent approach, using the
observed number of events, which is used as the quoted uncertainty in the measurements, and the
second uses the expected number of SM events, which is appropriate
when testing the observed cross-section against the SM prediction.
In the first approach,
3500 pseudo-datasets are generated by assigning random Poisson-distributed
weights of mean one to the data events, and taking the root mean square of the differences
between all the unfolded results obtained using the  pseudo-datasets.
The statistical uncertainties obtained in this way are equivalent to
frequentist confidence intervals in the large-sample limit, while
in the bins with few entries, the quoted bands are
known to be up to 10\%{} narrower than a frequentist
confidence interval.
In the second approach, 3500 Poisson-distributed pseudo-datasets are generated with a mean equal to the predicted reconstruction-level SM event yield in each bin,  unfolding each one and taking the root mean square of the differences between all the unfolded results.

The systematic uncertainties of the measured cross-sections are evaluated by
repeating the measurement after applying each associated variation and
comparing the unfolded result with the nominal one. Significant contributions
arise from uncertainties in the lepton reconstruction, identification,
isolation and track-to-vertex matching efficiencies, and momentum
resolution and scale. These uncertainties are derived from the
data-driven measurements used to determine the factors
applied to the simulation~\cite{EGAM-2018-01, PERF-2015-10}, as discussed in
Section~\ref{sec:theory}. Another important source of uncertainty
arises from the choice of generator in the simulation of the
\qqFourL{} process (the nominal \SHERPA{} prediction and the alternative
\POWHEG{} + \pythia{} prediction, both introduced in
Section~\ref{sec:theory}) used to unfold the results.
The predominant effect on the measured distributions comes from a known difference
between the two generators' modelling of the final-state radiation of photons.
When this difference is evaluated the \POWHEG{} + \pythia{} prediction is first
reweighted to match the  \SHERPA{} prediction for the distribution
being unfolded. This is to avoid double counting with the data-driven
closure test, described in Section~\ref{sec:unfolding}.
Particularly important in the tails is the uncertainty in the estimate of the background
from non-prompt leptons, as described in Section~\ref{sec:bkg}. A flat uncertainty of $\pm$1.7$\%$ is assigned as a result of the uncertainty in determining the luminosity for the Run~2 dataset~\cite{ATLAS-CONF-2019-021,LUCID2}.
Smaller uncertainties come from the slight non-closure in the data-driven test for the
unfolding, described in Section~\ref{sec:unfolding}, the statistical
uncertainty of the simulated samples, the  uncertainty in the inelastic
cross-section, and the other theory uncertainties, the last two both introduced in
Section~\ref{sec:theory}. The theory uncertainties have small effects
on the unfolding due to changes in the shape and normalisation of the
various SM contributions. They have a much larger effect on the
particle-level predictions that are compared with the data.
The dominant uncertainty comes from the scale variations for each
process.
Figure~\ref{fig:m4lunfoldedsys} shows the breakdown of the
uncertainties for the measured \mFourL{} distribution.
The statistical uncertainty of the data is the dominant source of
uncertainty in all but the third mass bin (at $\mFourL \approx
m_{Z}$), where the uncertainty in lepton efficiencies dominates.
The generator uncertainty shows some bin-to-bin fluctuations due to the
limited Monte Carlo sample sizes when comparing the unfolding performed
with the \SHERPA{} and
\POWHEG{} + \pythia{} predictions, as well as some real features due
to the \mFourL{}-dependence  of the differences in the generators'
modelling of the final-state radiation of photons.
 
\begin{figure}[hp]
\centering
\includegraphics[width = 0.65 \textwidth]{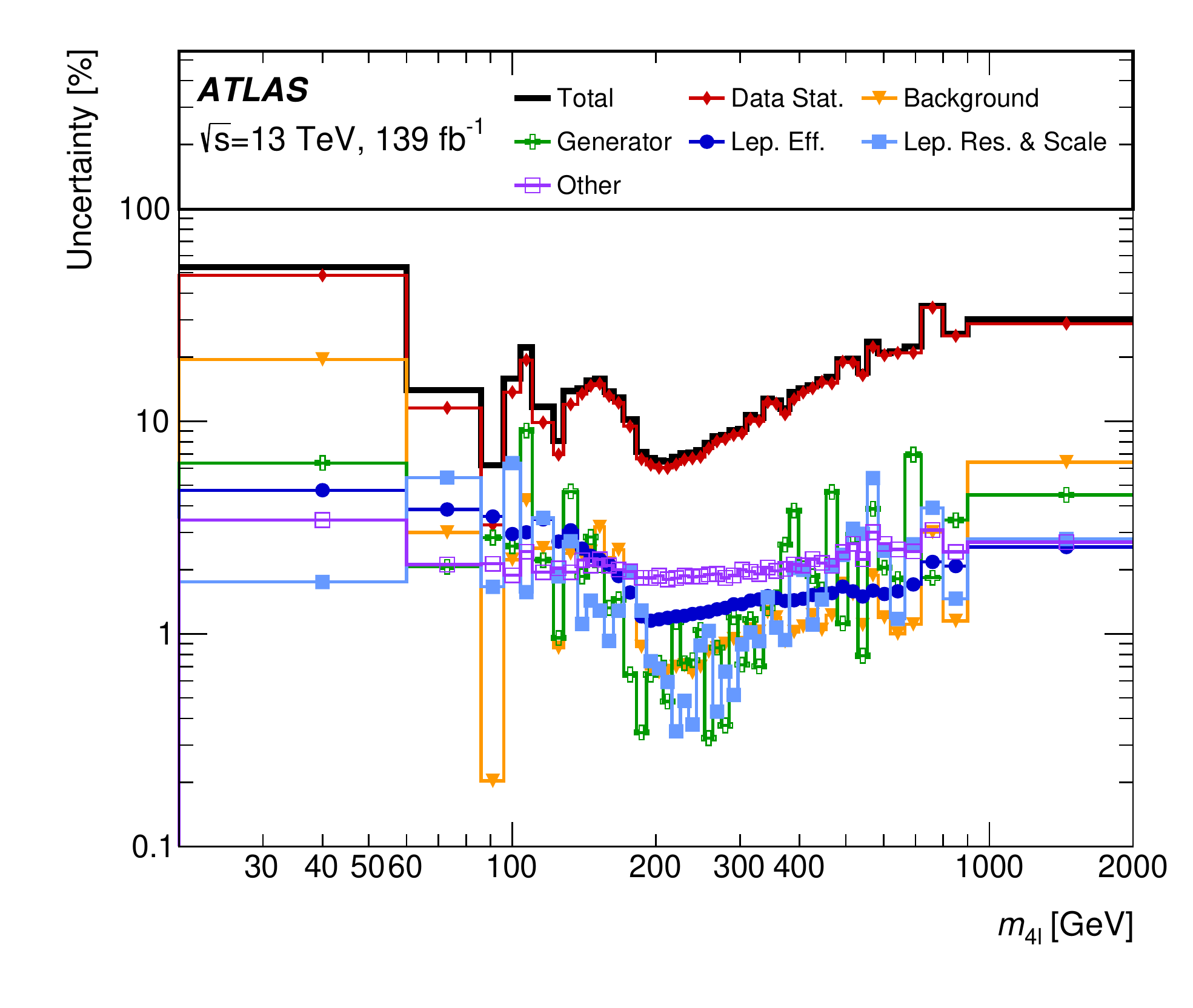}
\caption{Uncertainty breakdown of the measured cross-section as a
function of $\mFourL$. ``Lep.\ Eff.'' refers to the uncertainties in
the lepton efficiencies, ``Lep.\ Res.\ \& Scale'' refers to the
uncertainties in the lepton resolutions and scales, and the theory
uncertainties are included in the ``Generator''
uncertainty. Contributions from the Monte Carlo statistical
uncertainty and uncertainties in the
luminosity and the inelastic cross-section
are included in ``Other''.\label{fig:m4lunfoldedsys}}
\end{figure}
 
The covariance matrices for the  statistical and systematic
uncertainties are available in HEPData for each distribution.
The dominant parts of the lepton
uncertainties are highly correlated across bins. The background
uncertainties are mostly uncorrelated as they are driven by limited
sample size. The systematic uncertainty arising from the choice of generator is highly correlated
across bins.

For the  theoretical uncertainties of the particle-level predictions,
the uncertainties from each individual process are treated as
uncorrelated and an additional decorrelation of the scale  uncertainty
between the four  \mFourL{} regions for the \qqFourL{} process is introduced. This strategy is motivated by the fact that
each region probes a very different interaction scale, and a
different type of process is dominant. This correlation scheme also
leads to the most conservative results when interpreting the data.

\section{Results}
\label{sec:results}
 
\subsection{Measurements}
\label{sec:xsecs}
Table~\ref{tab:fidxs} gives the measured cross-sections in the full
fiducial phase space
and in the four \mFourL{} regions, each dominated by a different
process, compared with the
theoretical predictions described in Section~\ref{sec:theory}. Two
predictions are shown, one where the \qqFourL{} process is simulated
with \SHERPA{} at NLO accuracy in  QCD and one where it is simulated with \POWHEG{} +
\pythia{} normalised to a prediction at NNLO accuracy in QCD, as
described in Section~\ref{sec:theory}. All the other SM processes are the same in the two
predictions.
The \SHERPA{} prediction is generally higher than the \POWHEG{} +
\pythia{} prediction in all but the on-shell  region, where the
predictions are very close.
The cross-sections measured in data generally agree with both predictions
within the quoted uncertainties.
The data central values are above the  \POWHEG{} +
\pythia{} predictions in all regions, and
in all but the $\Z\rightarrow 4\ell$ region for \SHERPA. In the
on-shell region the \SHERPA{} prediction
is a bit more than $1\sigma$ below the data.
In Ref.~\cite{ATLAS:2020wny} the \HFourL{}  cross-section is measured
by ATLAS
in a fiducial phase space that differs slightly from the \HFourL{} region measured here.
The phase space is designed to minimise the contribution
from non-\HFourL{} processes.
In the dedicated Higgs boson measurement the cross-section is
found to be slightly below the SM prediction. The dedicated Higgs
boson measurement
differs from the present measurement in using a slightly different phase space, in subtracting
non-Higgs boson processes using a data-driven approach, and in including a $\sim1\%$
contribution from  Higgs boson production in association with a $b$-quark pair in the prediction.

\begin{table}[t]
\centering
\caption{Fiducial cross-sections in fb in the full fiducial phase space and in the
following regions of
$\mFourL$: \ZFourL{}  ($60 < \mFourL < 100$~\GeV), \HFourL{}  ($120 <
\mFourL < 130$~\GeV), off-shell $\Z\Z$  ($20 <
\mFourL < 60$~\GeV\ or $100 <
\mFourL < 120$~\GeV\ or $130 <
\mFourL < 180$~\GeV) and  on-shell \Z\Z{} ($180 <
\mFourL < 2000$~\GeV), compared with particle-level predictions and their
uncertainties as described in Section~\ref{sec:theory}. Two
predictions are shown for the
\qqFourL{} process simulated with
\SHERPA{} or with \POWHEG{} + \pythia{}. All other SM processes are the
same for the two predictions. \label{tab:fidxs} }
\begin{tabular} {c c c c c c }
\hline
& \multicolumn{5}{c}{Region} \\
& Full   & $Z\rightarrow 4\ell$  & \HFourL{}  & Off-shell $ZZ$  & On-shell $ZZ$   \\
\hline
Measured        & 88.9              & 22.1              & 4.76                & 12.4                & 49.3 \\
fiducial & $\pm$1.1 (stat.\,)    & $\pm$0.7 (stat.\,)    &  $\pm$0.29 (stat.\,)  & $\pm$0.5 (stat.\,)     & $\pm$0.8 (stat.\,) \\
cross-section        & $\pm$2.3 (syst.\,)    & $\pm$1.1 (syst.\,)    &   $\pm$0.18 (syst.\,) & $\pm$0.6 (syst.\,)     & $\pm$0.8 (syst.\,) \\
$[$fb$]$ 			         & $\pm$1.5 (lumi.)    & $\pm$0.4  (lumi.)  & $\pm$0.08 (lumi.)	   & $\pm$0.2 (lumi.)	   &   $\pm$0.8 (lumi.) \\
& $\pm$3.0 (total\,)   & $\pm$1.3 (total\,)   &   $\pm$0.35  (total\,)   & $\pm$0.8 (total\,)    &   $\pm$1.3 (total\,) \\
\hline
\SHERPA{}                            & 86$\pm$5          & 23.6$\pm$1.5      & 4.57$\pm$0.21       & 11.5$\pm$0.7       & 46.0$\pm$2.9 \\
\POWHEG + \pythia{}         & 83$\pm$5          & 21.2$\pm$1.3      & 4.38$\pm$0.20       & 10.7$\pm$0.7       & 46.4$\pm$3.0 \\
\hline
\end{tabular}
\end{table}
The differential cross-section as a function of  \mFourL{} is shown in
Figure~\ref{fig:cross-sec-m4l}, in much
finer bins than those in Table~\ref{tab:fidxs}.
The breakdown of the contribution from different SM processes is also
shown.
The features seen in the reconstruction-level distribution in
Figure~\ref{fig:recoresults1} are also
present here.
The SM predictions agree well with the measurement within
uncertainties over the
entire \mFourL{} spectrum, with the same features seen as in the
comparisons in Table~\ref{tab:fidxs}.
For this distribution, and all the others shown below, two $p$-values for the observed data given the predicted SM cross-section
(using either \SHERPA{} or \POWHEG{} to model the \qqFourL{} contribution) are obtained
from the $\chi^2$. This is defined as $\chi^2 = \left[ \sigdata - \sigpred
\right] ^T C^{-1} \left[  \sigdata - \sigpred  \right]$, where \sigdata{} and $\sigpred$ are $k$-dimensional vectors from the measured and predicted differential
cross-sections of a given observable respectively, and $C$ is the $k\times k$ total covariance
matrix defined by the sum of the statistical and systematic
covariances in \sigdata{} and \sigpred.
The statistical
covariance on \sigdata{}  is obtained from the expected number of SM events, as
described in Section~\ref{sec:unc}.
The $p$-value is the probability for the
$\chi^2$, with $k$ degrees of
freedom,  to have at
least the observed value.
 
\begin{figure}[tb]
\centering
\includegraphics[width = 0.85\textwidth]{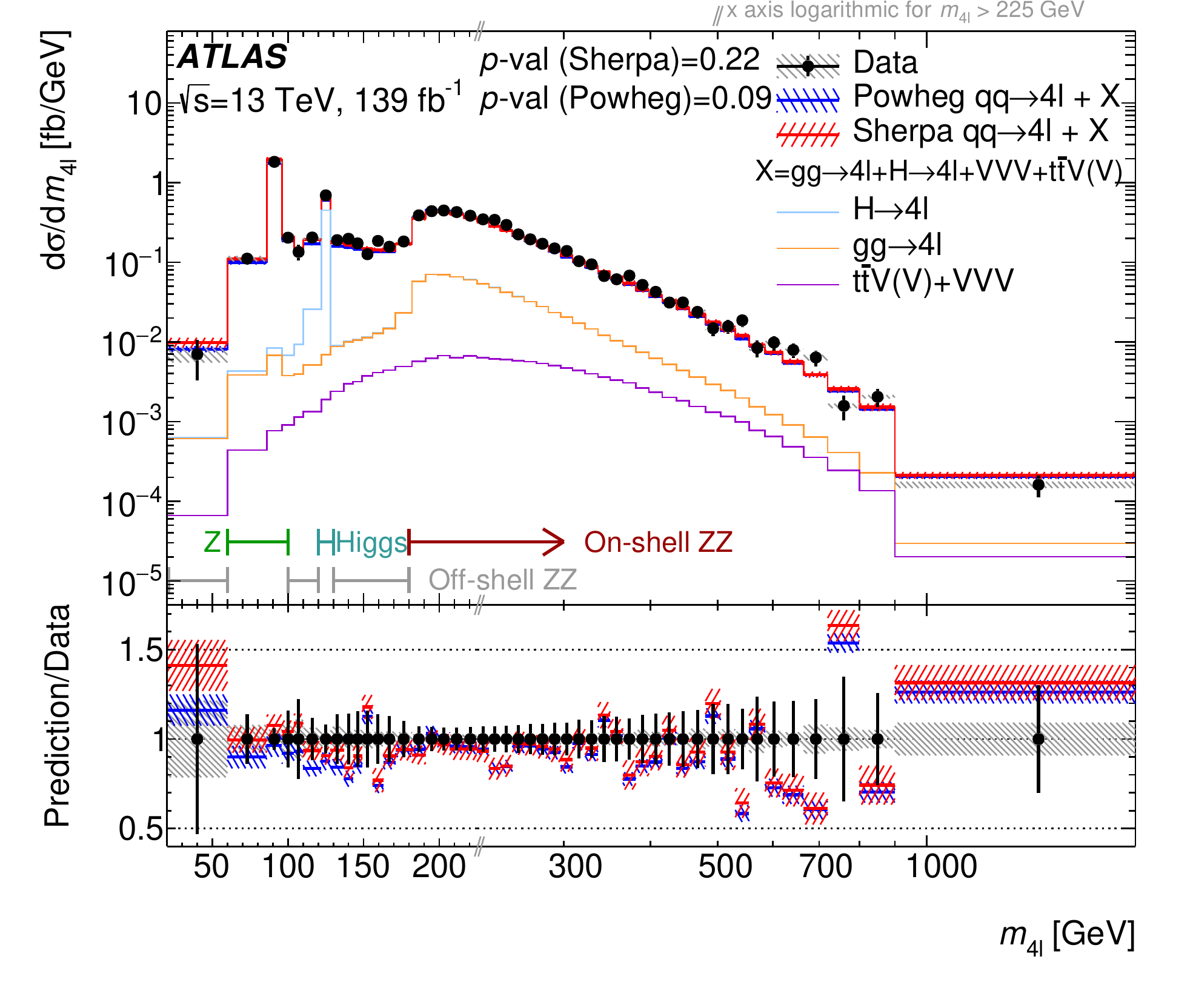}
\caption{Differential cross-section as a function of \mFourL. The measured data
(black points)  are compared with the SM
prediction using either \SHERPA{} (red, with red hashed
band for the uncertainty) or \POWHEG{} + \pythia{} (blue,
with blue hashed band for the uncertainty) to model the \qqFourL{} contribution. The error bars on the data points give the total uncertainty
and the grey hashed band gives the systematic uncertainty. The
breakdown of the contribution from different SM processes is also
shown in successive stacked histograms.
The short vertical lines terminating horizontal lines indicate the boundaries of the different
\mFourL{} regions in which the other variables are measured.
\Pvalue{}
The
lower panel shows the ratio of the SM predictions to the
data. The $x$-axis is on a linear scale until $\mFourL = 225$~\GeV,
where it switches to a logarithmic scale. \label{fig:cross-sec-m4l}}
\end{figure}
In order to study the different \mFourL{} regions in more detail,  Figures~\ref{fig:mz1res} and~\ref{fig:mz2res}
show the cross-section versus
\mZOne{} and \mZTwo{} respectively in each region.
In the \HFourL{} region, the contribution from Higgs boson production is
shown separately.
The different regions show peaks in different places due to the
kinematic constraints of the \mFourL{} requirements.
For all regions but \ZFourL{} there is a clear
enhancement at \mZ{} for \mZOne. Conversely,  \mZTwo{}
only has contributions from on-shell \Z{} bosons in the on-shell
region.
The \ZFourL{} and off-shell
regions are dominated by off-shell photon and \Z{} boson exchange, and
the \HFourL{} region is dominated by off-shell \Z{} production.
The data are generally well modelled by the SM predictions within
uncertainties,
although, as already discussed, the normalisation of the
predictions in the on-shell region is lower than the observed
measurement, and for  the \POWHEG{} +
\pythia{} prediction it is lower in all the regions, especially the
off-shell \Z\Z{} region. Combined with some statistical fluctuations
in the data this gives some low $p$-values for some of the
distributions, especially for the \POWHEG{} +
\pythia{} prediction. This is also true for the measurements in
Figures~\ref{fig:onshell}--\ref{fig:polarisation3}.
For \mZOne{} in the on-shell \Z\Z{} region, the predictions are
below the data for \mZOne{} below \mZ{}, and above the data for
\mZOne{} above \mZ{}. Here the shapes of the two SM predictions also
deviate from each other, indicating differences in the modelling, perhaps related to
the modelling of the final-state radiation of photons.
The \mZOne{} and \mZTwo{} measurements provide particular sensitivity to BSM
models in which the lepton pairs do not come from \Z{} boson
decays, as discussed in Section~\ref{sec:int}.
 
\begin{figure}[hbt!]
\centering
\subfigure[\ZFourL \ region]{\includegraphics[width = 0.48\textwidth]{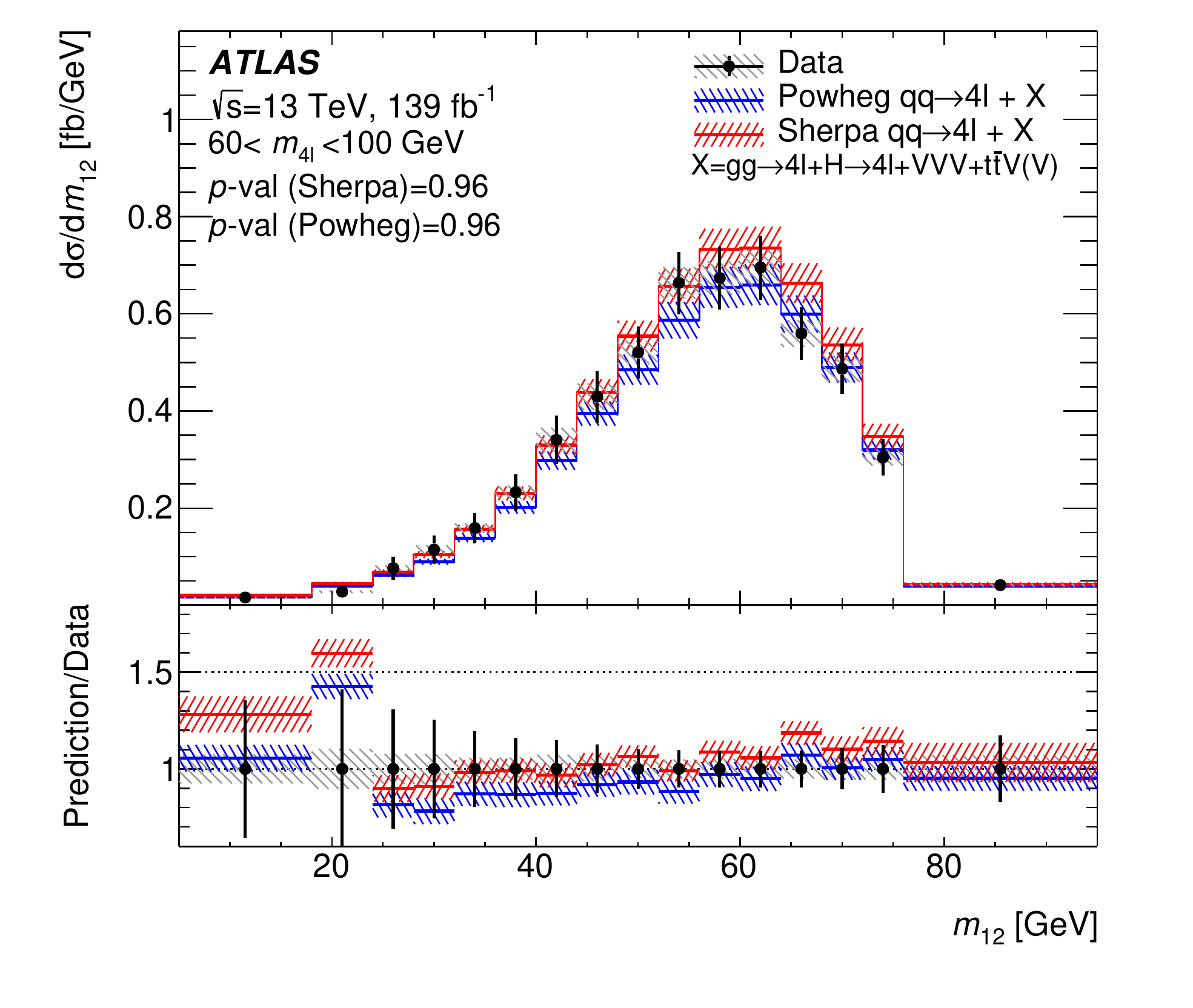}}
\subfigure[\HFourL \ region]{\includegraphics[width = 0.48\textwidth]{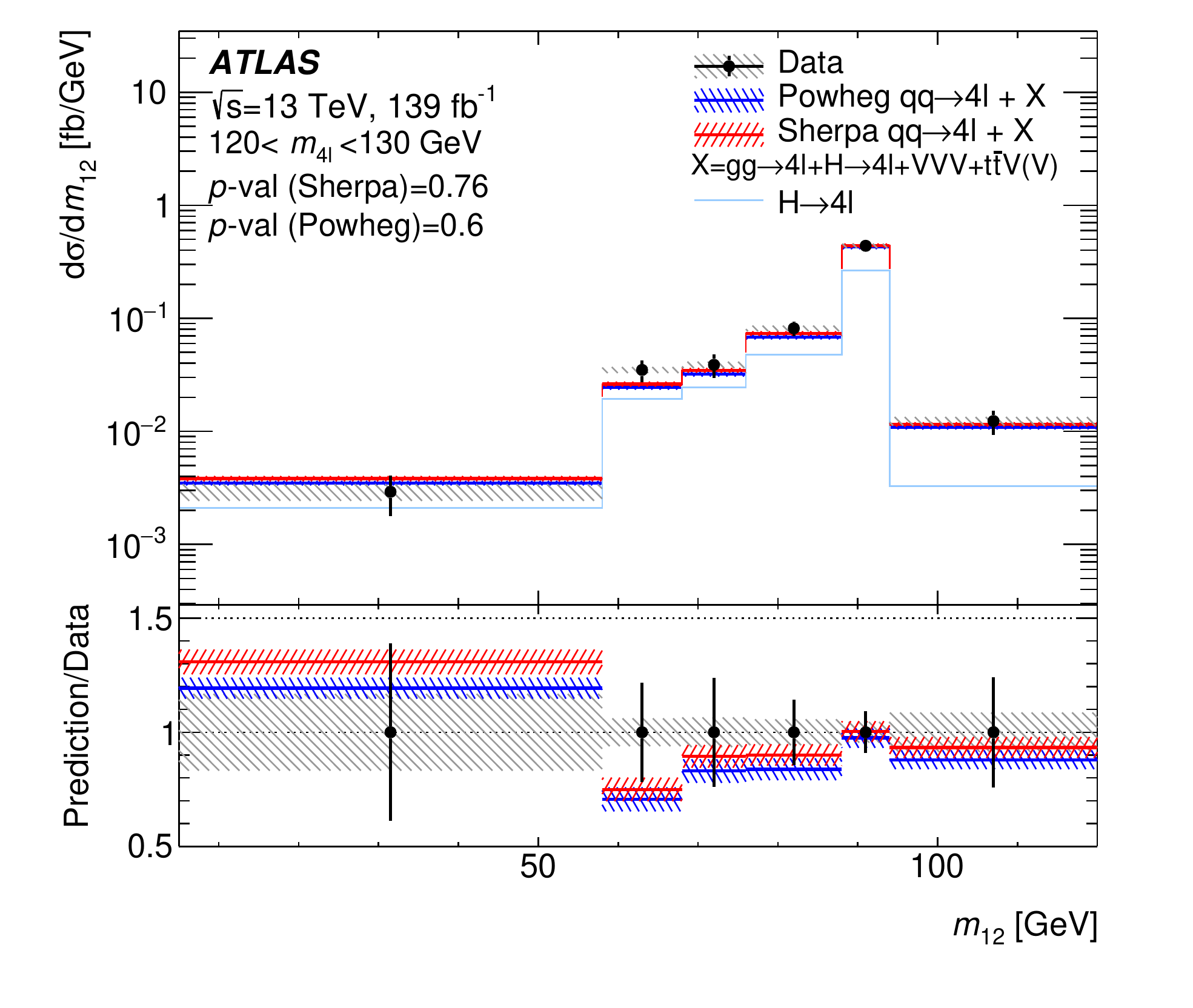}}\\
\subfigure[Off-shell $\Z\Z$ region]{\includegraphics[width = 0.48\textwidth]{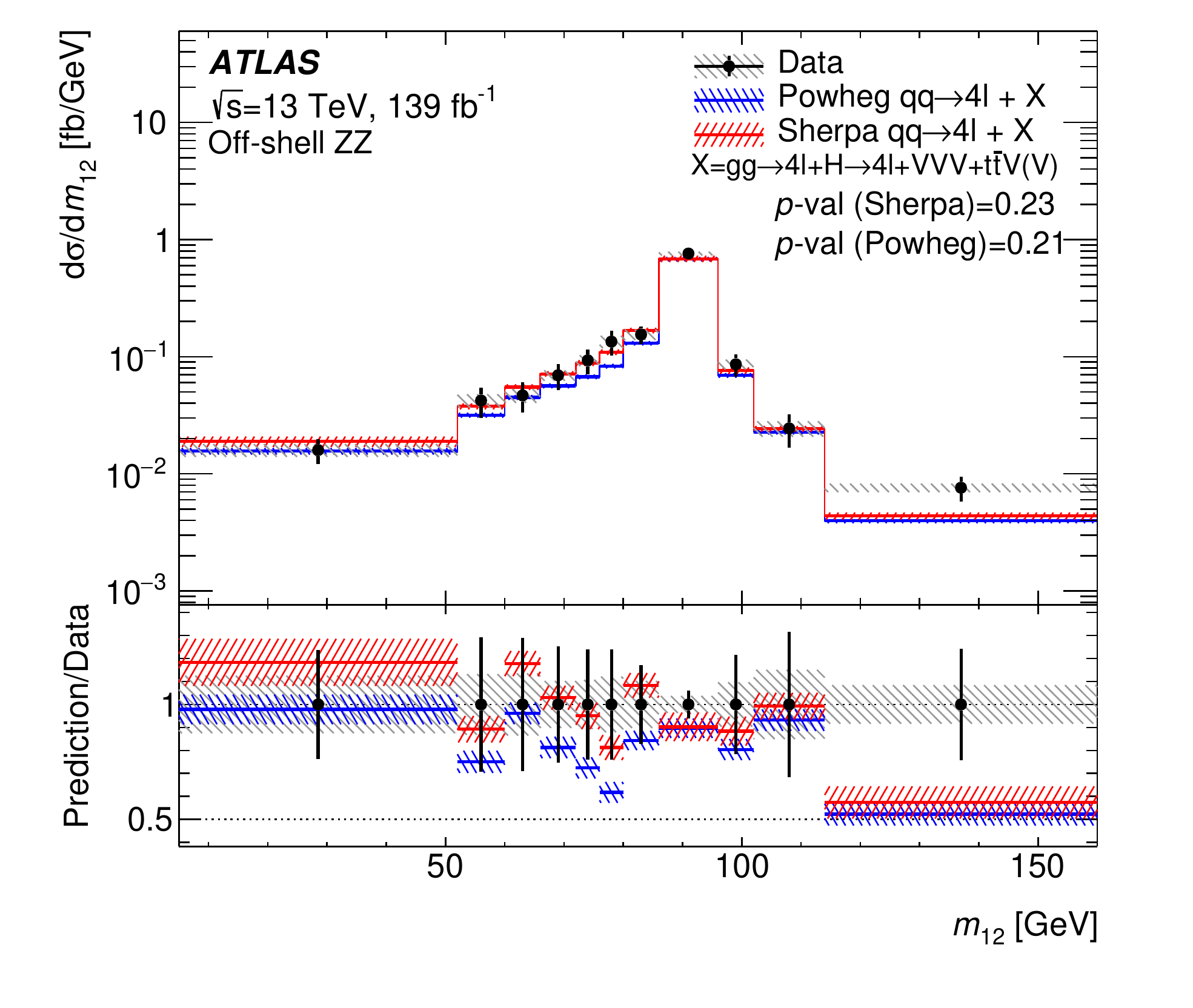}}
\subfigure[On-shell $\Z\Z$ region]{\includegraphics[width =0.48\textwidth]{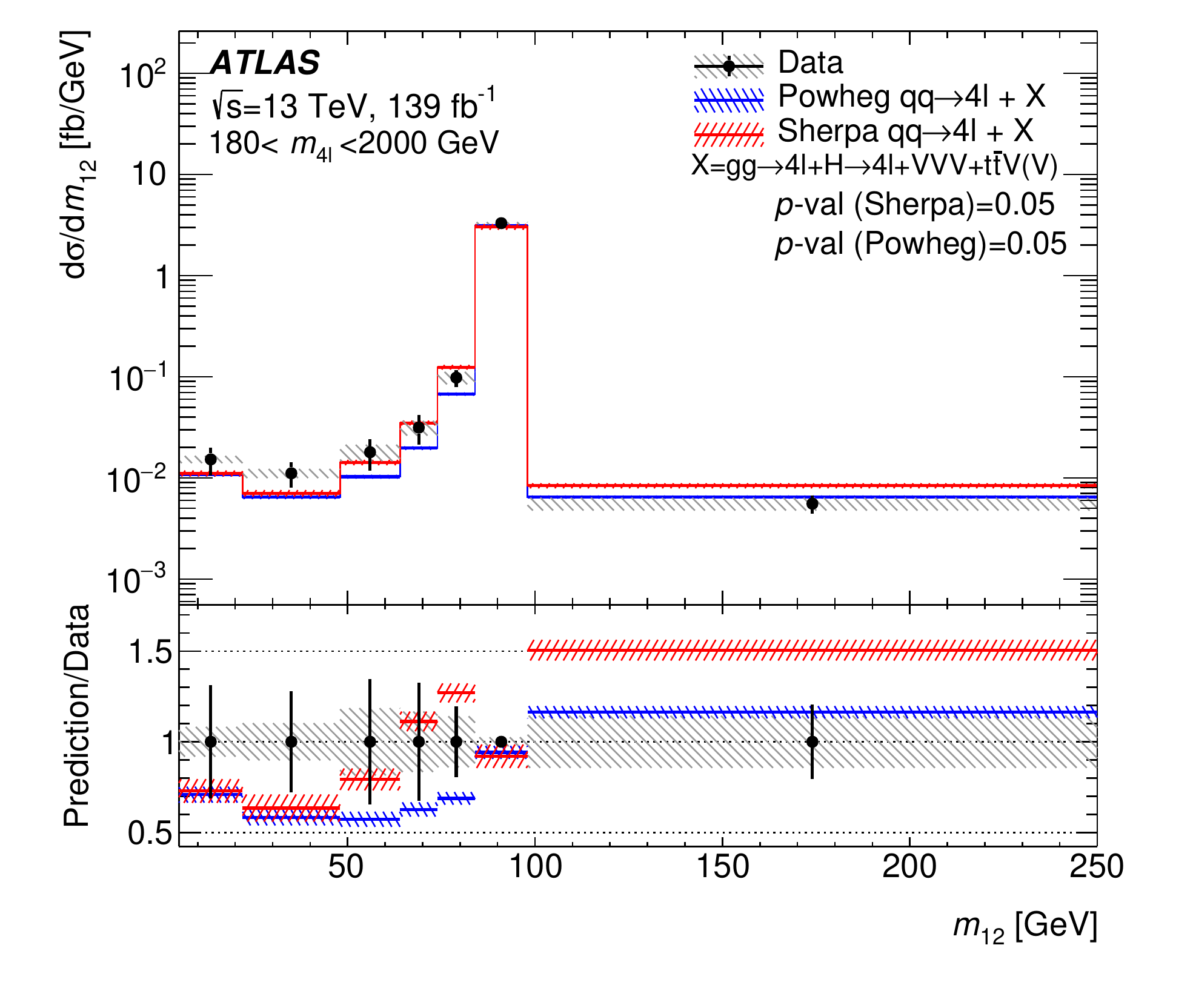}}\\
\caption{Differential cross-section as a function of \mZOne{} in the four
\mFourL{} regions. The measured data (black points) are  compared with the SM
prediction using either \SHERPA{} (red, with red hashed
band for the uncertainty) or \POWHEG{} + \pythia{} (blue,
with blue hashed band for the uncertainty) to model the \qqFourL{}
contribution.
In (b) the contribution from Higgs production is shown in addition to
the total SM prediction.
The error bars on the data points give the total uncertainty
and the grey hashed band gives the systematic uncertainty.
\Pvalue{}
The  lower panel shows the ratio of the SM predictions to the   data. \label{fig:mz1res}}
\end{figure}
 
\begin{figure}[htb!]
\centering
\subfigure[\ZFourL \ region]{\includegraphics[width = 0.48\textwidth]{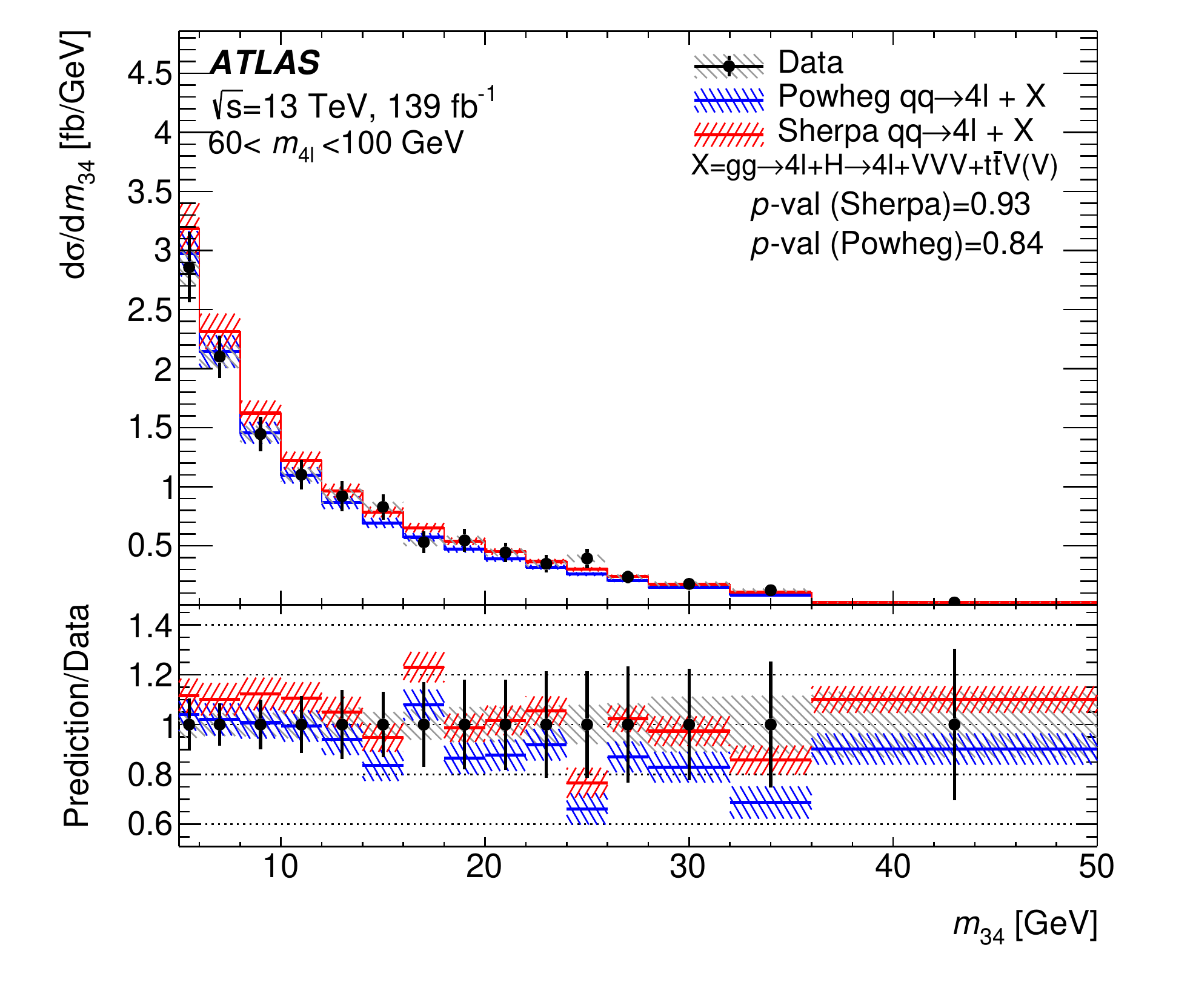}}
\subfigure[\HFourL \ region]{\includegraphics[width = 0.48\textwidth]{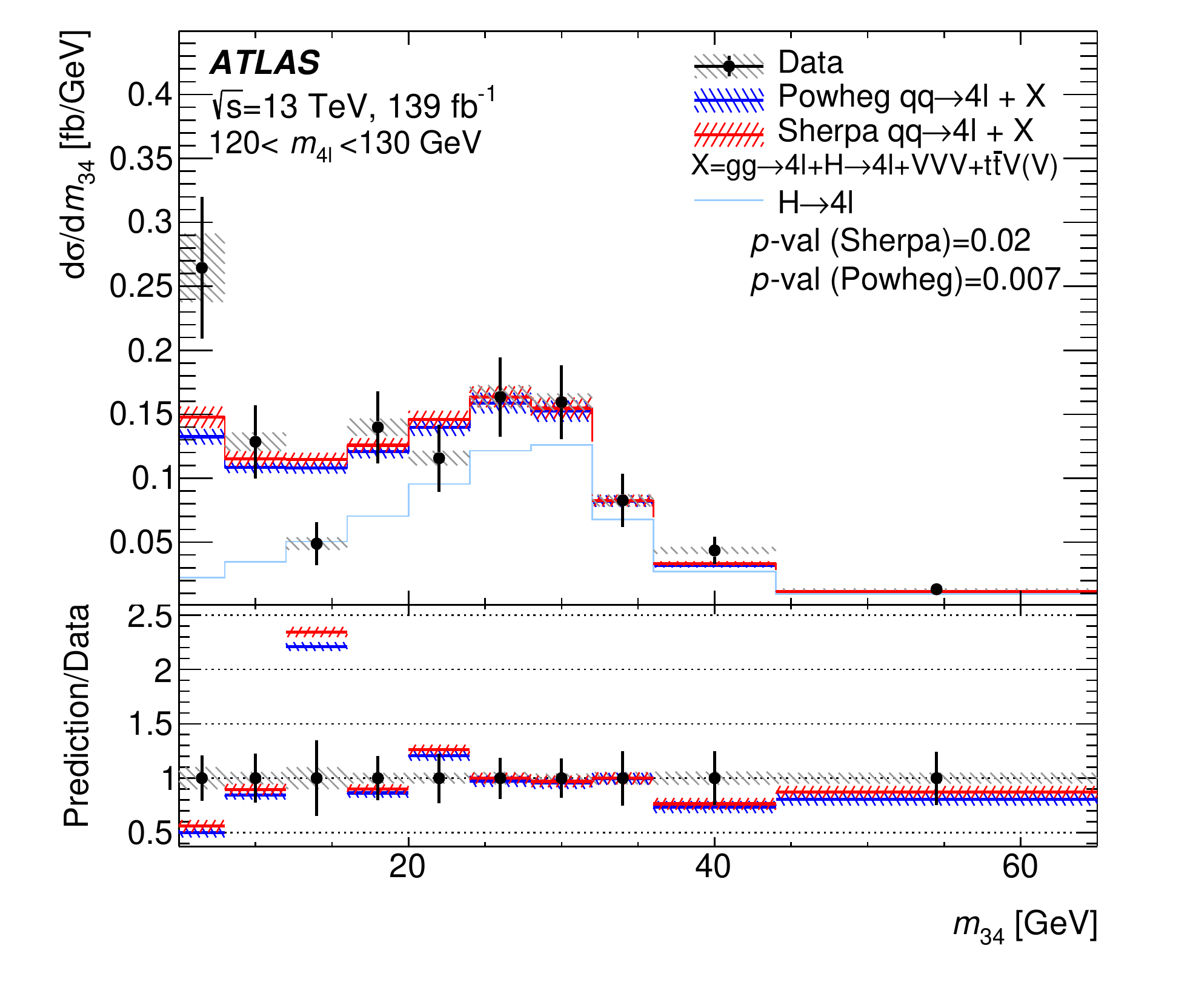}}\\
\subfigure[Off-shell $\Z\Z$ region]{\includegraphics[width = 0.48\textwidth]{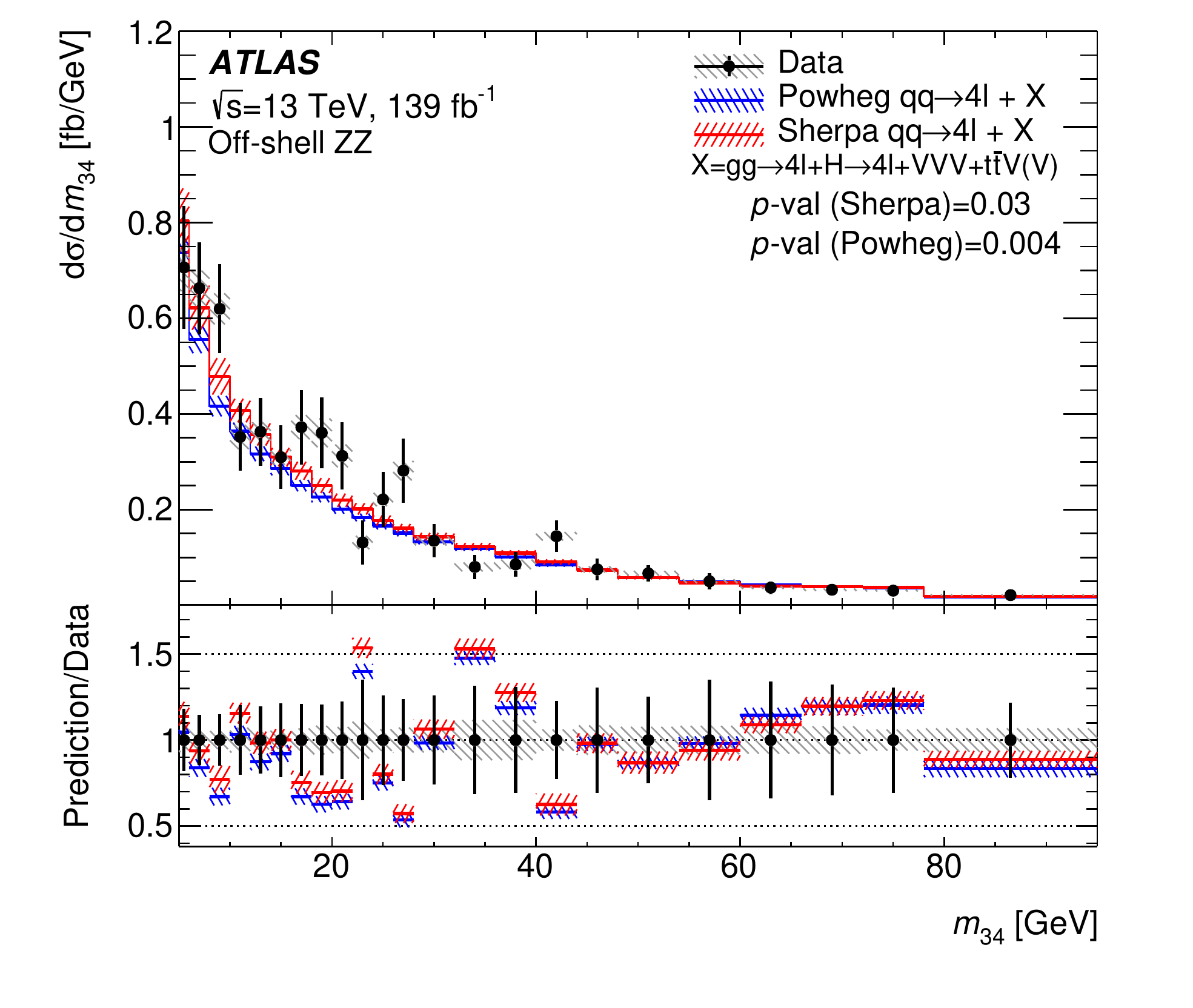}}
\subfigure[On-shell $\Z\Z$ region]{\includegraphics[width = 0.48\textwidth]{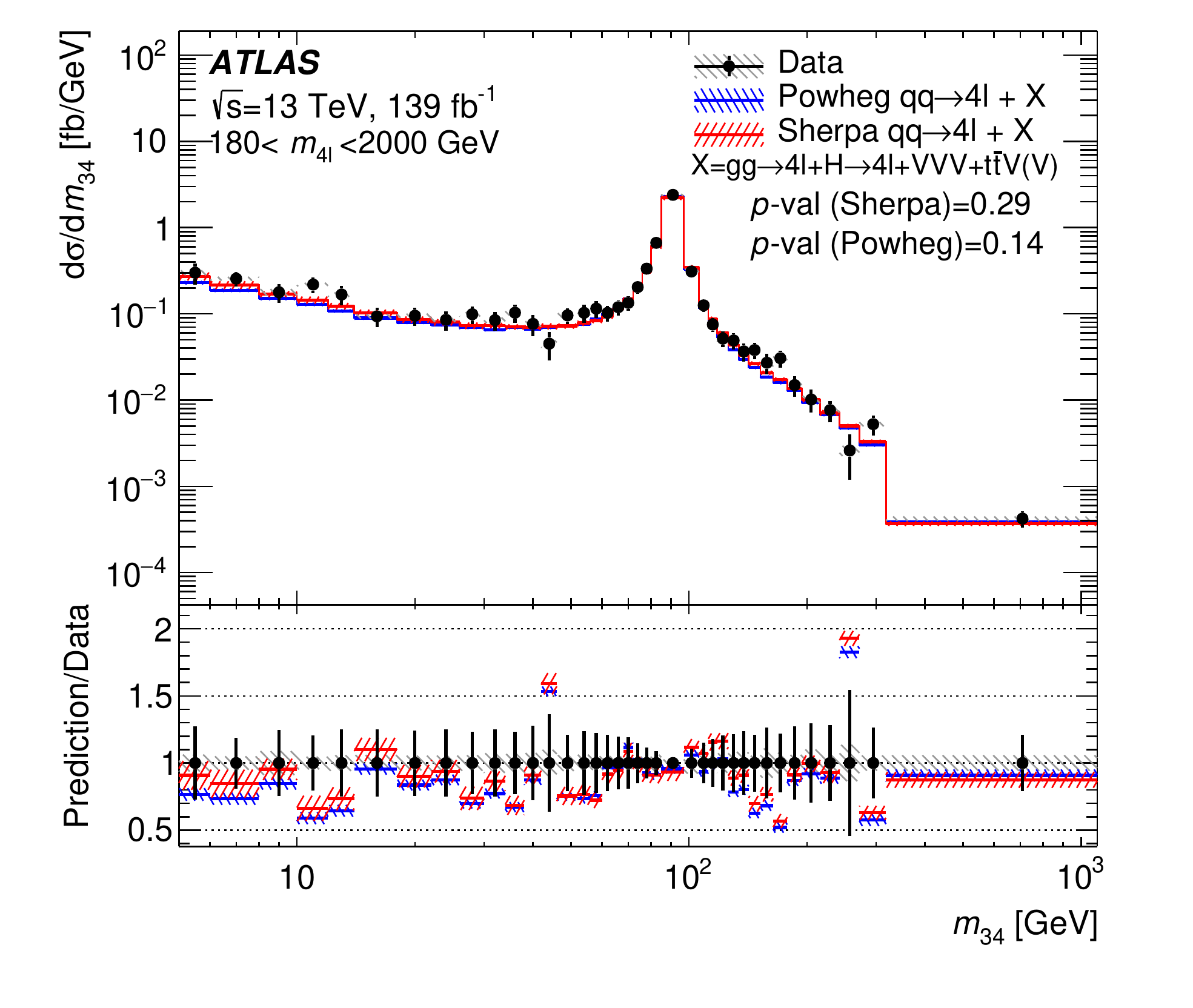} \label{fig:mz2res-onshell}}\\
\caption{Differential cross-section as a function of \mZTwo{} in the four \mFourL{}
regions. The measured data (black points)  are compared with the SM
prediction using either \SHERPA{} (red, with red hashed
band for the uncertainty) or \POWHEG + \pythia{} (blue,
with blue hashed band for the uncertainty) to model the \qqFourL{}
contribution.
In (b) the contribution from Higgs production is shown in addition to
the total SM prediction.
The error bars on the data points give the total uncertainty
and the grey hashed band gives the systematic uncertainty.
\Pvalue{}
The  lower panel shows the ratio of the SM predictions to the   data. \label{fig:mz2res}}
\end{figure}

Figure~\ref{fig:onshell} shows the differential cross-sections as a function
of \dPhiPairs, \dYPairs, \ptZOne, and \ptZTwo{} in the
highest cross-section, on-shell \Z\Z{}
region. The  \dPhiPairs{} distribution peaks at $\pi$, with the
dilepton pairs back-to-back. The $\dYPairs$ distribution peaks at
zero, with a tail going out to five. The  \ptZOne{} and \ptZTwo{}
distributions peak at around 40~\GeV.
Overall, the SM gives a reasonable description of the kinematics in
this region,
although the SM prediction
is about 20\%{} lower than  the data for $2.6 < \dYPairs{} < 3.2$ and
50\%{} lower for  $\dYPairs{} > 3.2$, indicating mis-modelling by
the simulation in this region of phase space.

\begin{figure}[htb!]
\centering
\subfigure[\dPhiPairs]{\includegraphics[width = 0.48\textwidth]{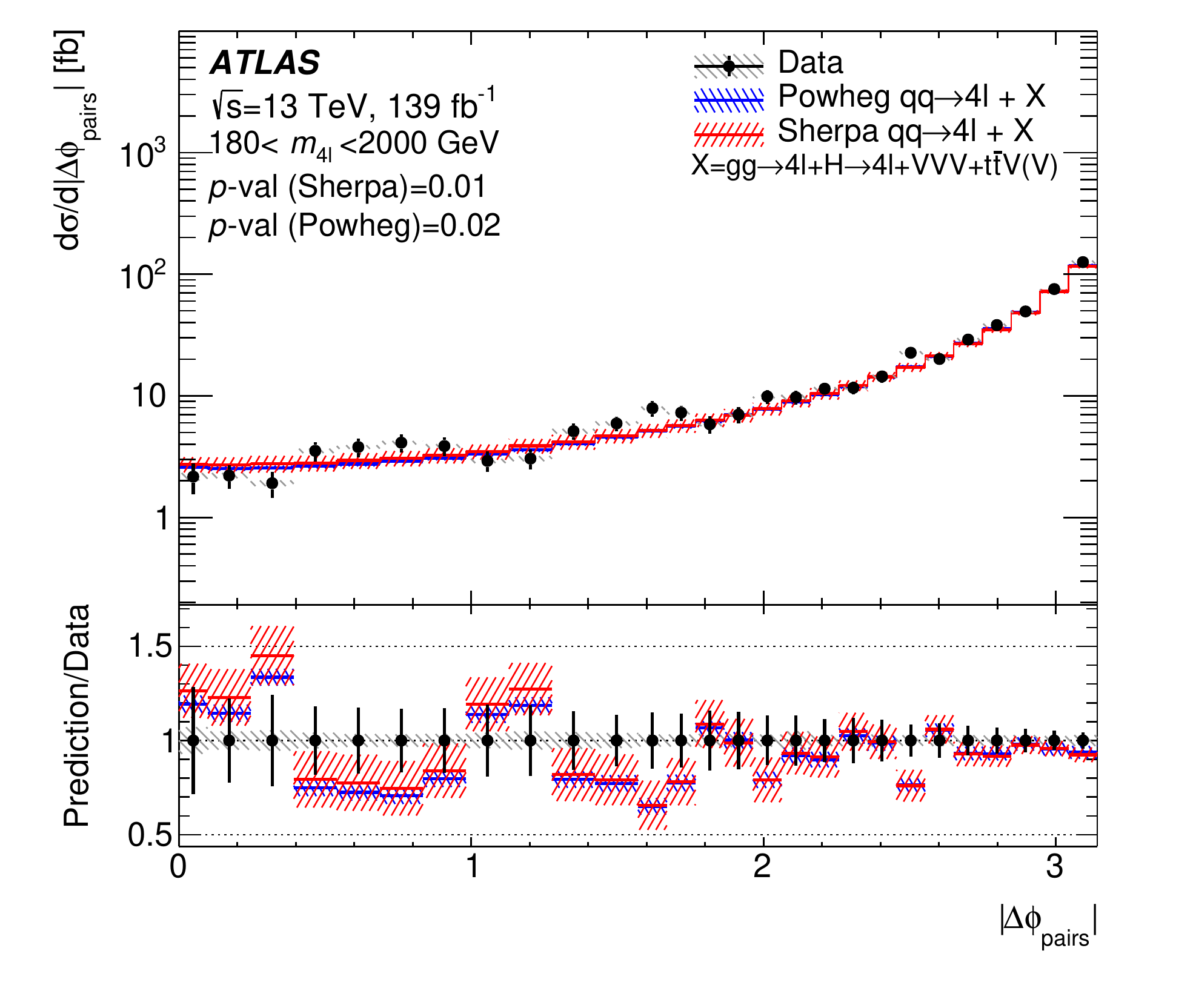}}
\subfigure[\dYPairs]{\includegraphics[width = 0.48\textwidth]{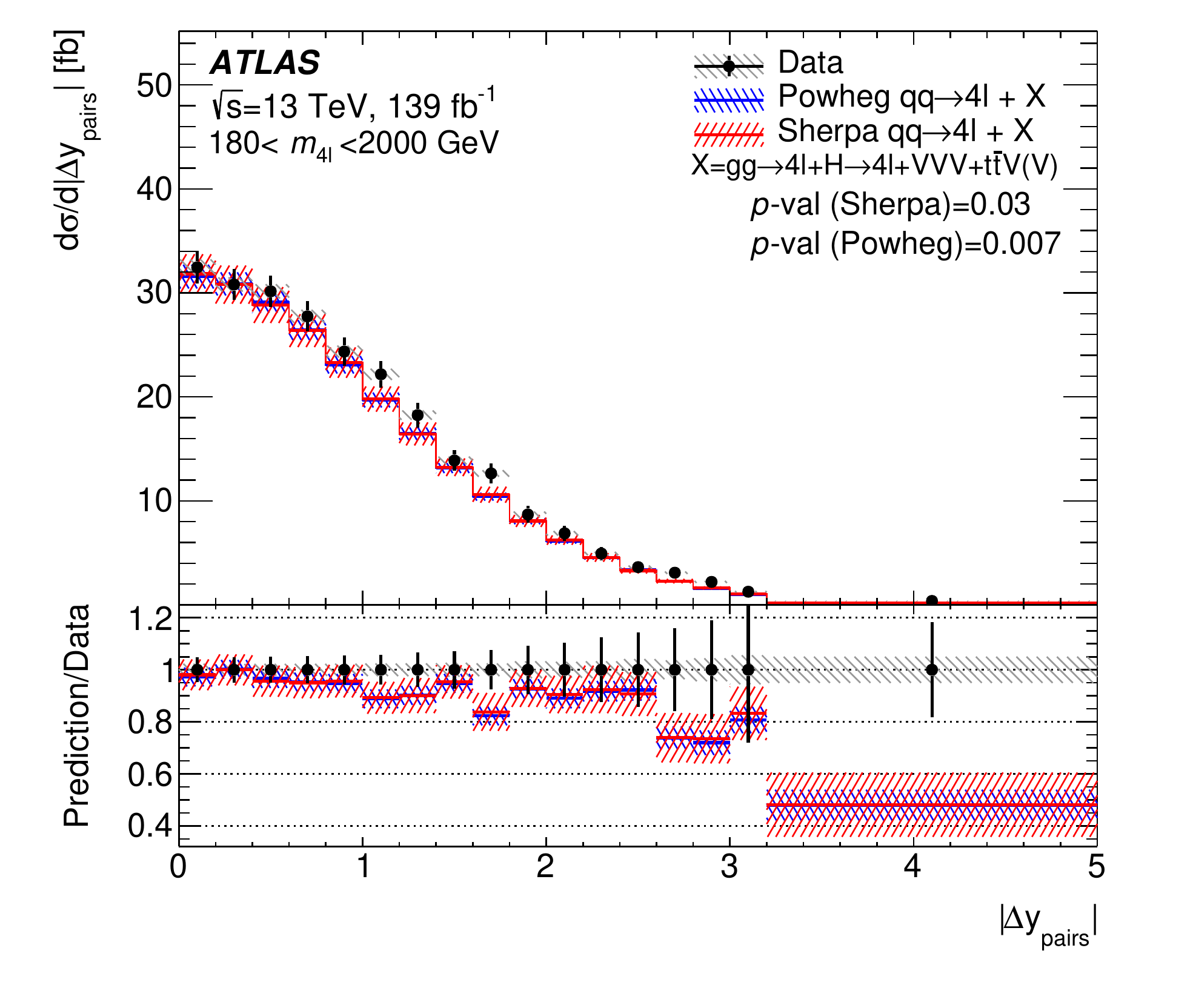}}\\
\subfigure[\ptZOne]{\includegraphics[width = 0.48\textwidth]{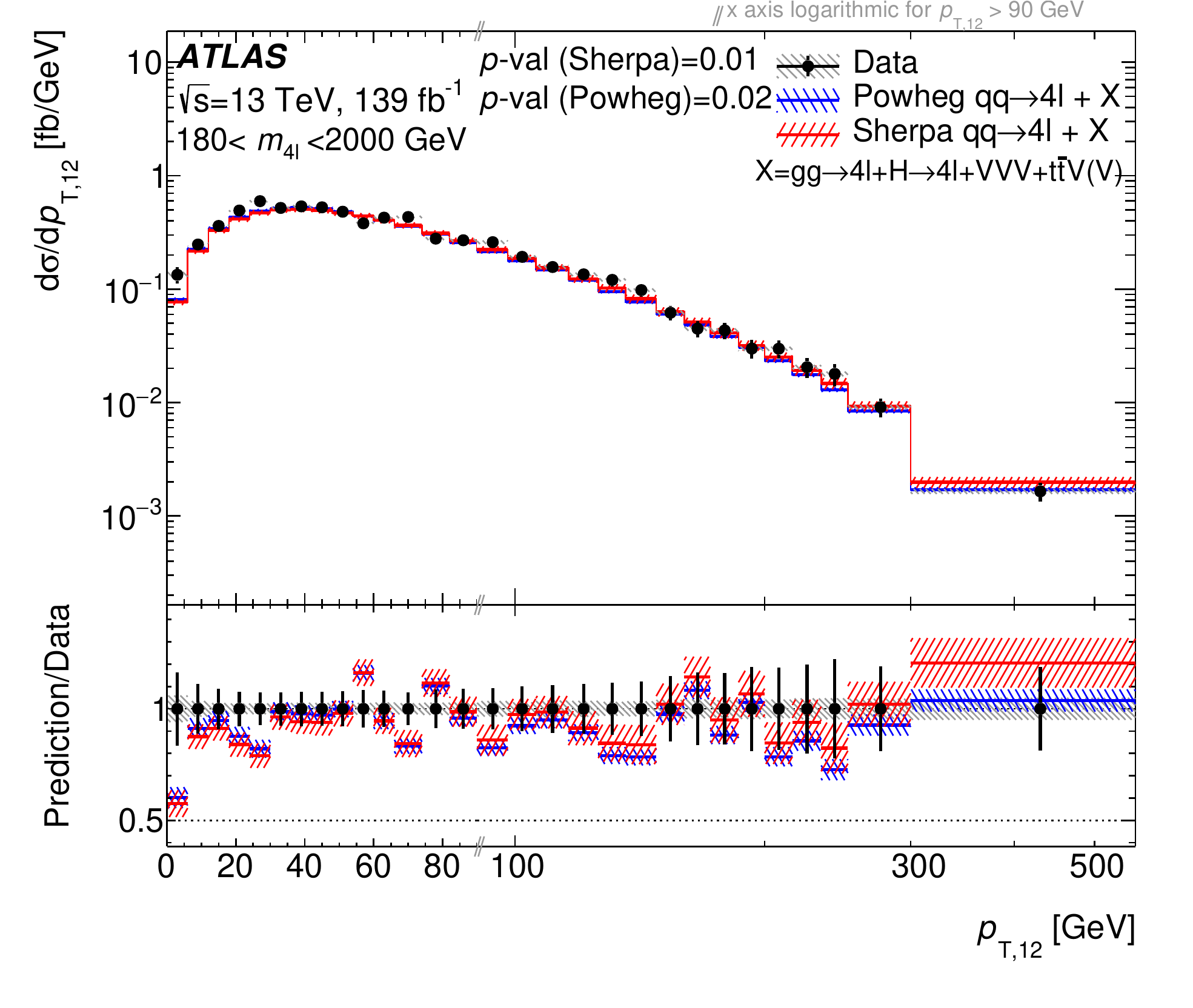}}
\subfigure[\ptZTwo]{\includegraphics[width = 0.48\textwidth]{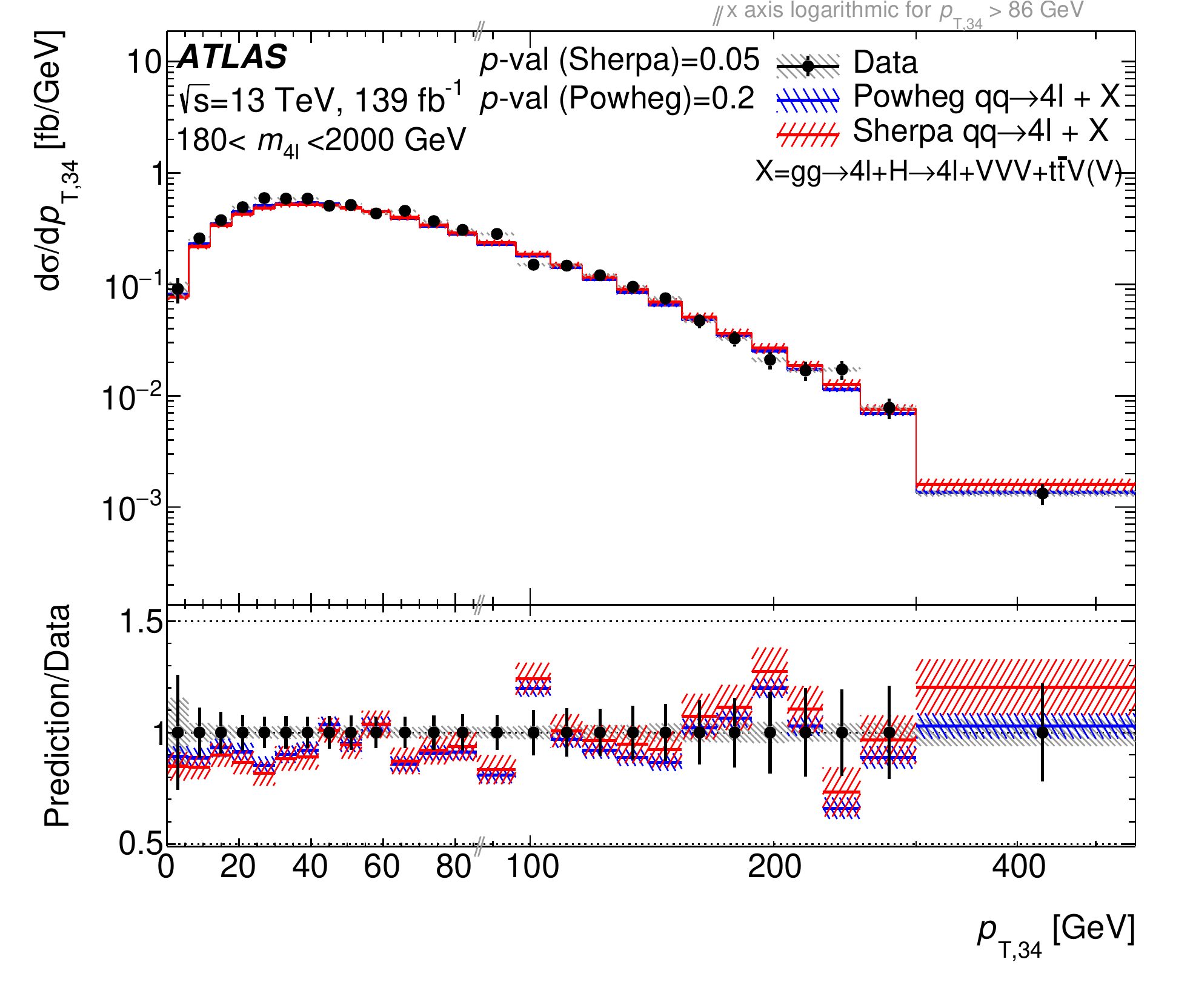}}\\
\caption{Differential cross-sections as a function of (a) \dPhiPairs, (b)
\dYPairs, (c) \ptZOne, and (d) \ptZTwo{} in the  on-shell $\Z\Z$ region.
The measured data (black points) are  compared with the SM
prediction using either \SHERPA{} (red, with red hashed
band for the uncertainty) or \POWHEG{} + \pythia{} (blue,
with blue hashed band for the uncertainty) to model the \qqFourL{} contribution. The error bars on the data points give the total uncertainty
and the grey hashed band gives the systematic uncertainty.
\Pvalue{}
The
lower panel shows the ratio of the SM predictions to the
data. In (c) the $x$-axis is on a linear scale until $\ptZOne = 90$~\GeV,
where it switches to a logarithmic scale; in (d) it switches at $\ptZTwo = 86$~\GeV.
\label{fig:onshell}}
\end{figure}
 
Figure~\ref{fig:dphires} shows the cross-section versus \dPhill{} for each region, and good
agreement with the SM prediction is seen.
In each region
the cross-section peaks when the two leading leptons are
back-to-back. This variable is sensitive to
electroweak corrections to the four-lepton final state~\cite{Gutschow:2020cug}.

\begin{figure}[htb!]
\centering
 
\subfigure[\ZFourL \ region]{\includegraphics[width = 0.48\textwidth]{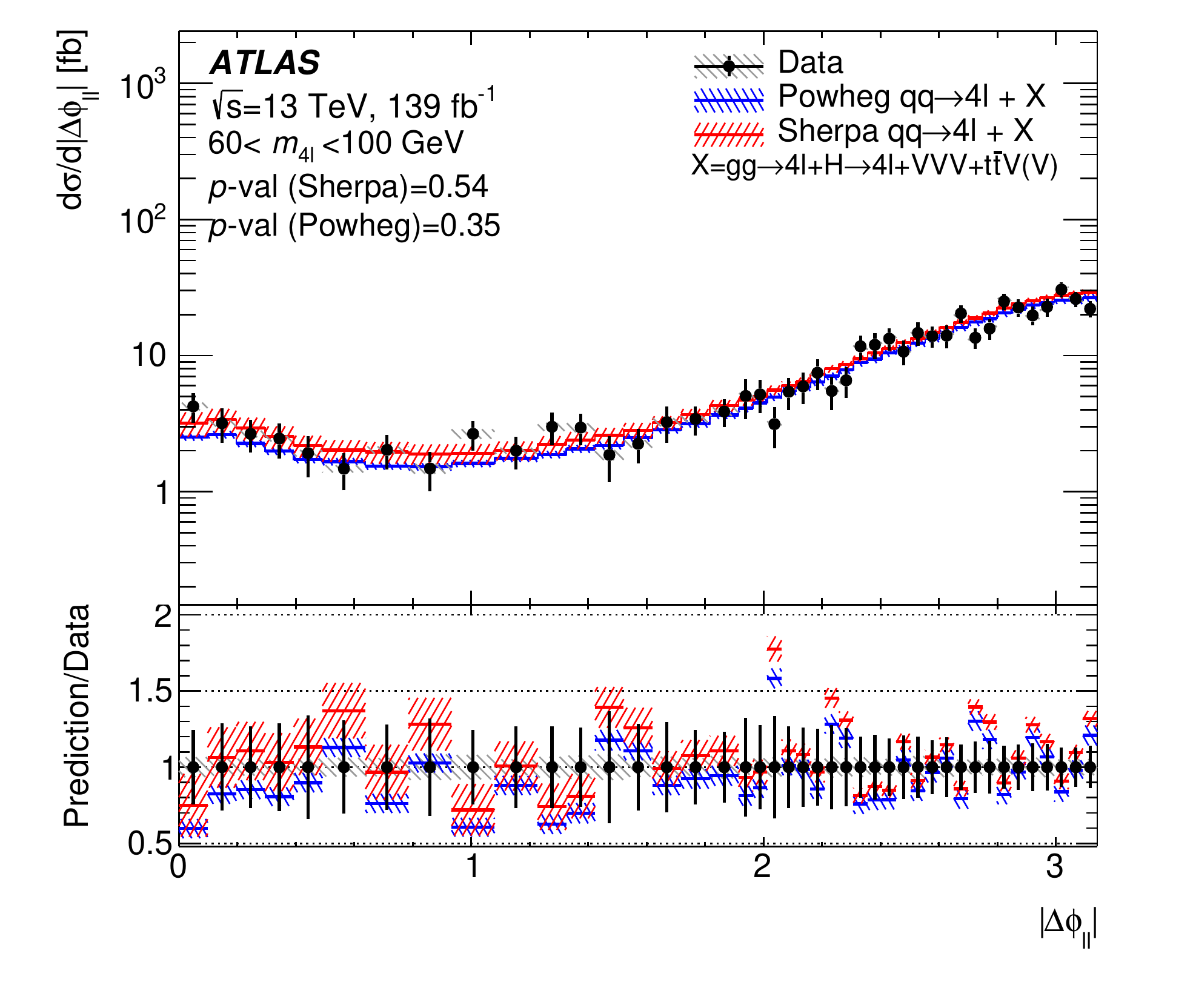}}
\subfigure[\HFourL \ region]{\includegraphics[width = 0.48\textwidth]{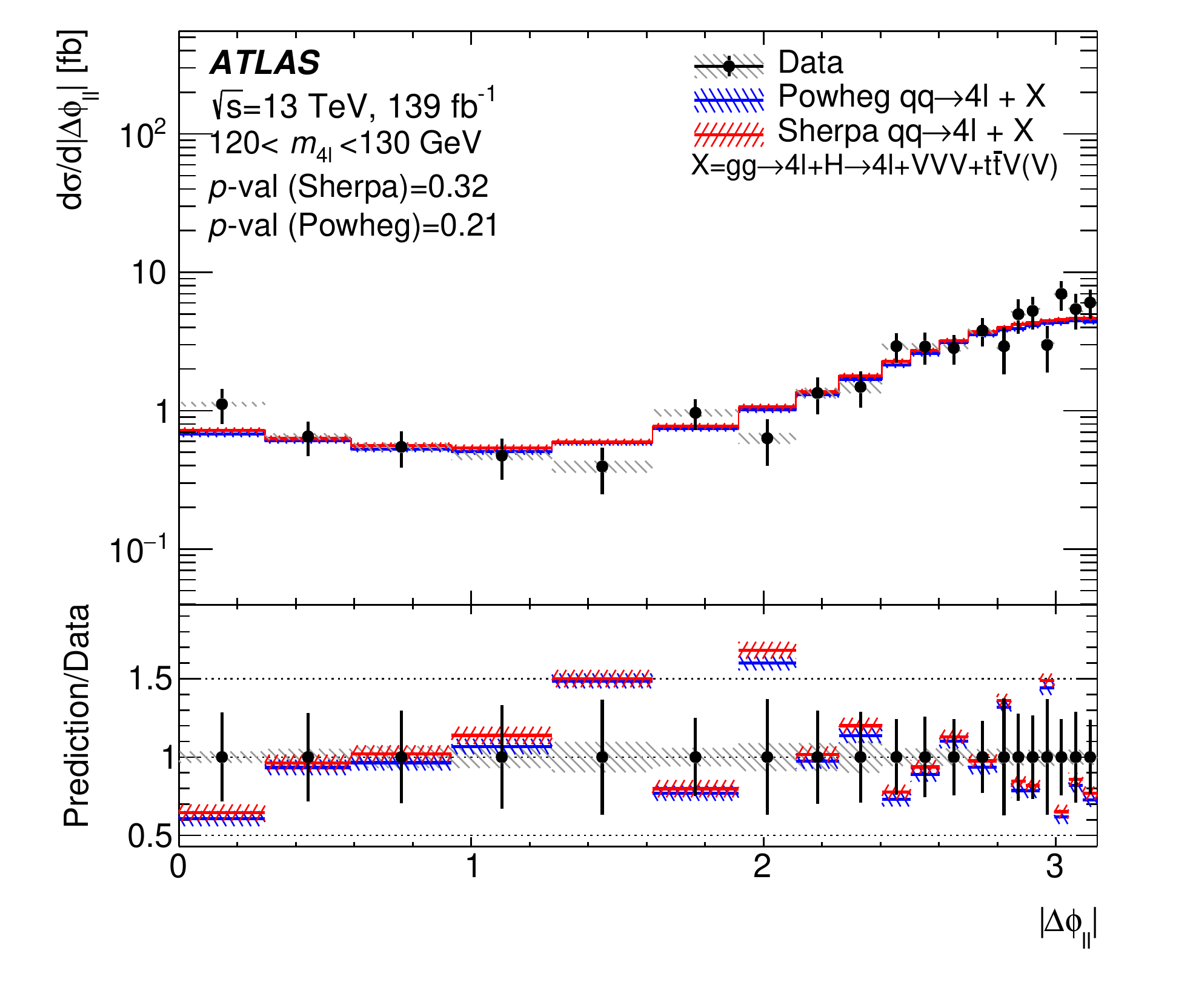}}\\
\subfigure[Off-shell $\Z\Z$ region]{\includegraphics[width = 0.48\textwidth]{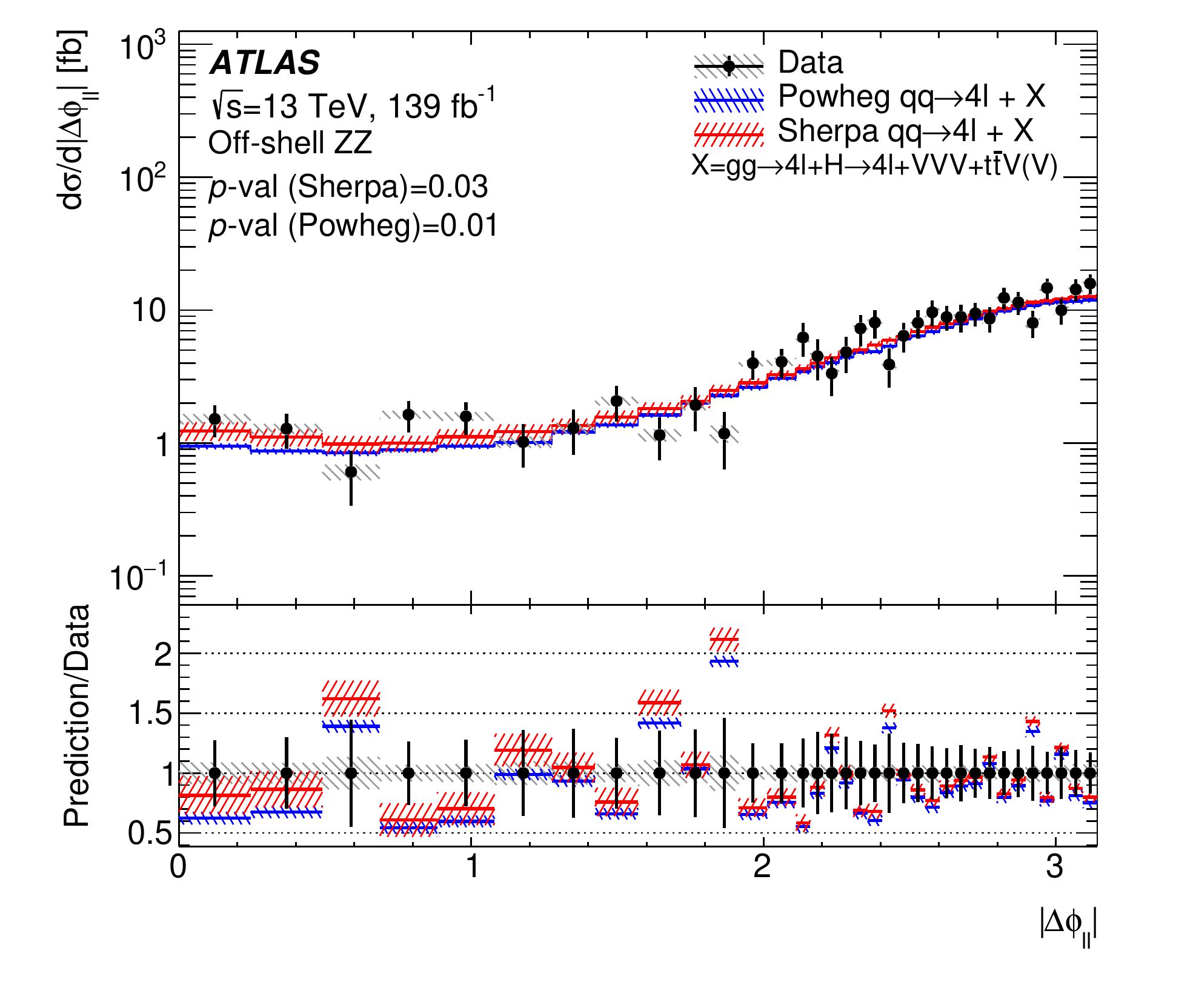}}
\subfigure[On-shell $\Z\Z$ region]{\includegraphics[width = 0.48\textwidth]{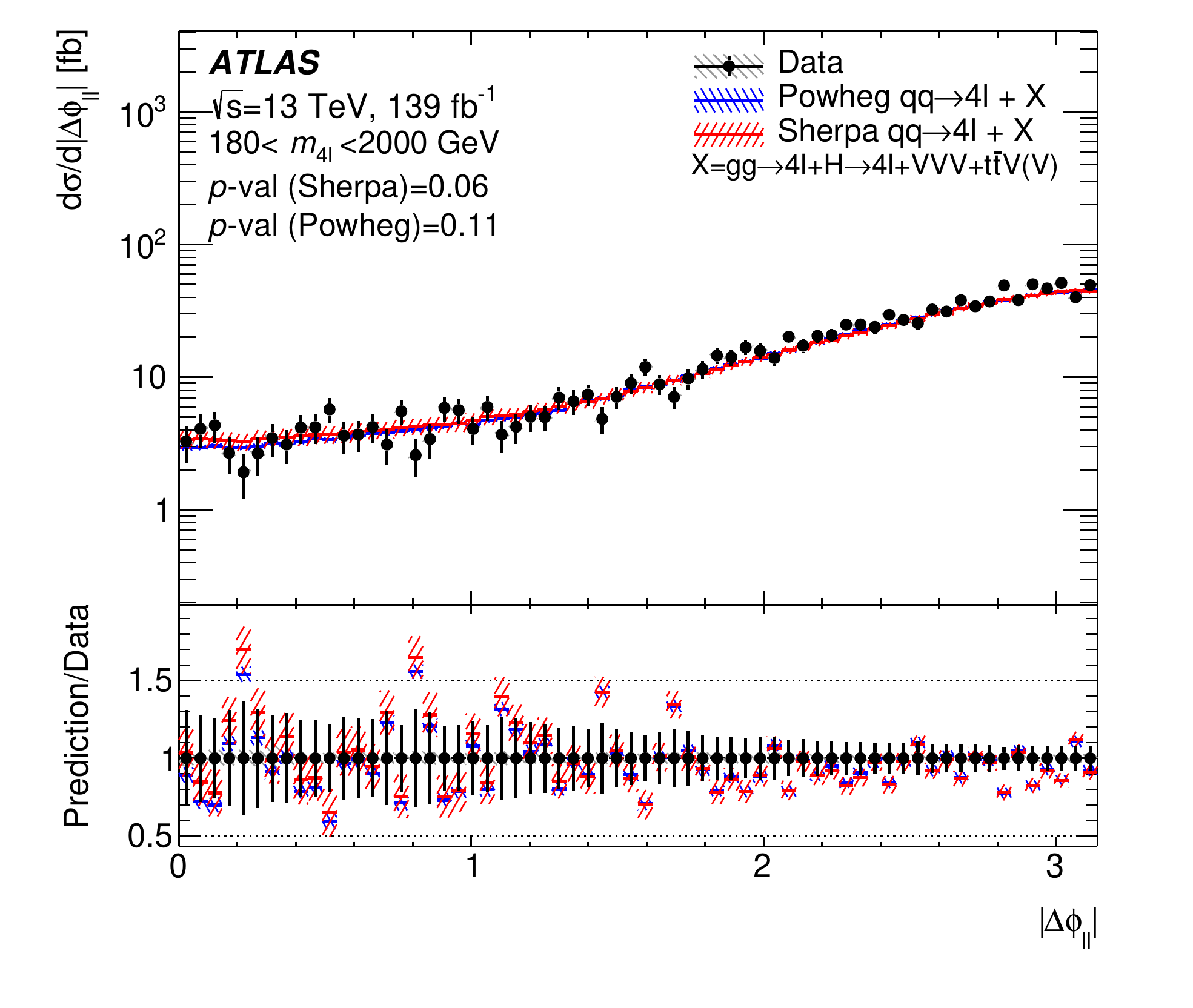}}\\
\caption{Differential cross-section as a function of \dPhill{} in the four \mFourL{}
regions. The measured data (black points)  are compared to the SM
prediction using either \SHERPA{} (red, with red hashed
band for the uncertainty) or \POWHEG + \pythia{} (blue,
with blue hashed band for the uncertainty) to model the \qqFourL{} contribution. The error bars on the data points give the total uncertainty
and the grey hashed band gives the systematic uncertainty.
\Pvalue{}
The  lower panel shows the ratio of the SM predictions to the data. \label{fig:dphires}}
\end{figure}
 
Figures~\ref{fig:polarisation1} and~\ref{fig:polarisation3} show $\costhetastar_{12}$ and
$\costhetastar_{34}$ respectively, in all four regions. These variables are sensitive to the polarisation of the
decaying bosons.
Good agreement with the SM prediction is seen in all regions.
 
The remaining measured differential cross-sections introduced in
Section~\ref{sec:vars} are shown in the Appendix.
 
\begin{figure}[htb!]
\centering
\subfigure[\ZFourL \ region]{\includegraphics[width = 0.48\textwidth]{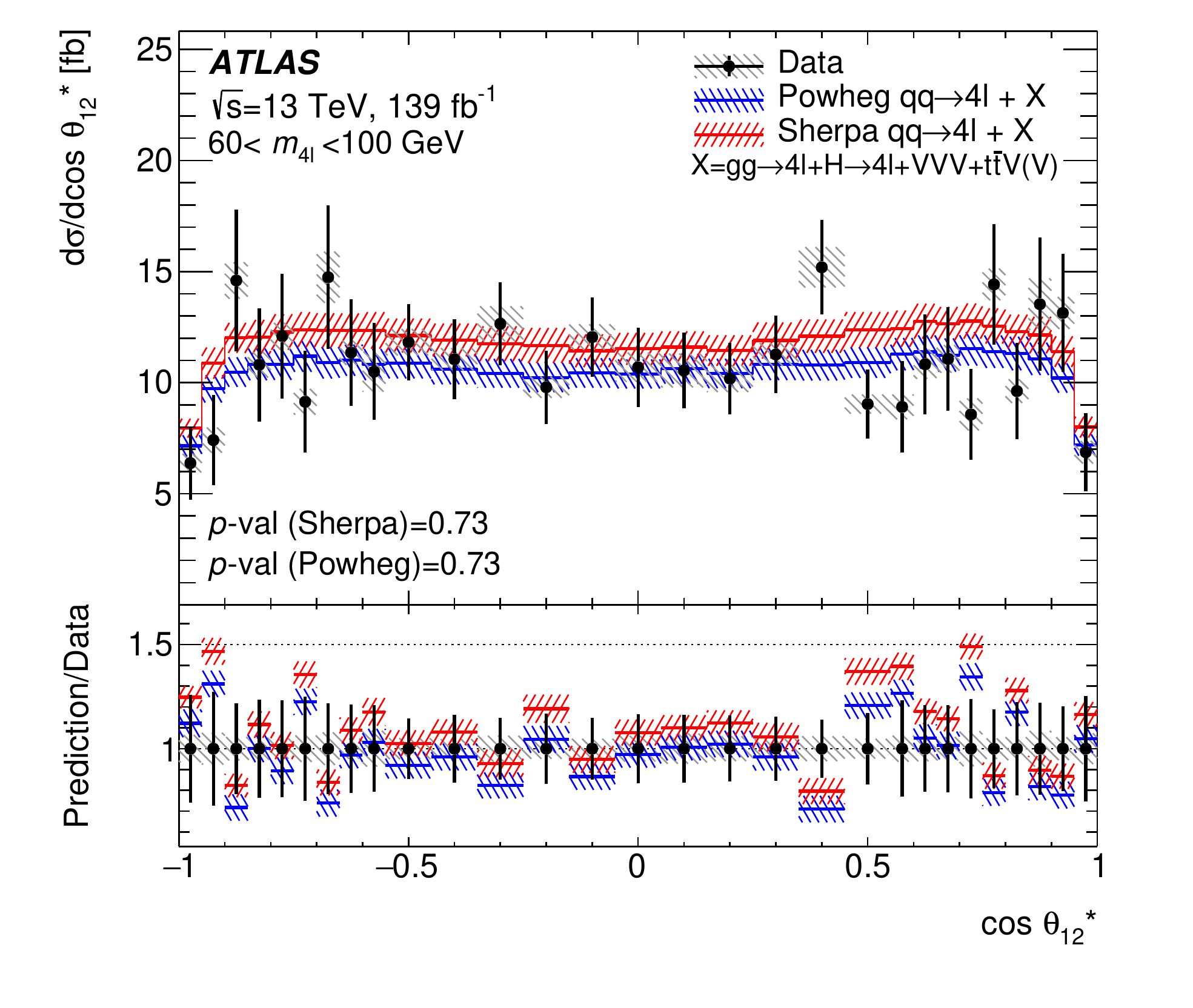}}
\subfigure[\HFourL \ region]{\includegraphics[width = 0.48\textwidth]{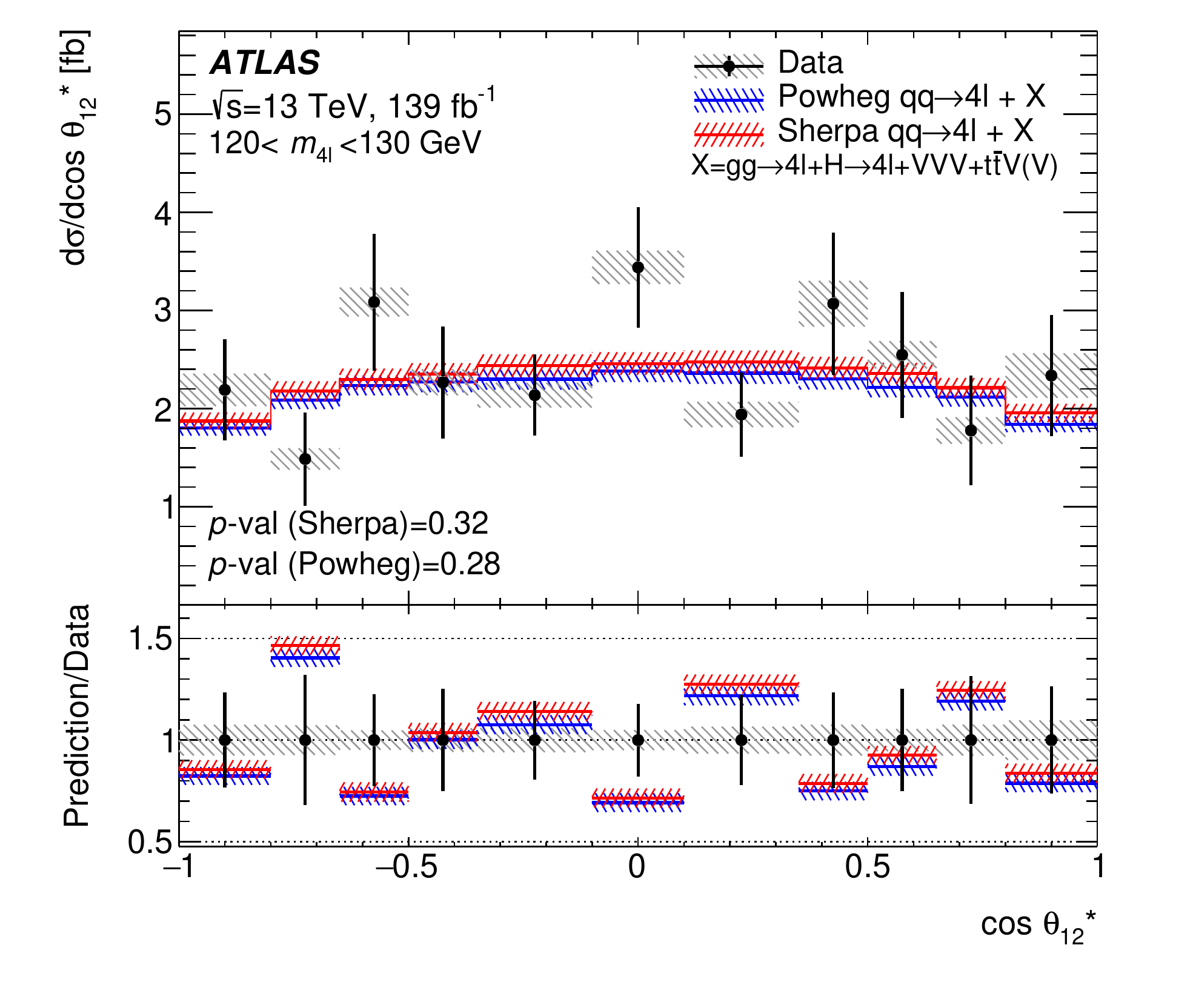}}\\
\subfigure[Off-shell $\Z\Z$ region]{\includegraphics[width = 0.48\textwidth]{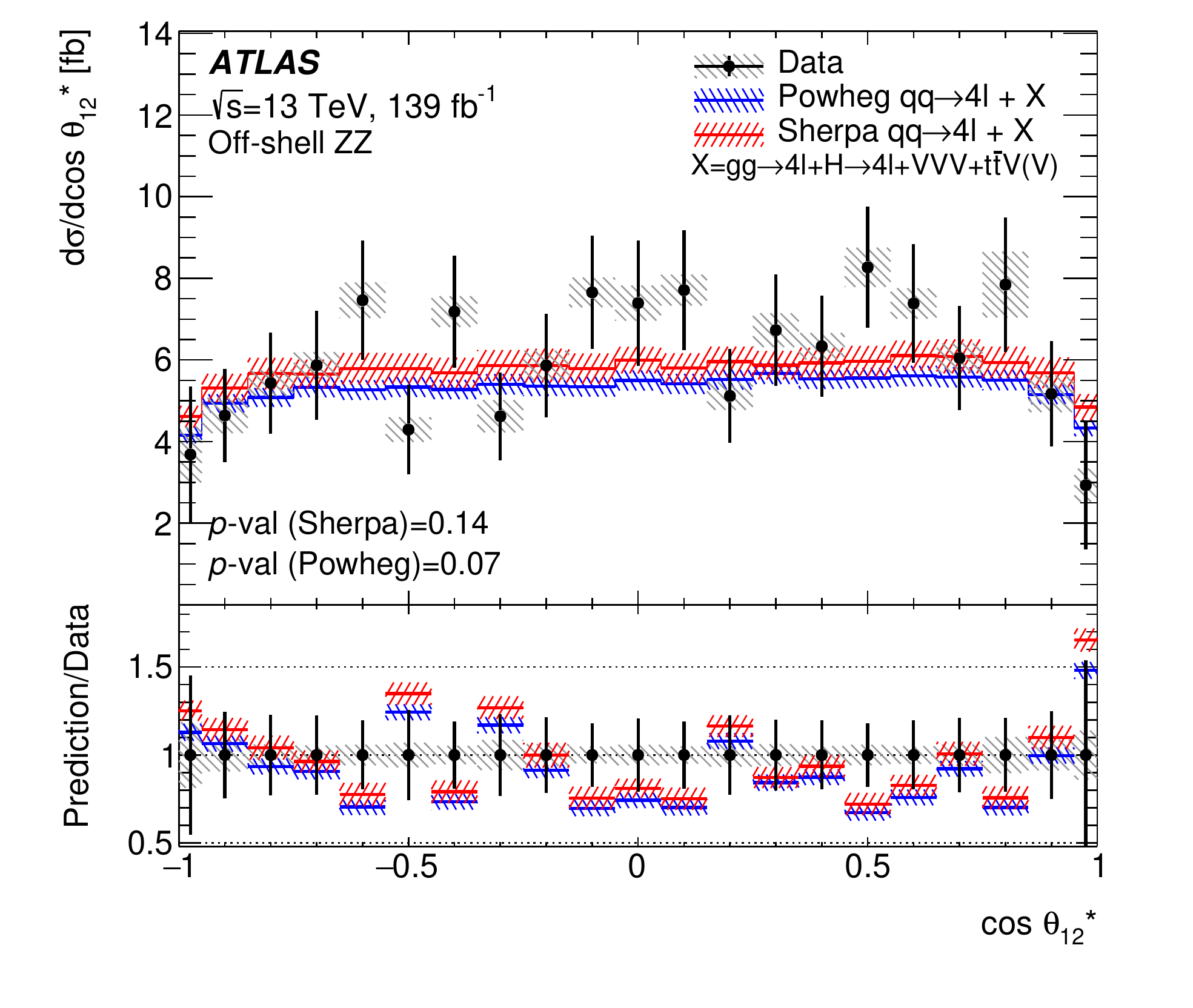}}
\subfigure[On-shell $\Z\Z$ region]{\includegraphics[width = 0.48\textwidth]{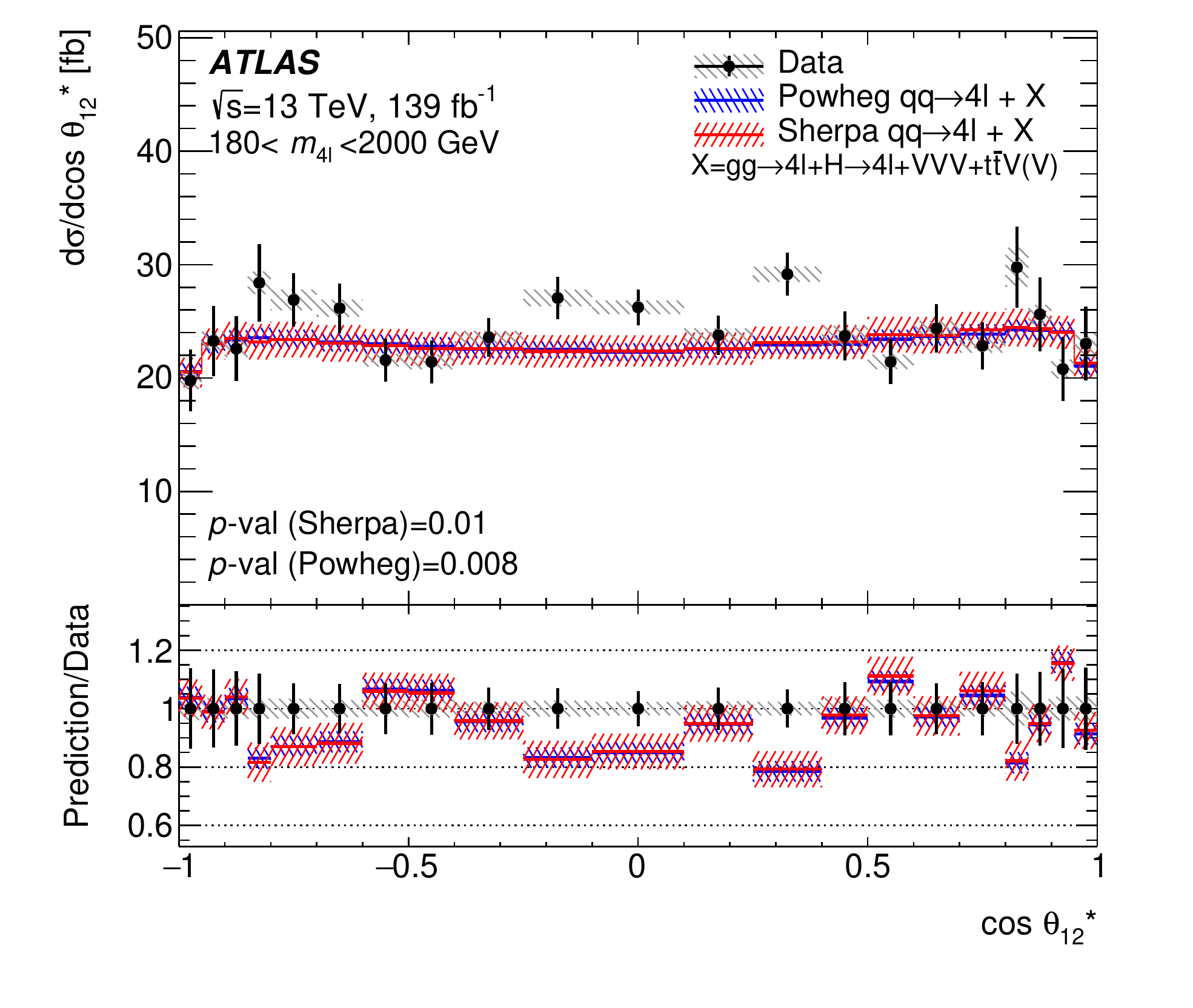}}\\
 
\caption{Differential cross-section as a function of $\costhetastar_{12}$ in the four \mFourL{} regions. The measured data (black
points)  are compared with the SM
prediction using either \SHERPA{} (red, with red hashed
band for the uncertainty) or \POWHEG + \pythia{} (blue,
with blue hashed band for the uncertainty) to model the \qqFourL{} contribution. The error bars on the data points give the total uncertainty
and the grey hashed band gives the systematic uncertainty.
\Pvalue{}
The  lower panel shows the ratio of the SM predictions to the data. \label{fig:polarisation1}}
\end{figure}
 
\begin{figure}[htb!]
\centering
\subfigure[\ZFourL \ region]{\includegraphics[width = 0.48\textwidth]{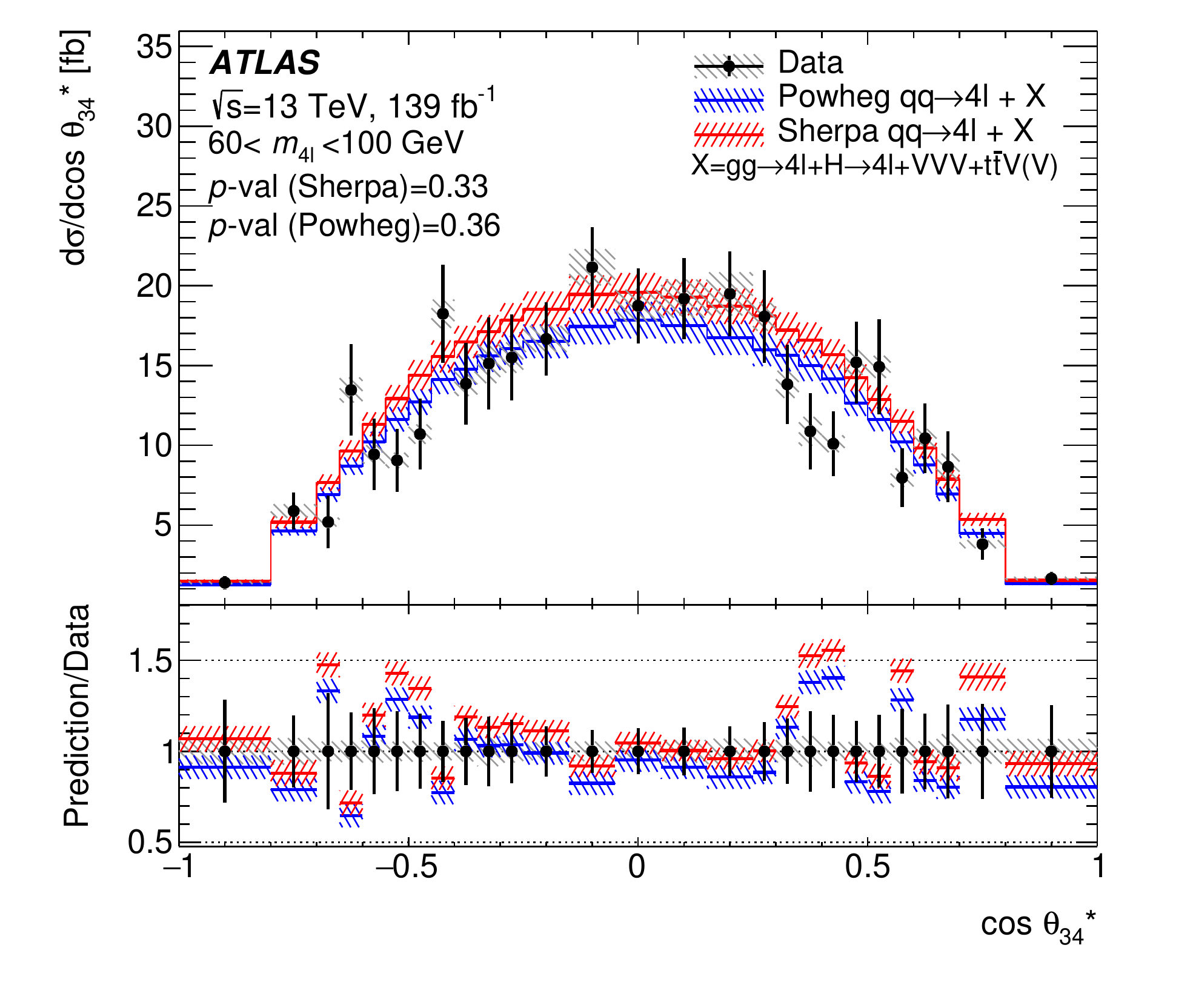}}
\subfigure[\HFourL  \ region]{\includegraphics[width = 0.48\textwidth]{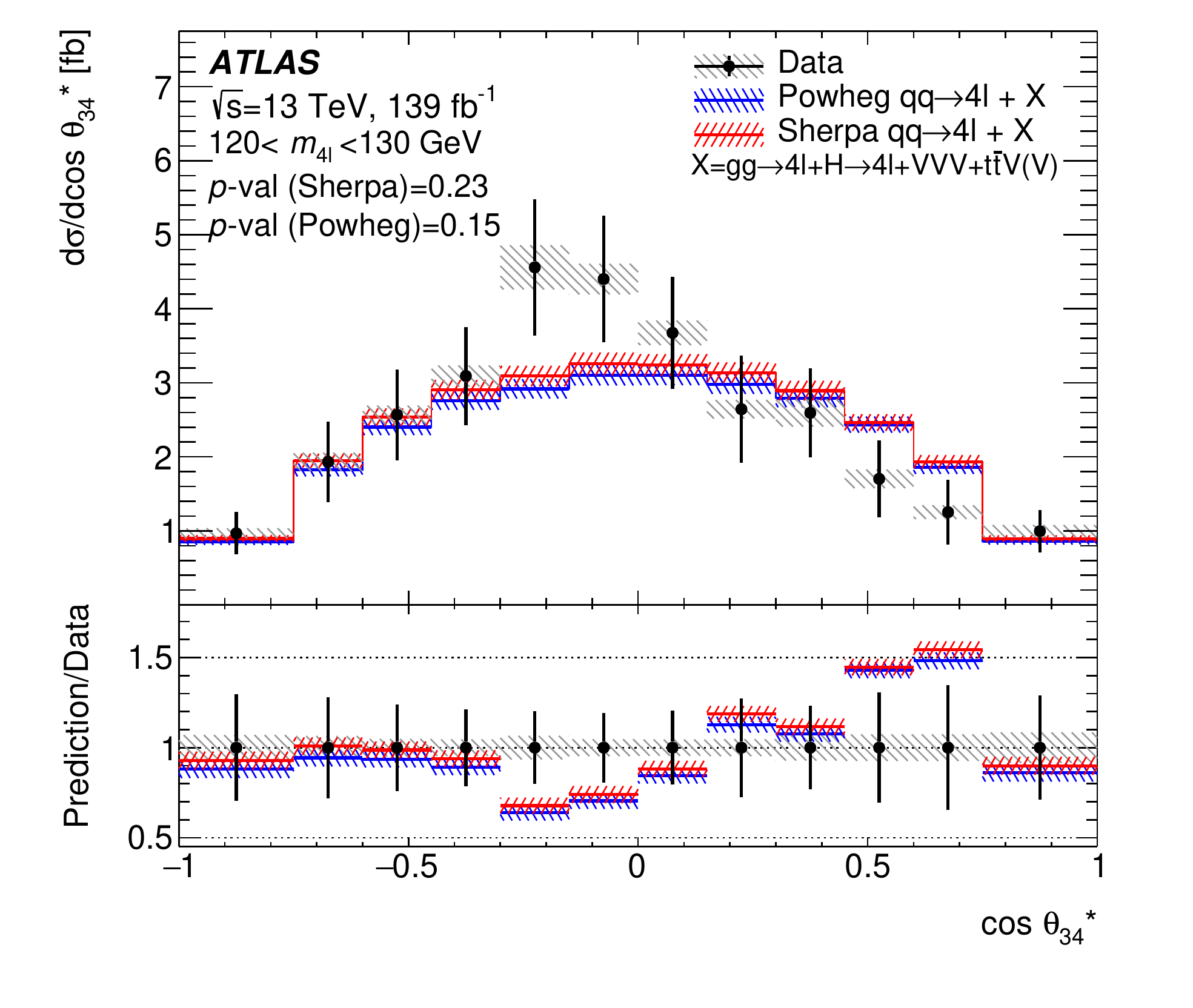}}\\
\subfigure[Off-shell $\Z\Z$ region]{\includegraphics[width = 0.48\textwidth]{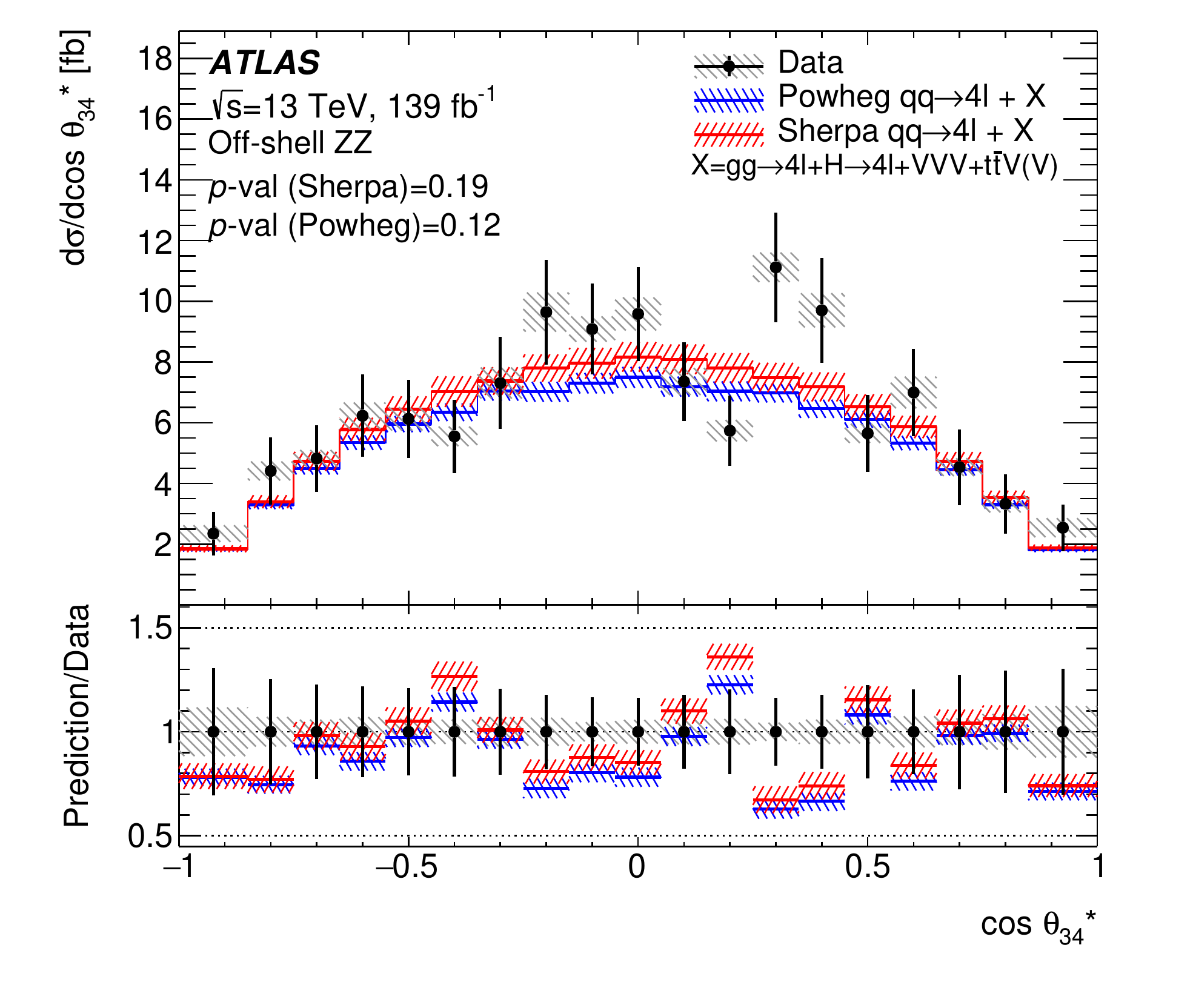}}
\subfigure[On-shell $\Z\Z$ region]{\includegraphics[width = 0.48\textwidth]{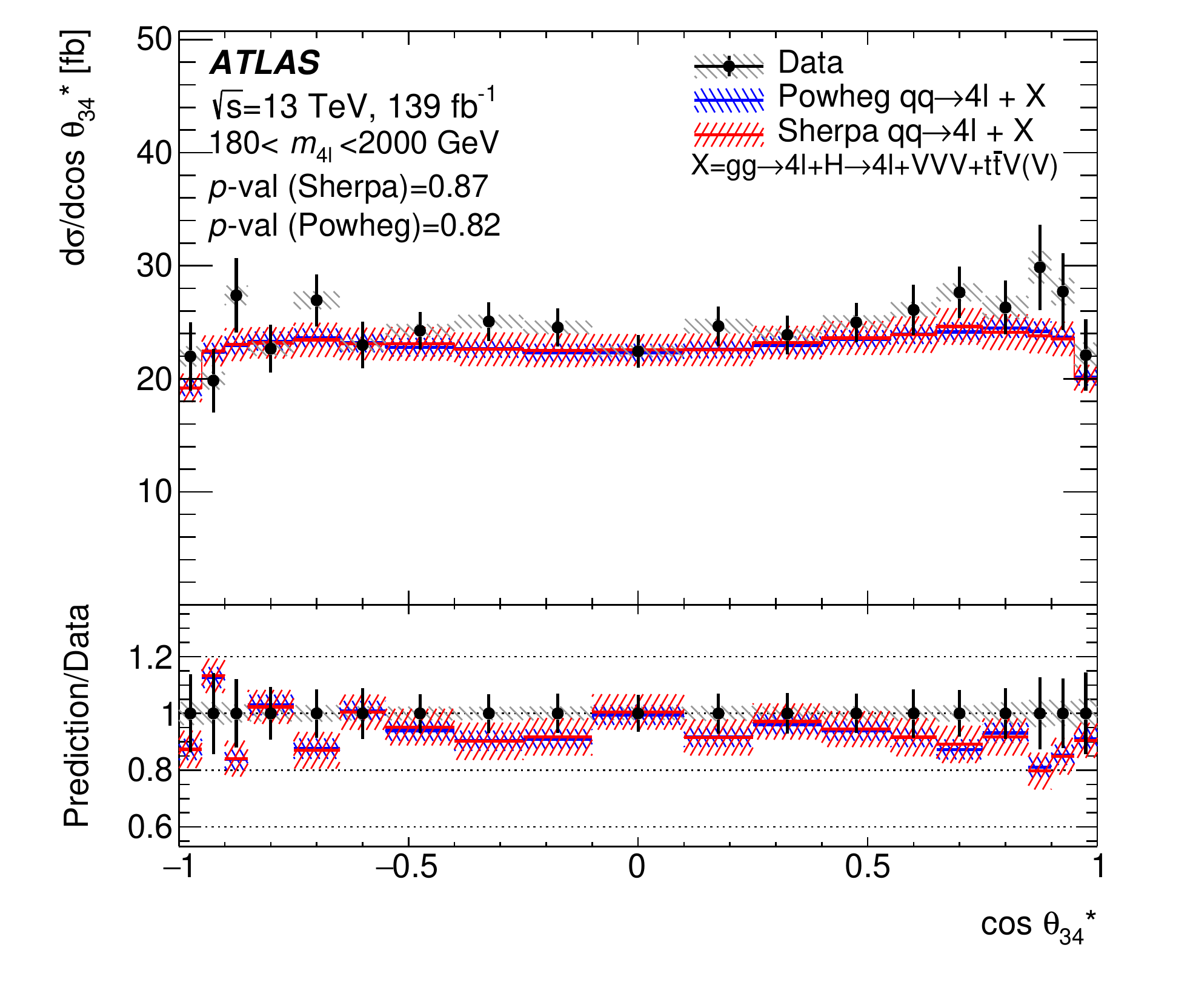}}\\
 
\caption{Differential cross-sections as a function of
$\costhetastar_{34}$ in the four \mFourL{} regions. The measured data (black
points)  are compared with the SM
prediction using either \SHERPA{} (red, with red hashed
band for the uncertainty) or \POWHEG + \pythia{} (blue,
with blue hashed band for the uncertainty) to model the \qqFourL{} contribution. The error bars on the data points give the total uncertainty
and the grey hashed band gives the systematic uncertainty.
\Pvalue{}
The  lower panel shows the ratio of the SM predictions to the   data. \label{fig:polarisation3}}
\end{figure}
 
\clearpage
\subsection{Extraction of the \ZFourL{} branching fraction}
\label{sec:branching}
The \ZFourL{} branching fraction is extracted using the
measured cross-section in the \ZFourL{} region, $\sigMeas{}
= 22.1 \pm 1.3$~fb, shown in Table~\ref{tab:fidxs}.
An extrapolation is performed using particle-level simulation to a
phase space that was used in previous measurements of the
branching fraction~\cite{CMS-SMP-16-017,STDM-2013-08,STDM-2017-09}, allowing direct comparison.
This phase space is defined by only two requirements: $80 < \mFourL < 100$~\GeV\
and $\mll > 4$~\GeV, with no requirements made
on the kinematics of the individual leptons. The definition of the
final state is based on leptons before photon  radiation (Born level),
with no isolation applied and excludes leptons  from $\tau$ decays. An acceptance factor
$\Accep = 0.0852 \pm 0.0015$ is defined
as the ratio of simulated particle-level events passing the fiducial
selection used for  \sigMeas{} to those in the extended
phase space, calculated using \SHERPA. The uncertainty comes from the
same sources as discussed in Section~\ref{sec:theory} and is dominated by the scale variations.
 
The branching fraction is calculated as
\begin{equation}
\BF = \frac{\left(\sigMeas{}-\sigPred{} \right) \times \fZ \times \fnonTau}{\sigZ \times \Accep},
\end{equation}
where $\sigPred = 0.22\pm 0.04$~fb is the predicted fiducial cross-section from sources other than
\qqFourL, obtained from the respective simulations;
$\fZ = 0.952 \pm 0.005$ is the fraction of \qqFourL{} coming from single \Z{}
production rather than the $t$-channel \Z\Z{} process, obtained from \POWHEGBOXV{v2};
$\fnonTau = 0.99186 \pm 0.00014$  is
the fraction of events where no leptons
originate from $\tau$ decays, obtained from \SHERPA; and \sigZ{} is the total cross-section for
single-\Z{} production, taken from the ATLAS measurement in
Ref.~\cite{STDM-2015-03} using 81~\ipb, with a correction of $0.935
\pm 0.001$ to account for the
different \Z{} mass window.
The uncertainties in \sigPred{} and \fZ{} come from the
sources discussed in Section~\ref{sec:theory}, and are dominated by scale
variations, while the uncertainty in \fnonTau{} is statistical only.

Together, these yield a result of
\begin{equation*}
\BF = \left( 4.41 \pm 0.13 \left(\text{stat.}\right) \pm 0.23 \left(\text{syst.}\right) \pm 0.09 \left(\text{theory}\right)\pm 0.12 \left(\text{lumi.}\right) \right) \times 10^{-6} =  \left( 4.41 \pm 0.30 \right) \times 10^{-6},
\end{equation*}
where the statistical, systematic and luminosity uncertainties come from  the
measurements of \sigMeas{} and \sigZ{}, and the theory uncertainty
includes contributions from \Accep, \fZ, \fnonTau{} and \sigPred{}.
Since the measurement of  \sigZ{} was performed with only 81~\ipb{} of
$pp$ collision data, all detector-related systematic uncertainties as
well as the luminosity uncertainty are conservatively treated as
uncorrelated between \sigMeas{} and \sigZ{}. The measurement is compatible with previous
measurements~\cite{CMS-SMP-16-017,STDM-2013-08,STDM-2017-09} and with
the SM prediction of $\left(4.50 \pm 0.01 \right) \times 10^{-6}$,
calculated with \powheg~\cite{STDM-2013-08}.
The current measurement is the most precise to date, and benefits from
an acceptance gain of 130\%{} relative to the previous ATLAS
measurement in Ref.~\cite{STDM-2017-09}.
 
\subsection{BSM interpretation}
\label{sec:int}
The measured cross-sections can be used to constrain BSM models by using the central values and covariance matrices made available in HEPData.
For convenience, the SM particle-level predictions described in this
paper are also included, so that comparisons  can be made
simply by adding BSM simulations to the provided SM predictions.
In this section, two well-motivated interpretations are given as
examples: constraints on effective field theory parameters and
constraints on a model with a \Zp{}
boson, where the \Zp{} bosons may be produced in pairs.
Limits are set by constructing a likelihood function,
\begin{equation}\label{eq:likelihoodEFT}
\mathcal{L} = \frac{1}{\sqrt{ \left(2\pi\right)^k | C |}} \exp \left\{ - \frac 1 2 \left[ \sigdata - \sigpred\left(\vec{\theta}\right) \right]^T C^{-1} \left[  \sigdata - \sigpred\left(\vec{\theta}\right)  \right] \right\}  \times \prod_{i} \mathcal{G}\left(\theta_{i}, 0, 1\right)  \, ,
\end{equation}
where \sigdata{} and $\sigpred$ and $C$ are defined in
Section~\ref{sec:xsecs}. In this case only the uncertainty in the SM
contribution to \sigpred{} is included in $C$.
In order to avoid an underestimation of the statistical uncertainty in
bins where the data has fluctuated downward, the statistical
uncertainty of the expected SM yield is used.
It was shown that using the observed statistical uncertainty leads to
test statistics with reduced exclusion power and with probability distributions that are far from asymptotic.
The BSM theoretical uncertainties are
included as nuisance parameters, $\vec{\theta}$, with Gaussian
constraints, $\mathcal{G}\left(\theta_{i}, 0, 1\right)$.
Since there are a number of measured differential cross-sections, and
the statistical correlation between them is not determined, the
variable providing the best expected sensitivity is chosen to set the limits
for a given point in BSM parameter space. A variable measured in slices of another variable
counts as one observable.
 
\subsubsection{Effective field theory constraints}
A SM effective field theory (SMEFT) formalism using the Warsaw
basis~\cite{Grzadkowski:2010es} is considered.
The effective Lagrangian is defined as
$\mathcal L_{\mathrm{SMEFT}} = \mathcal L_{\mathrm{SM}} + \sum_{i} \left(C_i^{d}/\Lambda^{(d-4)}\right)\mathcal O_i^{d},$
where $\mathcal O_i^{d} $ correspond to operators of dimension $d$ describing
the new interactions, the coefficients $C_i^{d}$ specify the strength of
the interactions and are known as
Wilson coefficients, and $\Lambda$ is the scale of the new physics.
Only dimension-six operators are considered for this paper as they
are the lowest-dimension operators that conserve lepton and baryon
number, and the effect of higher-dimensional operators is expected to
be suppressed.
Since at energies well below  $\Lambda$ only the ratio $C_i^{(d=6)}/\Lambda^2$ is accessible, $\Lambda$ is absorbed into the Wilson coefficients,
and they are redefined as $c_i =
{C_i^{(d=6)}}/{\Lambda^{2}}$.
 
The \texttt{SMEFTsim} package~\cite{SMEFTsim} is used for the SMEFT implementations of \texttt{FeynRules}~\cite{Alloul:2013bka}.
Monte Carlo samples were simulated with \MGMCatNLOV{\!2.6.5} + \PYTHIAV{\!8.243}
at LO precision in QCD~\cite{Alwall:2014hca}, with the \nnpdfnlo{} set of
PDFs. Uncertainties come from missing higher-order QCD corrections,
PDFs and from the uncertainty in $\alphas$,
using the same strategy as is used for the SM \qqFourL{} predictions, described
in Section~\ref{sec:theory}.
 
The set-up allows the separate generation
of the SM  only, BSM only and
interference contributions, with the total predicted cross-section
given by
\begin{equation}
\sigpred = \sigSM \times \left( 1 + c_i \cdot \sigINT  / \sigSMLO + c_i^{2}
\cdot \sigQUAD /   \sigSMLO\right),
\label{eq:linear-quad}
\end{equation}
where  \sigSM{}  is the most accurate SM predicted cross-section, described in Section~\ref{sec:theory}, using \SHERPA{} for the \qqFourL{} prediction,
$c_i \cdot
\sigINT{} $ is the
interference term between the SM and BSM contributions, also referred to as
the linear term, and $c_i^{2} \cdot \sigQUAD$ is the BSM-only
cross-section, also  referred to as the quadratic term.
Since the BSM contributions are LO cross-sections, they are scaled by the ratio of
\sigSM{}  to  \sigSMLO, which is the  LO SM prediction
given in the above set-up, where the assumption is that higher-order
effects are the same for SM contributions as they are for
contributions from new physics.
A minimal flavour-violating scenario is assumed and
the full list of 59 dimension-six operators  can be found in
Ref.~\cite{Grzadkowski:2010es}.
Only the following 22 coefficients
that give non-negligible contributions to the four-lepton final state are considered:
\begin{itemize}
\item three affecting Higgs couplings: \chg, \chgtil, \chdd;
\item one affecting gauge boson couplings: \chwb;
\item seven affecting the $Z\to \ell\ell$ vertex: \chd, \chu, \che, \chlone, \chlthr, \chqone, \chqthr;
\item eleven from four-fermion interactions (contact terms): \ced, \cee, \ceu, \cld, \cle, \cll, \cllone, \clqone, \clqthr, \clu, \cqe.
\end{itemize}
Each one is considered separately with the others set to zero.

The test statistic is based on a ratio of profiled likelihoods~\cite{Cowan:2010js},
\begin{equation}
q = - 2 \ln \frac{\mathcal{L}( c, \hat{\hat{\vec{\theta}}}(c)) } {\mathcal{L}(\hat{{c}}, \hat{{\vec{\theta}}} )}\;,
\label{eq:lambda_simplified}
\end{equation}
where  $c$ is the Wilson coefficient, $\hat{c}$ and $\hat{\vec{\theta}}$ are the unconditional maximum-likelihood estimators, and
$\hat{\hat{\vec{\theta}}}$ is the conditional maximum likelihood estimated under the $c$ hypothesis.
The value of the test statistic is used to compare the different probabilities of hypotheses to construct a likelihood scan over
$c$. For simplicity the
uncertainties are taken to be fixed at the SM value, $c=0$, rather
than scaling with $c$. It has been shown that this leads to very similar results and is much
less computationally expensive.
In the case that the signal results in a small enhancement relative to
the SM yield, with a dominant linear (interference) term, the use of
uncertainties fixed at the SM value results in an asymptotic test statistic. For large, local enhancements or cases where the quadratic term dominates, the behaviour becomes non-asymptotic. For this reason Monte Carlo toys are used to  evaluate the distribution of the test statistic when obtaining 95\%{} confidence level (CL) intervals.

Two sets of results are presented: those where the full predicted
cross-section from Eq.~(\ref{eq:linear-quad}) is used, referred to as the full EFT model, and
those where the quadratic term in Eq.~(\ref{eq:linear-quad}) is neglected, referred to as the
linear-only model. Since the linear terms of the missing
dimension-eight operators in the effective Lagrangian
expansion are at the same order in an expansion in  $1 / \Lambda$ as the quadratic terms of the
dimension-six operators, large differences indicate that
the neglected dimension-eight operators may play a non-negligible role.
Tables~\ref{tab:eft} and~\ref{tab:eft-linear} show the observed and expected exclusion limits
at 95\%{} CL for the 22 Wilson coefficients, for the full EFT model
and for the model with only linear terms, respectively. Also shown in the
tables is the most sensitive observable that is used to set the limit.
The results are also presented graphically in Figure~\ref{fig:EFT}.
When comparing the results in Tables~\ref{tab:eft} and~\ref{tab:eft-linear},
the coefficients fall into four categories: those (\chdd, \chwb,
\che, \chlone, \chqthr{}, \chlthr{} and \cllone{}) that have a negligible
contribution from the quadratic term and therefore have very similar
limits whether or not the quadratic term is included;
those (\chqone{} and \clqthr{}) that, because of the quadratic term, make small positive contributions to the
total prediction, enhancing (lessening) the positive (negative)
linear contribution at positive (negative) values of the coefficient, and hence shifting both the lower and upper limits to slightly
lower values;
those
(\chg{} and \chu) that have more stringent
limits when using the linear-only terms because of the presence of double
maxima in the likelihood scans when the quadratic term is included; and the remaining 11
coefficients that have non-negligible contributions from the quadratic
term and hence more stringent limits when they are included (including
\chgtil{}, which has
zero contribution from the linear term). The differences seen in the
limits  in the last category indicate that dimension-eight terms may
not be negligible.
 
In general the observed limits are compatible with the expected
limits.
One  exception is  the observed lower limit on \clqone{}, which is significantly less stringent than
the expected limit for the results using linear-only terms.  This is an artifact of the simple model of the scale uncertainty affecting
the EFT prediction.
In general, for a large EFT signal that is incompatible with the data,
the EFT scale uncertainty's nuisance parameter is pulled in order to
bring the prediction closer to the observation. The nuisance parameter's constraint term
penalises this behaviour in the likelihood.
However, for large negative values of \clqone{} the size and shape of
the scale uncertainty's effect on the signal prediction
are related in a way that produces a prediction precisely imitating the
statistical fluctuations in the measured cross-section, increasing the likelihood.
This leads to a larger than expected observed lower limit on \clqone.
 
Limits have already been placed on \chg, \chgtil{} and \chwb{}, in
a measurement of the
\HFourL{} cross-section~\cite{Aad:2020mkp}. For \chg{}  the limits are
more stringent
in the  \HFourL{} cross-section analysis, but for \chgtil{} the limits are very similar,
with this paper providing slightly tighter constraints. For
\chwb{} the constraints from this paper are significantly more
stringent
with $[ -0.20, 0.21]$ expected and $[-0.29,0.13]$ observed compared
to $[-1.09,0.99]$ expected and $[-1.06,0.99]$ observed in the
\HFourL{} cross-section paper.
The improvement can be understood as being due to the fact that changes in
\chwb{}  affect the entire \mFourL{} spectrum, not just the region
close to  \mH{}.
Limits on the coefficients affecting the $Z\to \ell\ell$ vertex have been obtained
previously from a global fit to LEP and LHC data~\cite{Ellis:2018gqa}, and are generally
one or two orders of magnitude more stringent than the limits in this paper.

\begin{table}[t]
\centering
\caption{The expected and observed confidence intervals at 95\%{}
CL for the SMEFT Wilson coefficients, including both the linear and
quadratic terms. The most sensitive
observable indicated for each coefficient is used for the
constraints. Only one coefficient is fitted at a time, with all
others set to zero.\label{tab:eft} }
 
\begin{tabular} {c r c c }
\hline
\hline
Coefficient & Observable  & 95\% CL Expected [\TeV$^{-2}$] &   95\% CL Observed [\TeV$^{-2}$]  \\
\hline
\chg         & \mZTwo{} vs. \mFourL{}      & $[-0.18,-0.027] \cup [-0.014,0.011]$ & $[-0.20,-0.029] \cup [-0.010,0.0    12]$  \\
\chgtil      & \mZTwo{} vs. \mFourL{}      & $[-0.031,0.031]                    $ & $[-0.033,0.033]$  \\
\chdd        & \mZTwo{} vs. \mFourL{}      & $[-0.45,0.44]                      $ & $[-0.60,0.29]  $       \\
\hline
\chwb        & \mZTwo{} vs. \mFourL{}      & $[-0.20,0.21]                      $ & $[-0.29,0.13]  $        \\
\hline
\chd         & \ptZOne{} vs. \mFourL{}     & $[-4.9,9.8]                        $ & $[-2.6,8.3]    $     \\
\chu         & \dPhill{} vs. \mFourL{}     & ~$[-11, 2.8]                        $ & $ [-13,-6.9]    \cup  [-1.5,4.4]    $     \\
\che         & \dPhiPairs{} vs. \mFourL{}  & $[-0.46,0.49]                      $ & $[-0.70,0.21]  $       \\
\chlone      & \dPhiPairs{} vs. \mFourL{}  & $[-0.39,0.37]                      $ & $[-0.19,0.55]  $       \\
\chlthr      & \dPhill{} vs. \mFourL{}     & $[-0.28,0.29]                      $ & $[-0.47,0.12]  $       \\
\chqone      & \mZTwo{} vs. \mFourL{}      & $[-0.93,0.69]                      $ & $[-1.6,0.43]   $      \\
\chqthr      & \dPhiPairs{} vs. \mFourL{}  & $[-0.34,0.33]                      $ & $[-0.15,0.52]  $       \\
\hline
\ced         & \mZTwo{} vs. \mFourL{}      & $[-0.49,0.39]                      $ & $[-0.51,0.41]  $      \\
\cee         & \mZTwo{} vs. \mFourL{}      & $[-38,35]                          $ & $[-33,42]      $  \\
\ceu         & \mFourL{}~~~~~~             & $[-0.21,0.35]                      $ & $[-0.14,0.21]  $       \\
\cld         & \mZTwo{} vs. \mFourL{}      & $[-0.40,0.34]                      $ & $[-0.41,0.36]  $     \\
\cle         & \mZTwo{} vs. \mFourL{}      & $[-23,22]                          $ & $[-21,26]      $   \\
\cll         & \mZTwo{} vs. \mFourL{}      & $[-23,21]                          $ & $[-20,25]      $   \\
\cllone      & \dPhiPairs{} vs. \mFourL{}  & $[-0.34,0.33]                      $ & $[-0.17,0.50]  $       \\
\clqone      & \mFourL~~~~~~               & $[-0.14,0.28]                      $ & $[-0.086,0.17] $       \\
\clqthr      & \mZTwo{} vs. \mFourL{}      & $[-0.083,0.071]                    $ & $[-0.064,0.081]$       \\
\clu         & \mFourL~~~~~~               & $[-0.24,0.32]                      $ & $[-0.16,0.20]  $      \\
\cqe         & \mFourL~~~~~~               & $[-0.17,0.21]                      $ & $[-0.11,0.14]  $      \\
\hline
 
\hline
\hline

\end{tabular}
\end{table}
 
\begin{table}[t]
\centering
\caption{The expected and observed confidence intervals at 95\%{}
CL for the SMEFT Wilson coefficients, including only linear terms. The most sensitive
observable indicated for each coefficient is used for the
constraints. Only one coefficient is fitted at a time, with all
others set to zero.\label{tab:eft-linear} }
 
\begin{tabular} {c r c c }
\hline
\hline
Coefficient & Observable  & 95\% CL Expected [\TeV$^{-2}$] &   95\% CL Observed [\TeV$^{-2}$]  \\
\hline
\chg         & \mZTwo{} vs. \mFourL{}       &  $[-0.011,0.013]$ & $[-0.0090,0.015]$~~     \\
\chgtil      & \mZTwo{} vs. \mFourL{}       &  $-$              & $-$      \\
\chdd        & \mZTwo{} vs. \mFourL{}       &  $[-0.46,0.45] $  & $[-0.63,0.28]$     \\
\hline
\chwb        & \mZTwo{} vs. \mFourL{}       &  $[-0.21,0.20] $  & $[-0.29,0.13]$     \\
\hline
\chd         & \ptZOne{} vs. \mFourL{}      & $[-10,10]      $  & $[-3.0,18]$~    \\
\chu         & \dPhill{} vs. \mFourL{}      & $[-3.5,3.7]    $  & $[-1.6,6.1]$     \\
\che         & \dPhiPairs{} vs. \mFourL{}   & $[-0.47,0.46]  $  & $[-0.75,0.21]$     \\
\chlone      & \dPhiPairs{} vs. \mFourL{}   & $[-0.37,0.38]  $  & $[-0.19,0.57]$     \\
\chlthr      & \dPhill{} vs. \mFourL{}      & $[-0.29,0.29]  $  & $[-0.51,0.12]$     \\
\chqone      & \mZTwo{} vs. \mFourL{}       & $[-0.81,0.78]  $  & ~~$[-1.1,0.47] $    \\
\chqthr      & \dPhiPairs{} vs. \mFourL{}   & $[-0.34,0.35]  $  & $[-0.15,0.53]$     \\
\hline
\ced         & \mZTwo{} vs. \mFourL{}       & $[-1.3,1.8]    $  & $[-1.0,2.3] $~~    \\
\cee         & \mZTwo{} vs. \mFourL{}       & $[-59,65]      $  & ~~$[-25,100]   $\\
\ceu         & \mFourL{}~~~~~~                    & $[-0.62,0.45]  $  & $[-0.36,0.63]$     \\
\cld         & \mZTwo{} vs. \mFourL{}       & $[-1.8,2.5]    $  & $[-1.3,3.0]  $   \\
\cle         & \mZTwo{} vs. \mFourL{}       & $[-63,68]      $  & ~~$[-18,130]   $ \\
\cll         & \mZTwo{} vs. \mFourL{}       & $[-39,43]      $  & $[-17,70]    $ \\
\cllone      & \dPhiPairs{} vs. \mFourL{}   & $[-0.34,0.34]  $  & $[-0.18,0.50]$     \\
\clqone      & \mFourL~~~~~~                      & $[-0.76,0.40]  $  & ~~$[-4.1,0.53] $     \\
\clqthr      & \mZTwo{}  vs. \mFourL{}      & $[-0.059,0.083]$  & $[-0.050,0.098]$     \\
\clu         & \mFourL~~~~~~                      & $[-1.4,0.99]   $  & $[-0.78,1.4] $~~    \\
\cqe         & \mFourL~~~~~~                      & $[-1.1,0.83]   $  & $[-0.72,1.2] $~~    \\
\hline
\hline

\end{tabular}
\end{table}

\begin{figure}[htb]
\centering
\subfigure[]{ {\includegraphics[scale=0.39]{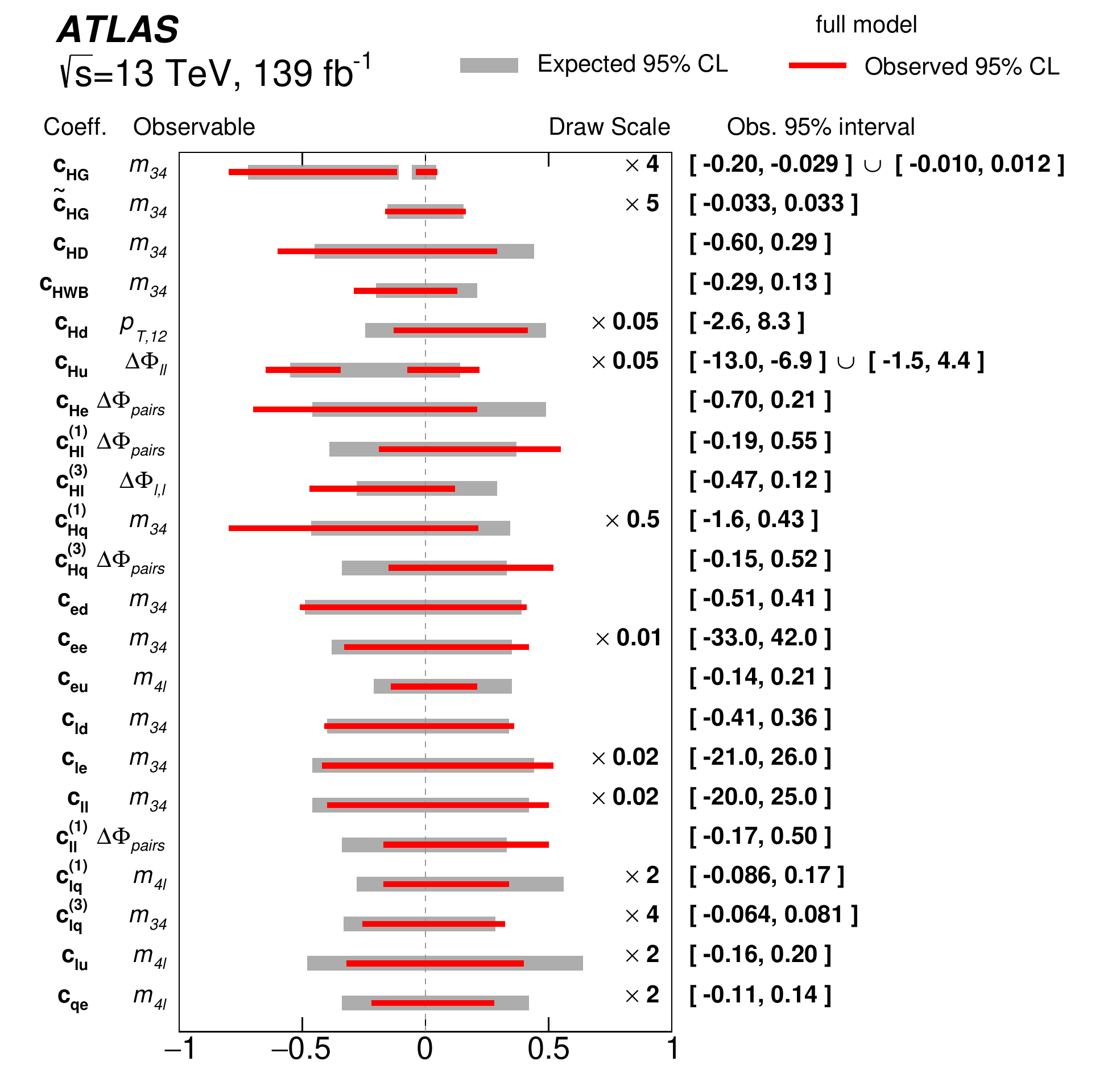}} \label{fig:EFT-full}}
\subfigure[]{ {\includegraphics[scale=0.39]{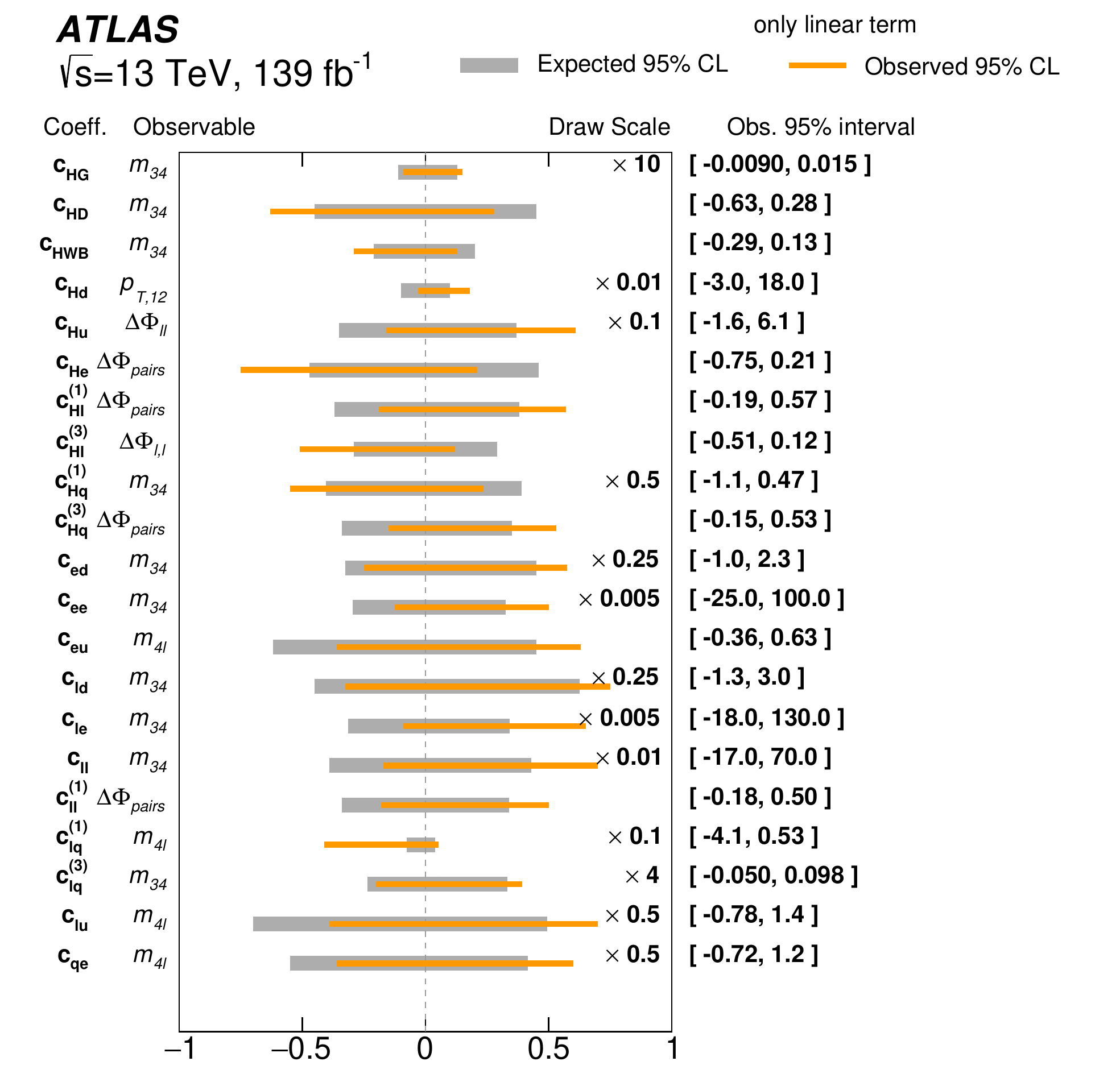}} \label{fig:EFT-linear}}
\caption{The expected and observed confidence intervals at 95\%{}
CL for the SMEFT Wilson coefficients, which are in units of
\TeV$^{-2}$, for (a) the full model and (b) only the linear terms. The multiplicative number shown to the right of plotted points indicates a scaling factor that is applied to the limit for the purpose of the plot, in order to show all  coefficients on the same scale.
The most sensitive
observable indicated for each coefficient is used for the
constraints. Only one coefficient is fitted at a time, with all
others set to zero. \label{fig:EFT}}
\end{figure}

\subsubsection{$B-L$ gauge model constraints}
 
The data are used to set limits on a model in which the global
baryon-number-minus-lepton-number $(B-L)$ symmetry is treated as a local gauge symmetry and spontaneously broken~\cite{Deppisch:2018eth}.
This model predicts a \Zp, the gauge boson of the new local symmetry,
and an exotic Higgs boson $h_{2}$, from the symmetry breaking, that
mixes with the SM Higgs boson, with a
mixing angle $\alpha$.
The new Higgs boson and the SM Higgs boson can decay into \Z\Z{} or \Zp\!\Zp, with the gauge
bosons decaying into lepton pairs.
 
The sensitivity of previous LHC measurements to this model was presented in Ref.~\cite{Amrith:2018yfb}, where
four-lepton measurements were seen to provide significant
constraints. One of the scans from that study is repeated here using
all the measured observables, to derive
exclusion limits in the plane of \sina{} versus the mass of $h_{2}$, \mht, for a low-mass (35~\GeV) \Zp{} which is
weakly coupled to the SM ($g^\prime = 10^{-3}$). In this scenario, depending on the other parameters, \Z\Z{} and \Zp\!\Zp{} production via
one of the Higgs bosons may lead to contributions to the
cross-sections measured in this paper.
 
BSM events were generated using Herwig~7.2~\cite{Bahr:2008pv,Bellm:2015jjp}.
Interference terms, and the impact of the model on SM decays of the SM
Higgs, are not taken into account. The impact of this on the limits is
expected to be negligible.
 
The likelihood function defined in Eq.~(\ref{eq:likelihoodEFT}) is
used to define
the test statistic, $\tilde{q}$, as
\begin{equation}
\tilde{q} =
\begin{cases}
- 2 \ln \frac{\mathcal{L}( {\mu}, \hat{\hat{\vec{\theta}}}({\mu})) } {\mathcal{L}(0,         \hat{\hat{\vec{\theta}}}(0) )} & \hat\mu < 0, \\
- 2 \ln \frac{\mathcal{L}( {\mu}, \hat{\hat{\vec{\theta}}}({\mu})) } {\mathcal{L}(\hat{\mu}, \hat{\vec{\theta}} )} & 0 \leq \hat\mu \leq \mu,  \\
0 & \hat\mu > \mu,
 
\end{cases}
\label{eq:lambda_qtilde}
\end{equation}
where $\mu$ determines the strength of the signal process, with $\mu =
0$ corresponding to the SM hypothesis and $\mu = 1$ being
the nominal signal hypothesis.
The $\mathrm{CL_s}$ method~\cite{Read:2002hq} is then used to derive
the 95\%{} CL exclusion contour in the $(\sina, \mht)$ plane as
shown in Figure~\ref{fig:BL1_limits_2D_limit}.\footnote{The asymptotic formulae from
Ref.~\cite{Cowan:2010js} are used, and their validity is checked by using pseudo-data simulations
for a number of points in the $(\sina, \mht)$ plane.}
Figure~\ref{fig:BL1_limits_2D_map} displays for each $(\sina, \mht)$ point the observable which has the
highest expected sensitivity and was therefore used to set the limit.
The most sensitive observable changes in the region $\mht \sim 125~\GeV$ since the
phenomenology of the model changes as the mass of the exotic Higgs boson approaches
that of the SM Higgs boson.
In order to demonstrate the power of the
different variables, a similar scan was performed using the \mFourL{}
distribution only, and the expected limits on \sina{} weakened from 0.28
to 0.46 at high \mht.
As can be seen in Figure~\ref{fig:BL1_limits_2D_map}, improvements at high \mht{} come mostly from including measurements of
\mZOne{}.
\begin{figure}
\centering
\subfigure[]{\includegraphics[scale=0.39]{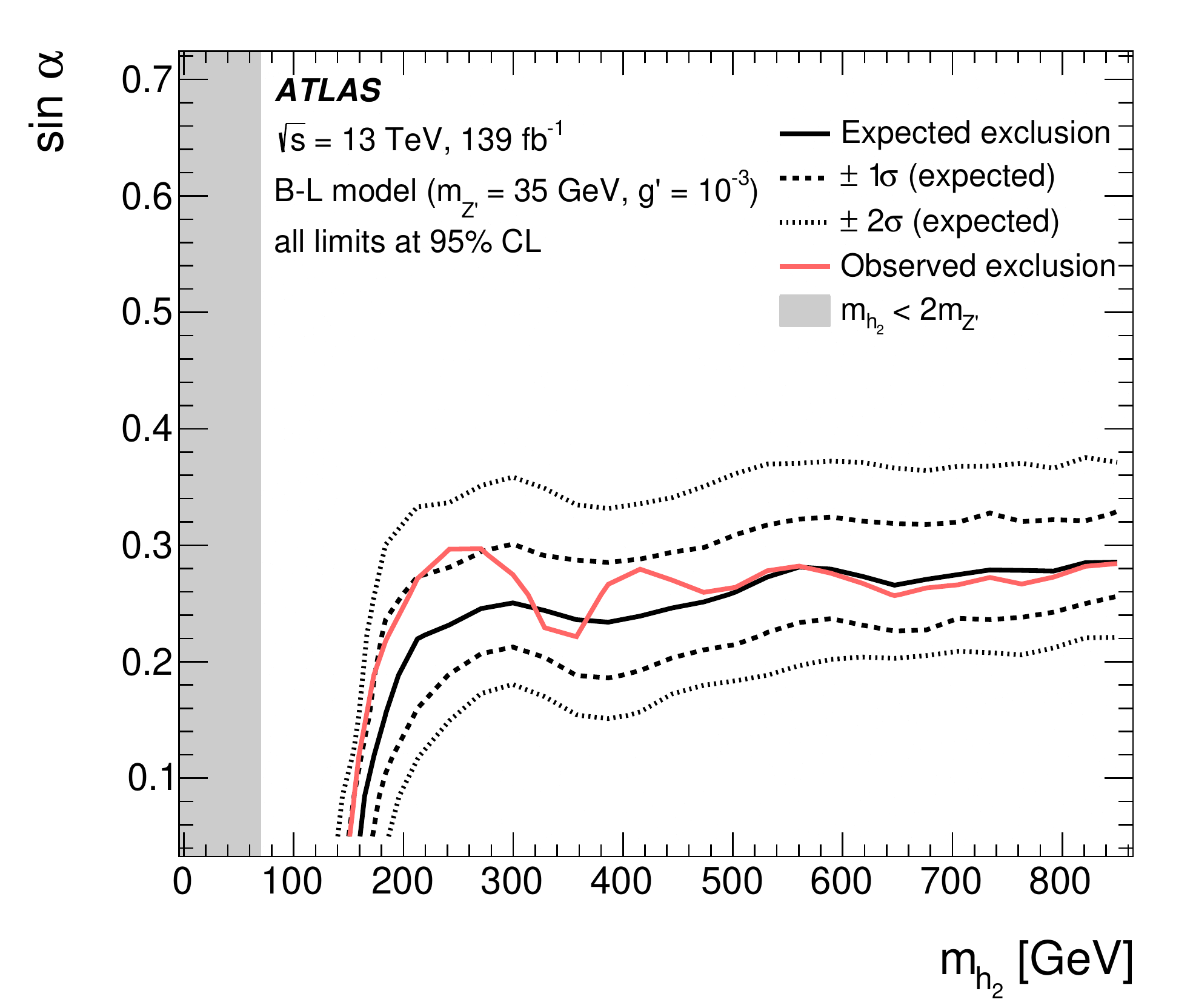} \label{fig:BL1_limits_2D_limit}}
\subfigure[]{\includegraphics[scale=0.39]{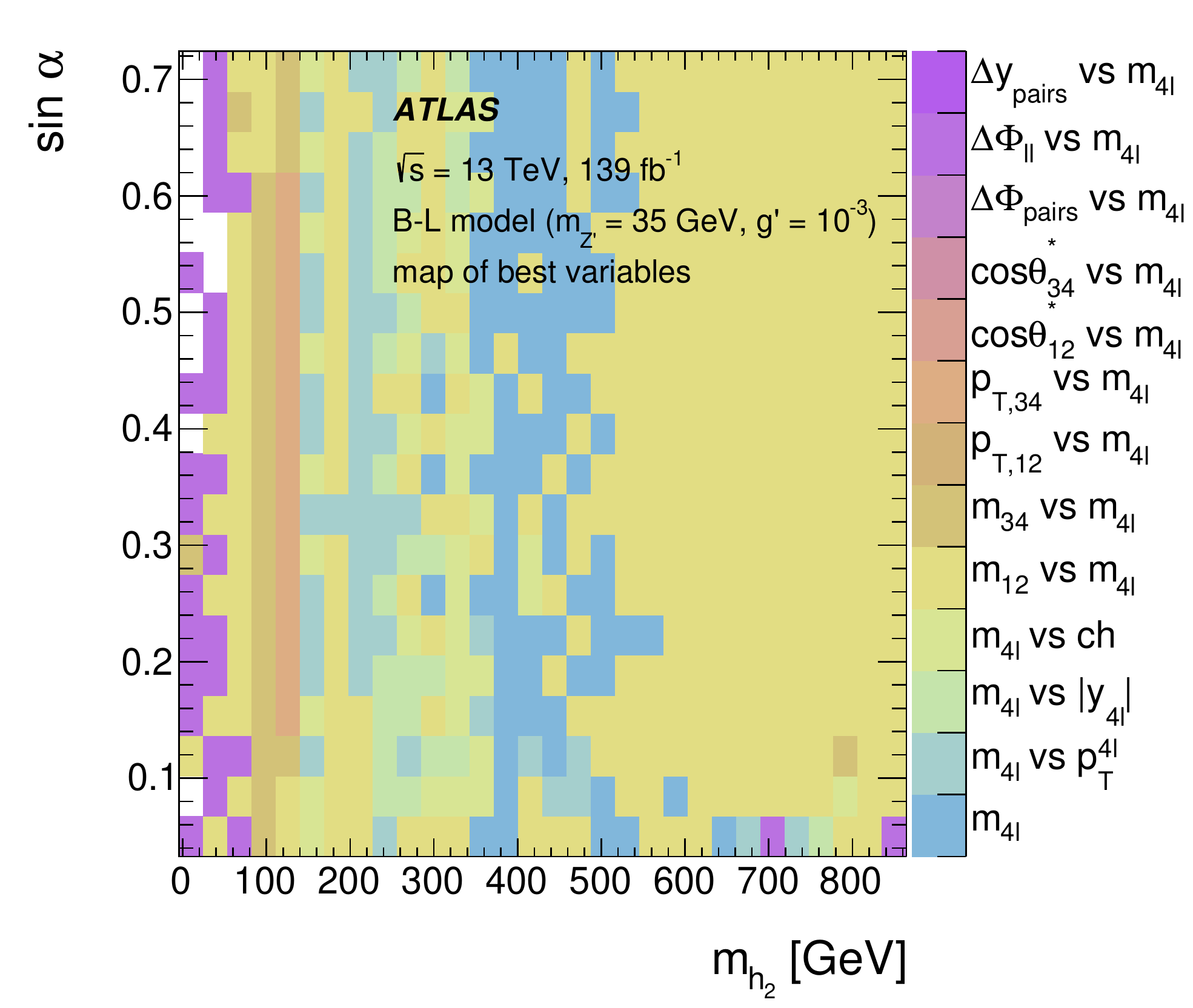}\label{fig:BL1_limits_2D_map}}
\caption[]{ (a) Exclusion contour at  95\% confidence level for the $B-L$ gauge model~\cite{Deppisch:2018eth} in the
plane of \sina{} versus \mht{} for $m_{\Zp} = 35$~\GeV\ and $g^\prime = 10^{-3}$.
The red line shows the observed exclusion and the solid black line
shows the expected exclusion, with the dashed black lines indicating
the $\pm 1\sigma$ and $\pm 2\sigma$ uncertainty bands. The region above and to the left
of the lines is excluded, up to the grey band, which shows the
region $\mht{} < 2m_{\Zp}$, where the four-lepton final state has no
sensitivity.
(b) Colour palette indicating which observable has the highest
expected sensitivity at each point in the parameter space and is
used to set the limit.  \label{fig:BL1_limits_2D}}
\end{figure}
The previous study~\cite{Amrith:2018yfb} only covers $\sina{}>0.4$
for most values of \mht{}, and provides no constraints for $\mht > 600$~\GeV.
The new overall limits significantly improve on this, excluding $\sina{}>0.28$
for most of the plane, even at high \mht{}, and,
for example, for $\mht =500$~\GeV, the limit on $\sina{}$ improves
from 0.5 to 0.26.
 
\FloatBarrier
\section{Conclusion}
\label{sec:conclusion}
Measurements of inclusive and differential fiducial cross-sections in
four-lepton events are presented using 139~\ifb{} of $\sqrt{s}=13$~\TeV\
proton--proton collisions, collected by the ATLAS detector during
Run~2 of the LHC.
Included are measurements of the
four-lepton invariant mass, the dilepton masses, and other kinematic variables of the
four leptons and dilepton pairs, including angular correlations.
The measured phase-space is significantly increased compared to
previous measurements with four leptons, in particular reducing
requirements that specifically enhance the contribution from SM \Z\Z{}
production. The methodology used to correct for detector effects was
developed to have minimal bias in the case of possible BSM contributions to the final-state.
The measured cross-sections are compared with state-of-the-art SM predictions and are
found to be in agreement.
 
The final state has contributions from a number of SM processes,
including \ZFourL{}, \HFourL{} and the dominant quark-induced
$t$-channel \qqFourL{} process.
The region dominated by \ZFourL{} production is used to extract the
most precise measurement of the \ZFourL{}
branching fraction to date, $\BF = \left( 4.41 \pm 0.13 \left(\text{stat.}\right) \pm 0.23 \left(\text{syst.}\right) \pm 0.09 \left(\text{theory}\right)\pm 0.12 \left(\text{lumi.}\right) \right) \times 10^{-6}$.  The result is consistent with previous
measurements and with the SM prediction.
 
Various BSM effects can contribute to the final state.
The data are used
to constrain SMEFT
coefficients and
a model based on a spontaneously broken $B-L$ gauge
symmetry.
All necessary information for a detailed reinterpretation has been
made public so it can be used to probe further BSM models in the future.
 
\clearpage
\appendix
\part*{Appendix}
\addcontentsline{toc}{part}{Appendix}
\label{app:more-results}
This Appendix shows the differential cross-sections that are not
included in Section~\ref{sec:xsecs}.
In each case the data are compared with the SM predictions. Two
predictions are shown, one where the \qqFourL{} process is simulated
with \SHERPA{} and one where it is simulated with \POWHEG{} +
\pythia{}. All the other processes are the same for both predictions.
 
Figure~\ref{fig:channelres} shows the \mFourL{} distribution
separately for the
three channels $4\mu$, $4e$ and  $2e2\mu$;
Figures~\ref{fig:pt4lres} and~\ref{fig:y4lres} show it
in slices of \ptFourL{} and $|\yFourL|$ respectively.
The SM predictions describe the data well for most of the
distributions. However,  the \POWHEG{} +
\pythia{} prediction in the highest-\pt{} slice in
Figure~\ref{fig:pt4lres}, and in the off-shell \Z\Z{} region in the $0.9 < |\yFourL| < 1.2$
slice in  Figure~\ref{fig:y4lres}, is below the data.
 
Figures~\ref{fig:ptz1res},~\ref{fig:ptz2res},~\ref{fig:dphipairs-rest},
and~\ref{fig:dy-rest} respectively show \ptZOne{}, \ptZTwo{},
\dPhiPairs{} and \dYPairs{} separately in the \ZFourL{}, \HFourL{} and
off-shell \Z\Z{} regions. The on-shell  \Z\Z{} region is shown in
Section~\ref{sec:xsecs}.
The data are well described by the SM predictions. The $p$-values for
the \dPhiPairs{} distribution in the off-shell \Z\Z{} region are low,
which is mainly driven by statistical fluctuations in the data,
together with predictions that are lower than the observations at low
\dPhiPairs{} values, especially for the \POWHEG{} +
\pythia{} prediction.

\begin{figure}[htb]
\centering
\subfigure[$4\mu$ channel ]{\includegraphics[width = 0.48\textwidth]{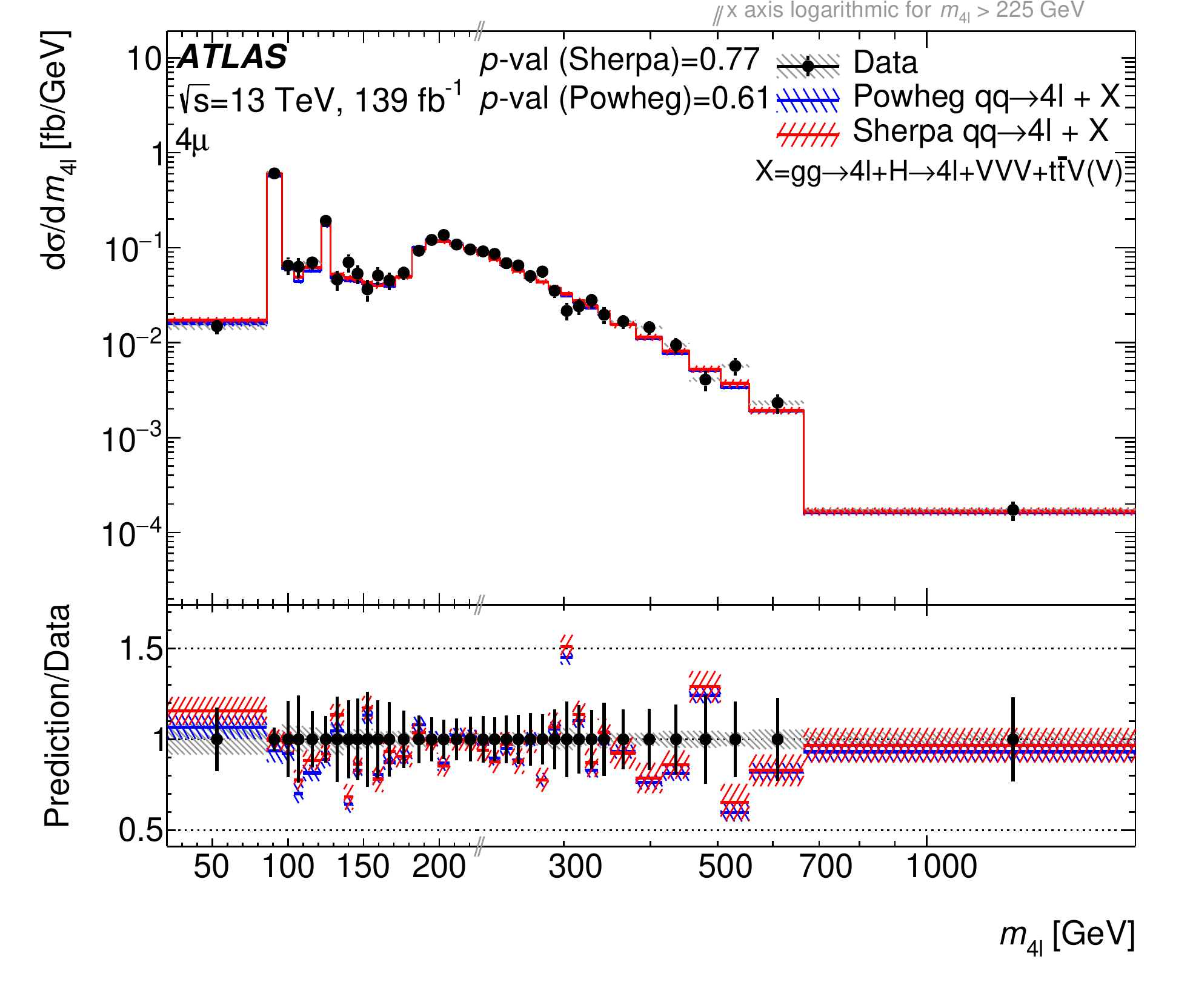}}
\subfigure[$4e$ channel]{\includegraphics[width = 0.48\textwidth]{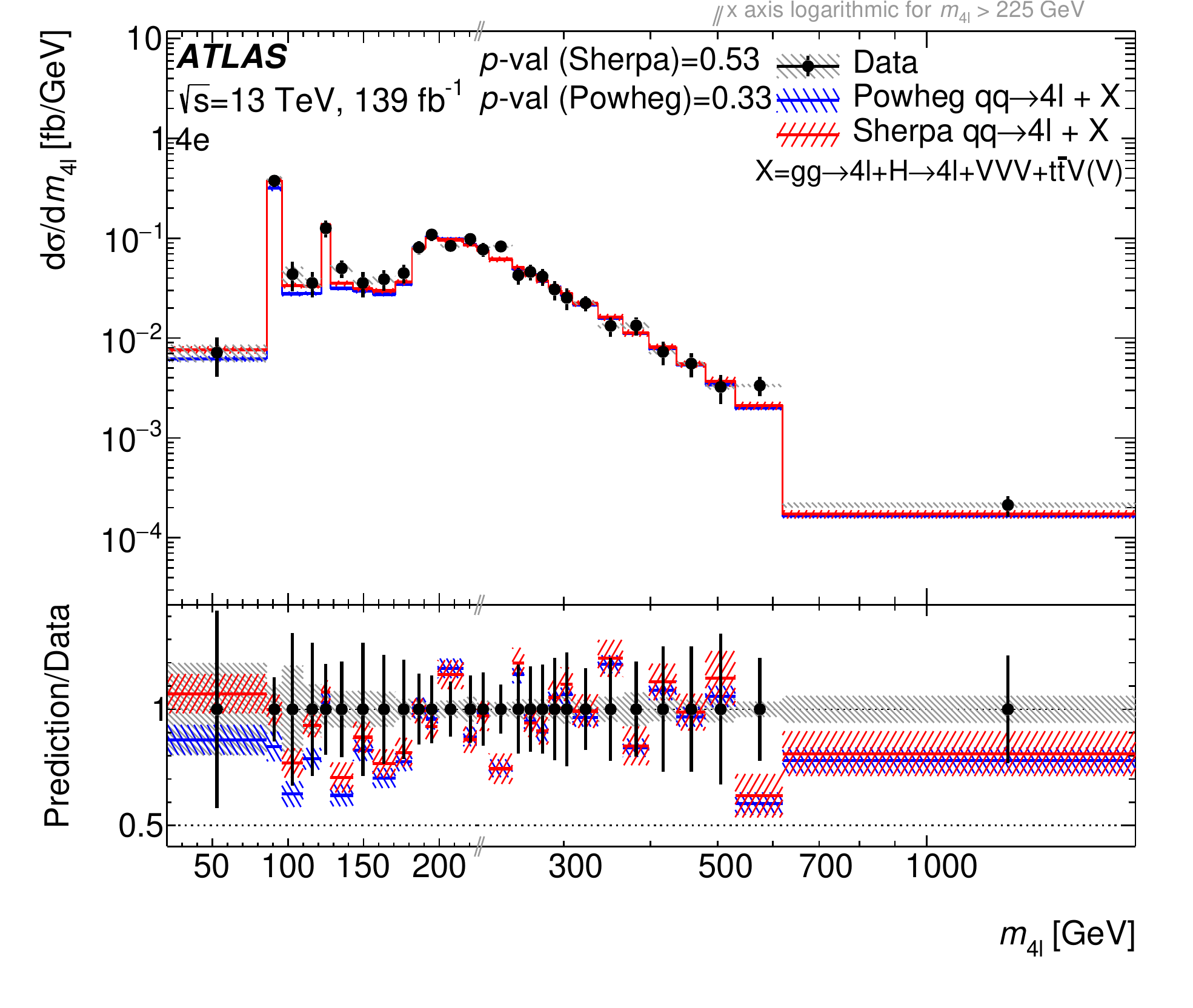}}\\
\subfigure[$2e2\mu$ channel]{\includegraphics[width = 0.48\textwidth]{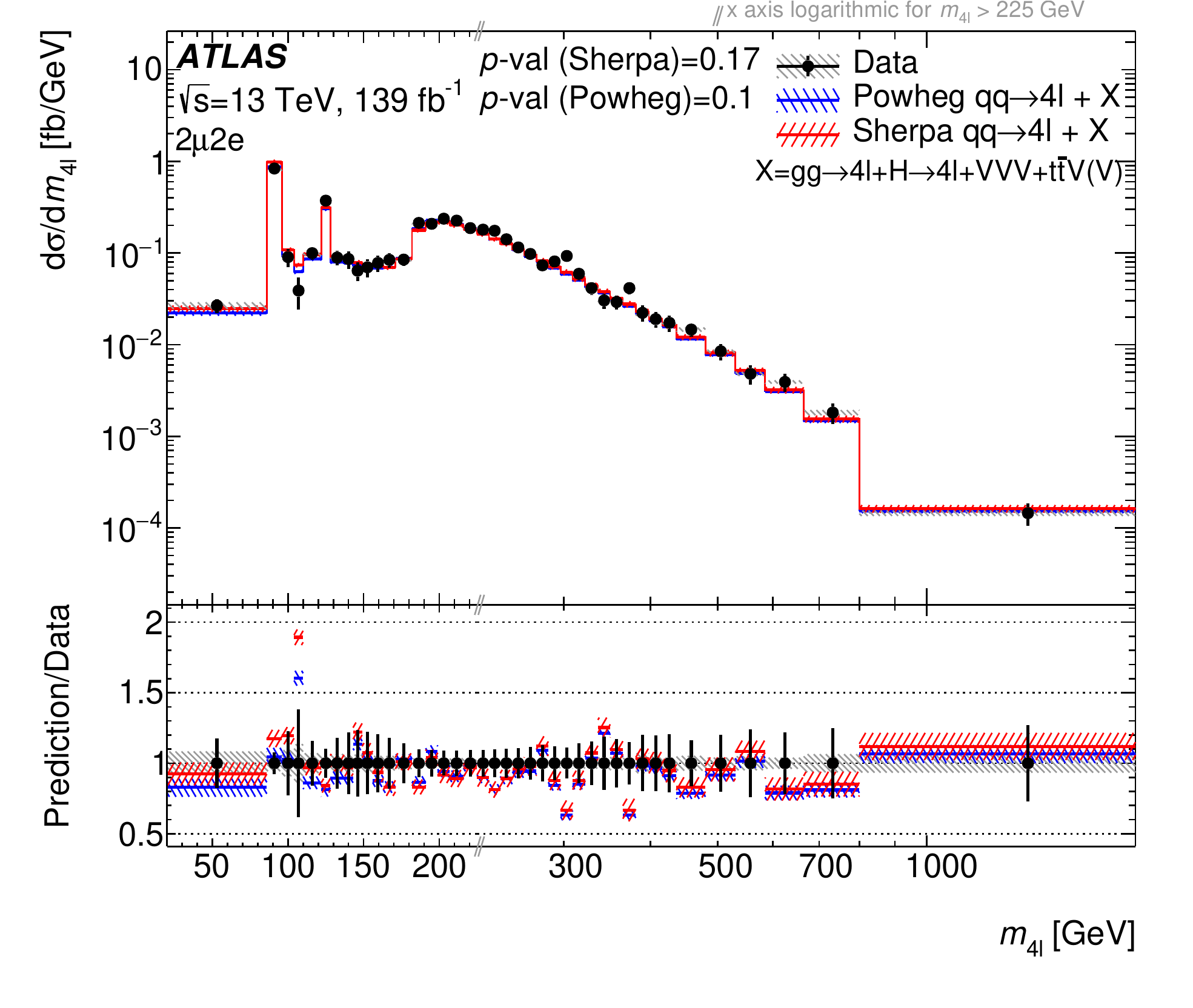}}
\caption{Differential cross-section as a function of \mFourL{} for each lepton
flavour channel. The measured data (black points)  are compared with the SM
prediction using either \SHERPA{} (red, with red hashed
band for the uncertainty) or \POWHEG{} + \pythia{} (blue,
with blue hashed band for the uncertainty) to model the \qqFourL{} contribution. The error bars on the data points give the total uncertainty
and the grey hashed band gives the systematic uncertainty.
\Pvalue{}
The
lower panel shows the ratio of the SM predictions to the
data.  The $x$-axis is on a linear scale until $\mFourL = 225$~\GeV,
where it switches to a logarithmic scale. \label{fig:channelres}}
\end{figure}
 
\begin{figure}[htb]
\centering
\subfigure[$\ptFourL  < 10~\GeV$
]{\includegraphics[width = 0.42\textwidth]{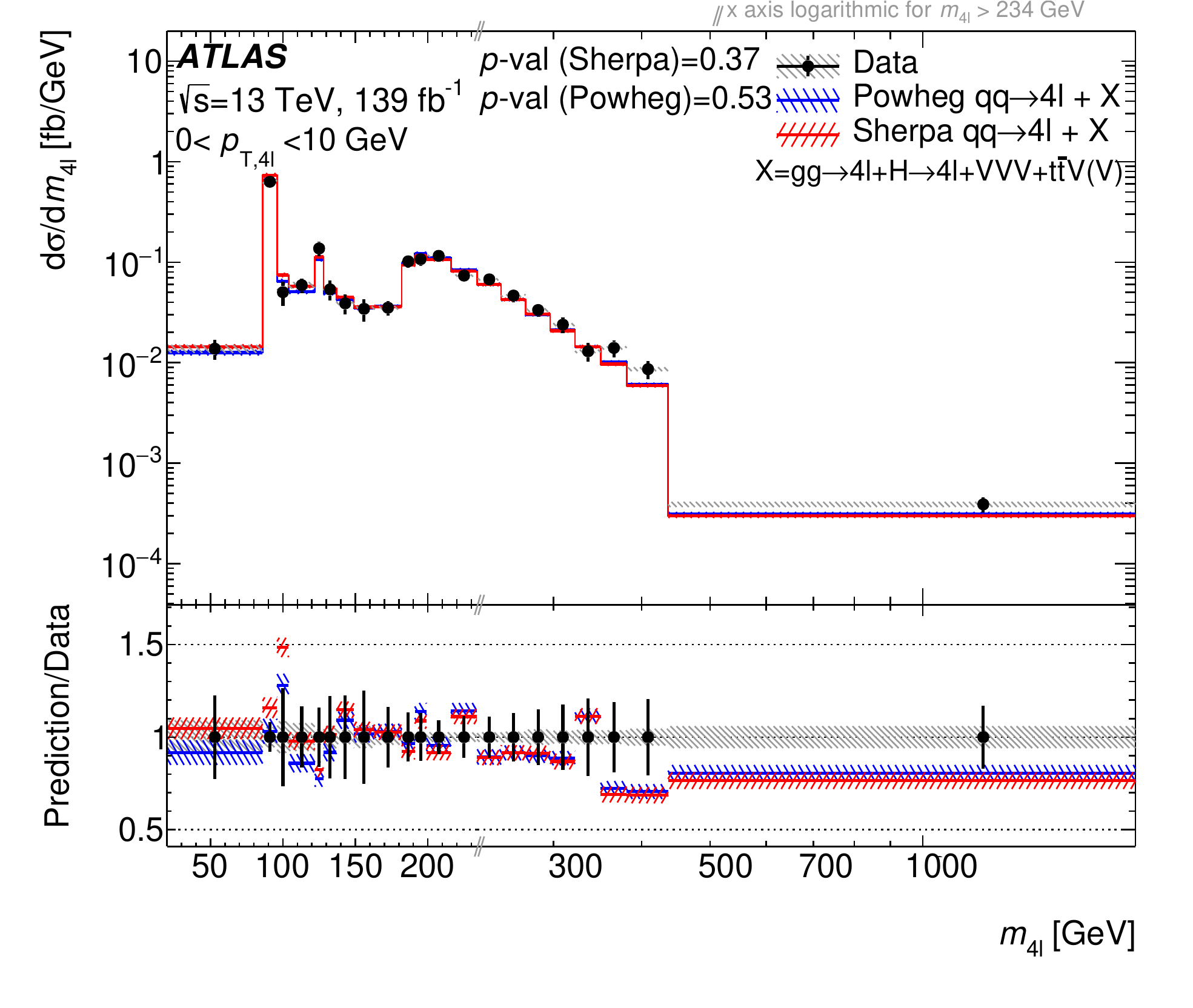}}
\subfigure[$10~\GeV <  \ptFourL  < 20~\GeV$
]{\includegraphics[width = 0.42\textwidth]{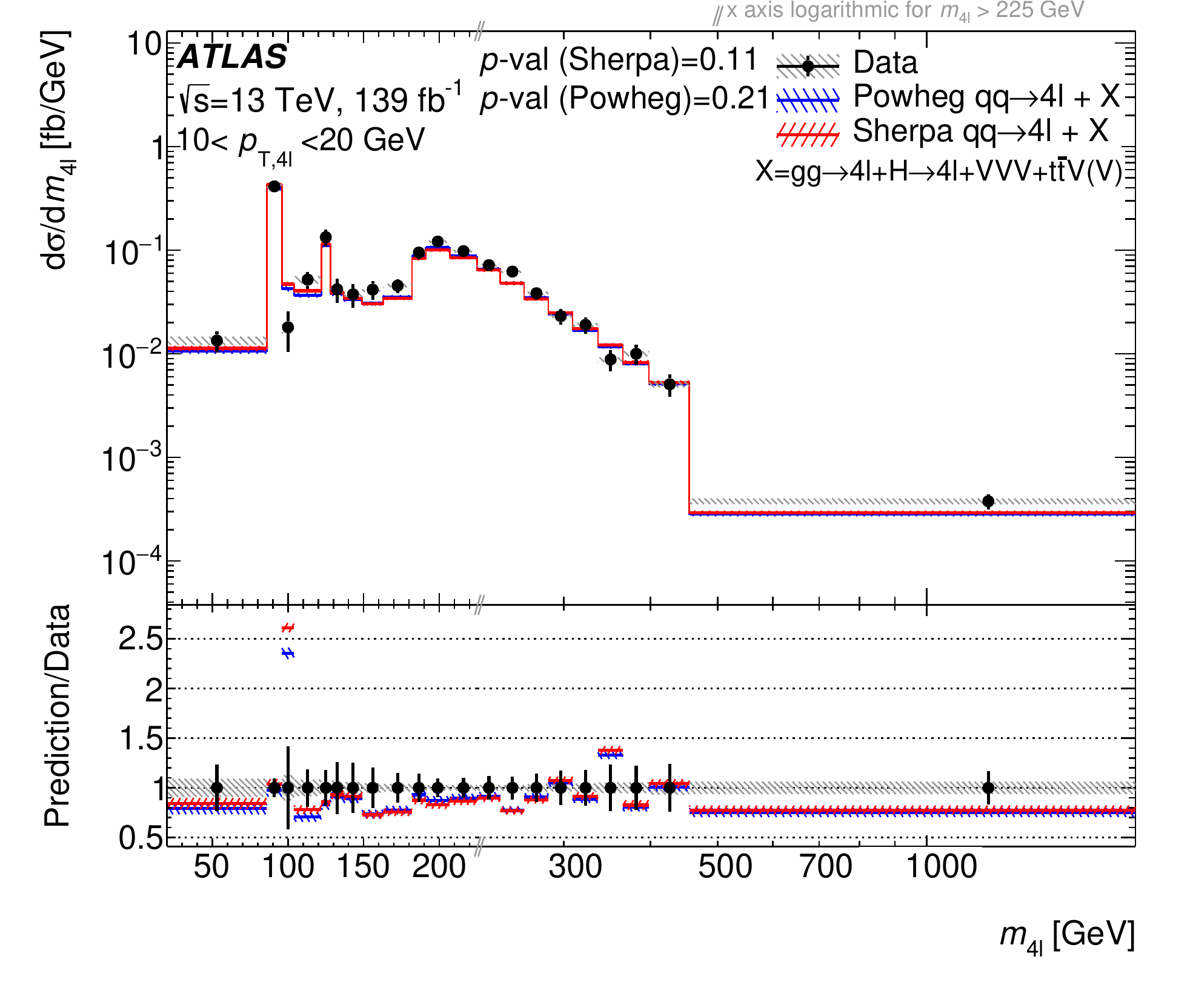}}\\
\subfigure[$20~\GeV <  \ptFourL < 50~\GeV$
]{\includegraphics[width = 0.42\textwidth]{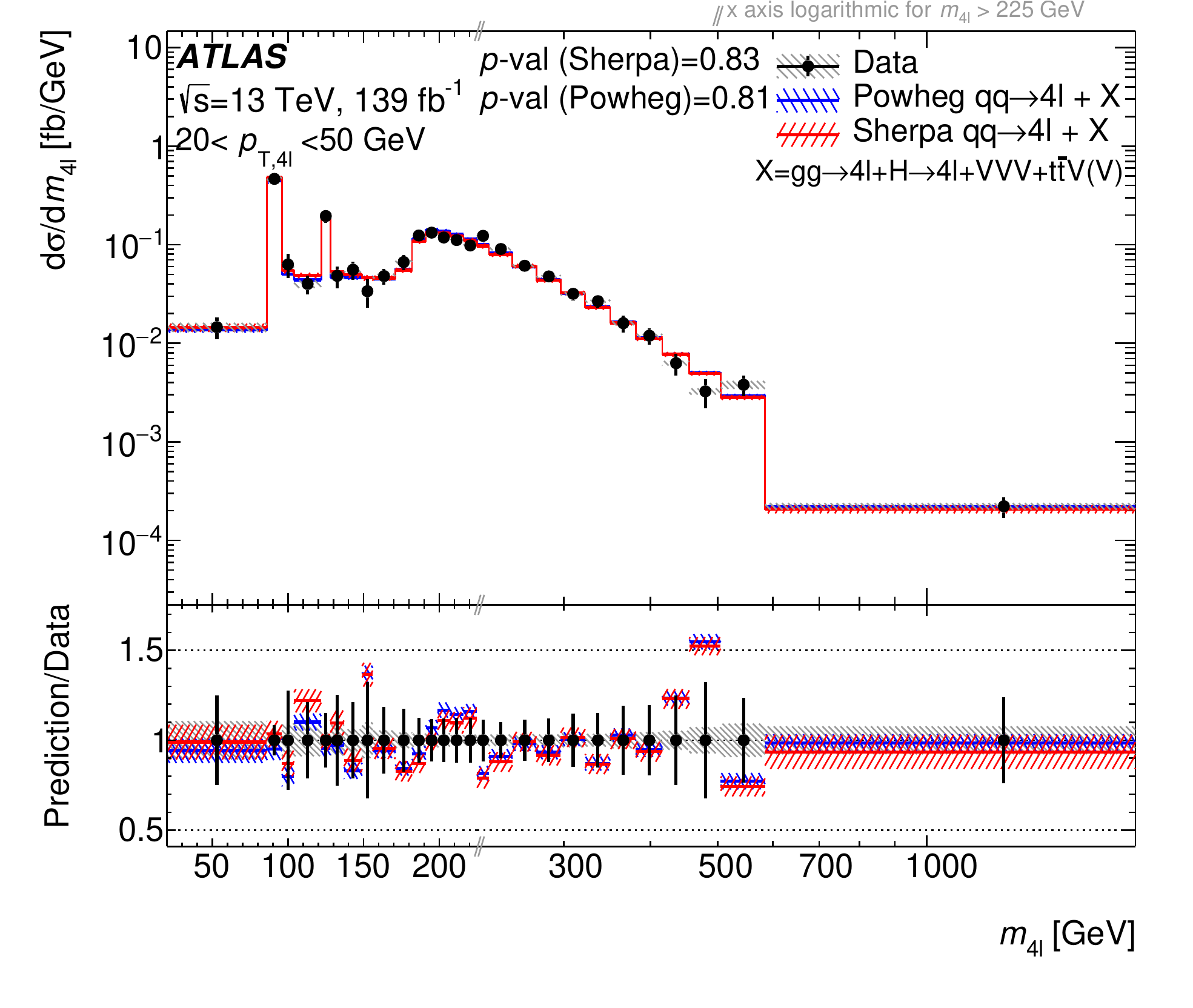}}
\subfigure[$50~\GeV <  \ptFourL  < 100~\GeV$
]{\includegraphics[width = 0.42\textwidth]{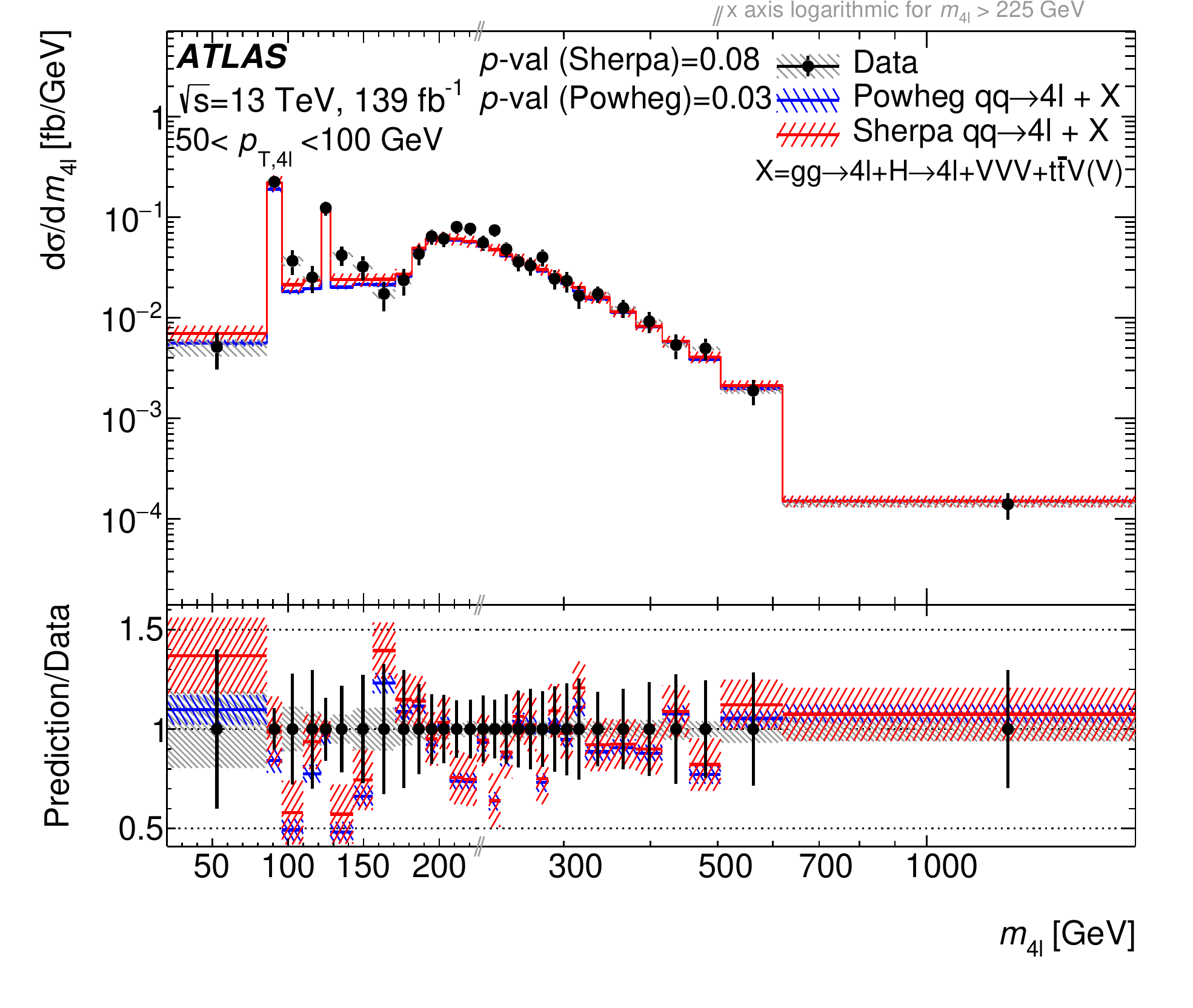}}\\
\subfigure[$100~\GeV <  \ptFourL  < 600~\GeV$
]{\includegraphics[width = 0.42\textwidth]{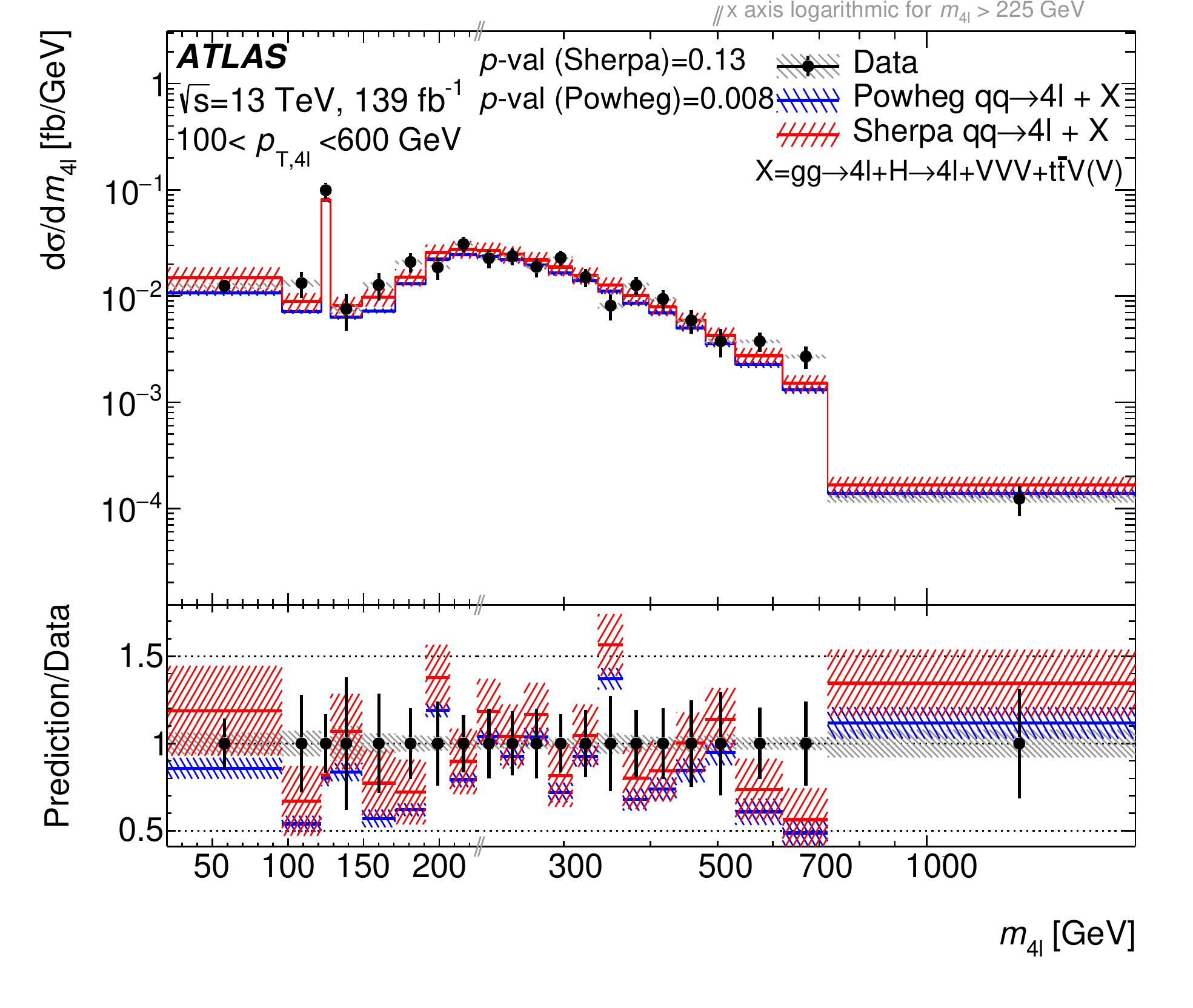}}
\caption{Differential cross-section as a function of \mFourL{} in   \ptFourL{}
slices. The measured data (black points)  are compared with the SM
prediction using either \SHERPA{} (red, with red hashed
band for the uncertainty) or \POWHEG{} + \pythia{} (blue,
with blue hashed band for the uncertainty) to model the \qqFourL{} contribution. The error bars on the data points give the total uncertainty
and the grey hashed band gives the systematic uncertainty.
\Pvalue{}
The
lower panel shows the ratio of the SM predictions to the
data.   The $x$-axis is on a linear scale until $\mFourL =
234~\GeV$  for the $\ptFourL  <
10~\GeV$ slice and $\mFourL = 225~\GeV$ for all other slices,
when it switches to a logarithmic scale. \label{fig:pt4lres}
}
\end{figure}
 
\begin{figure}[htb]
\centering
\subfigure[$|\yFourL|< 0.3$
]{\includegraphics[width = 0.42\textwidth]{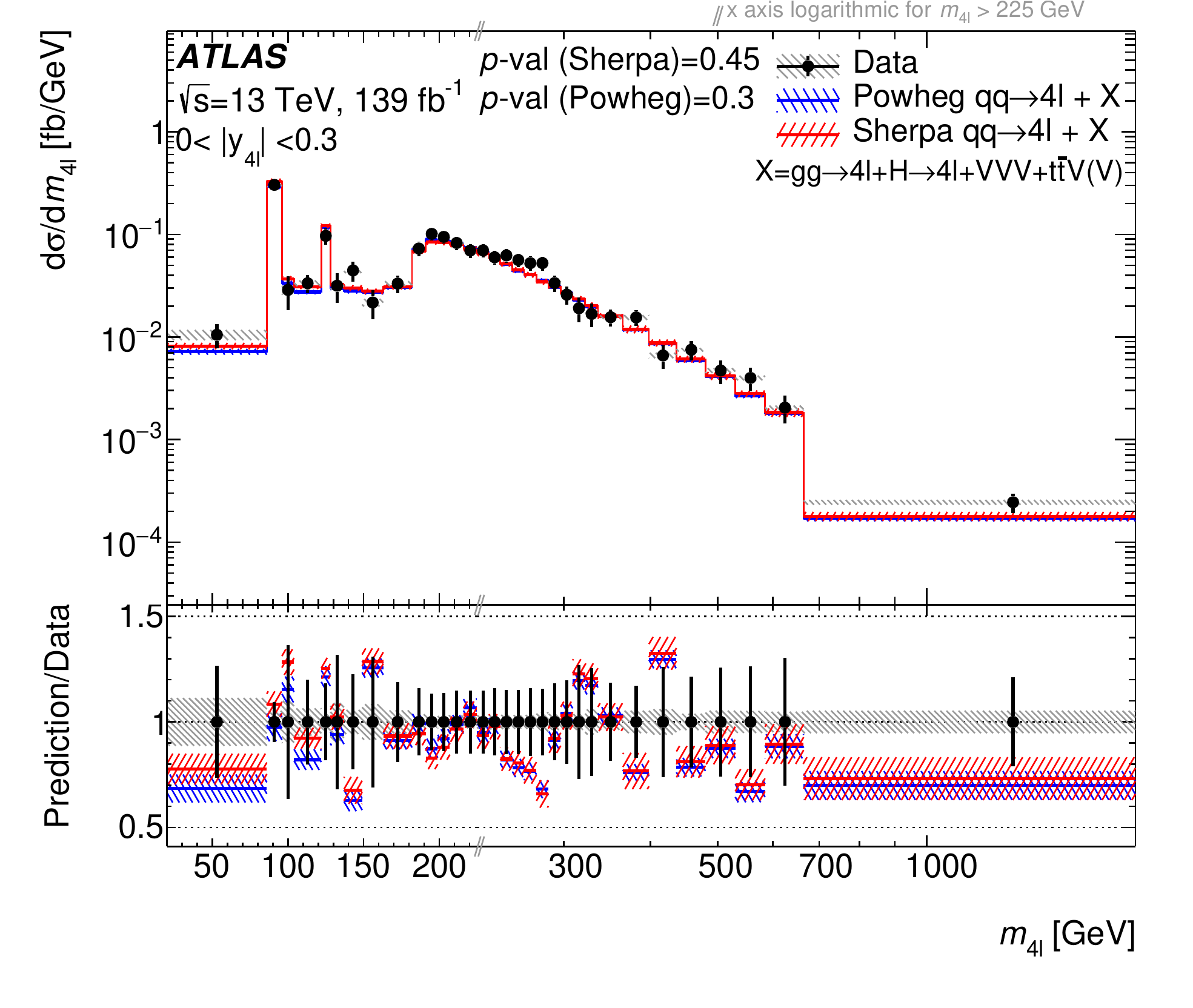}}
\subfigure[$0.3 < |\yFourL| < 0.6$
]{\includegraphics[width = 0.42\textwidth]{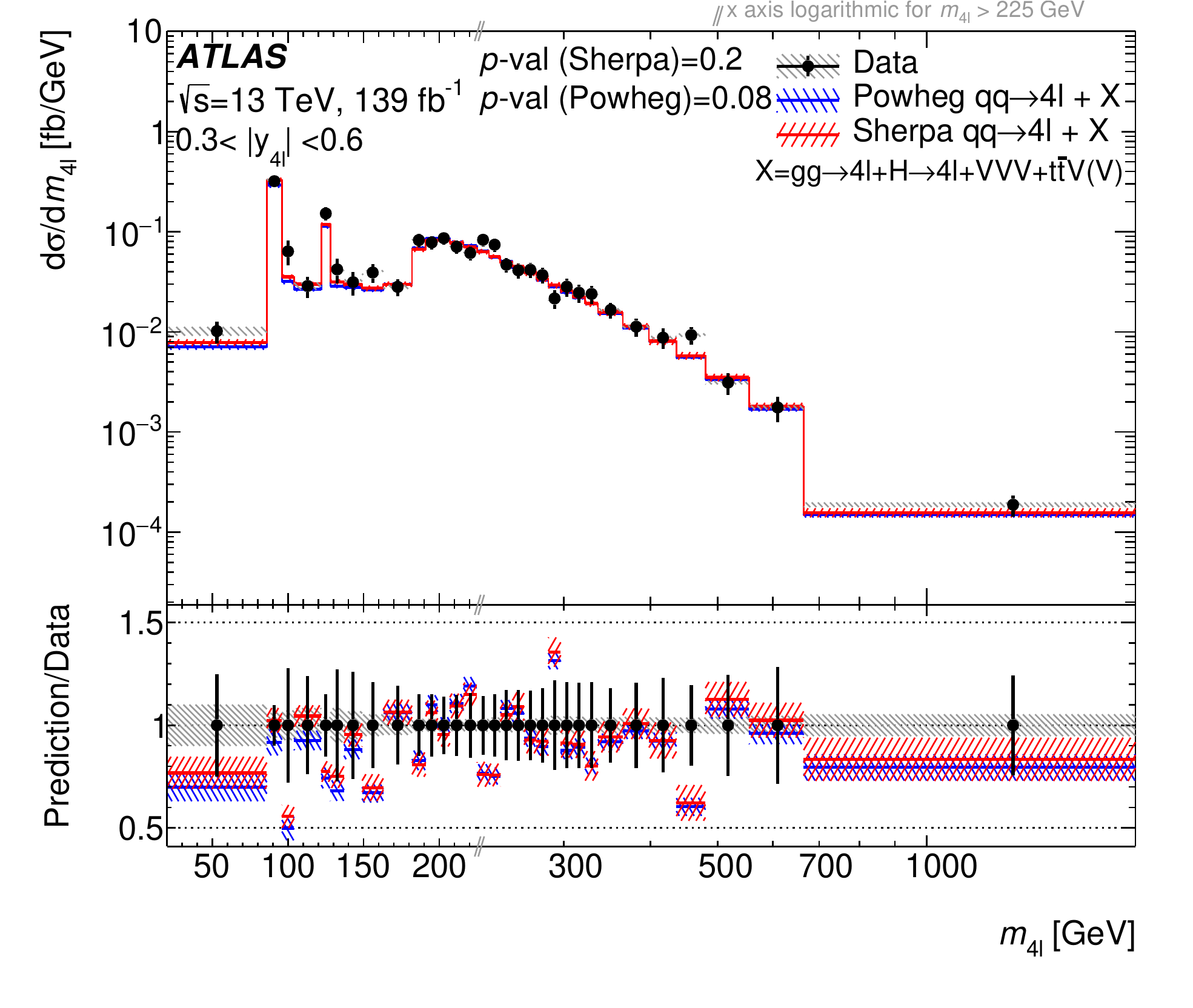}}\\
\subfigure[$0.6 < |\yFourL| < 0.9$
]{\includegraphics[width = 0.42\textwidth]{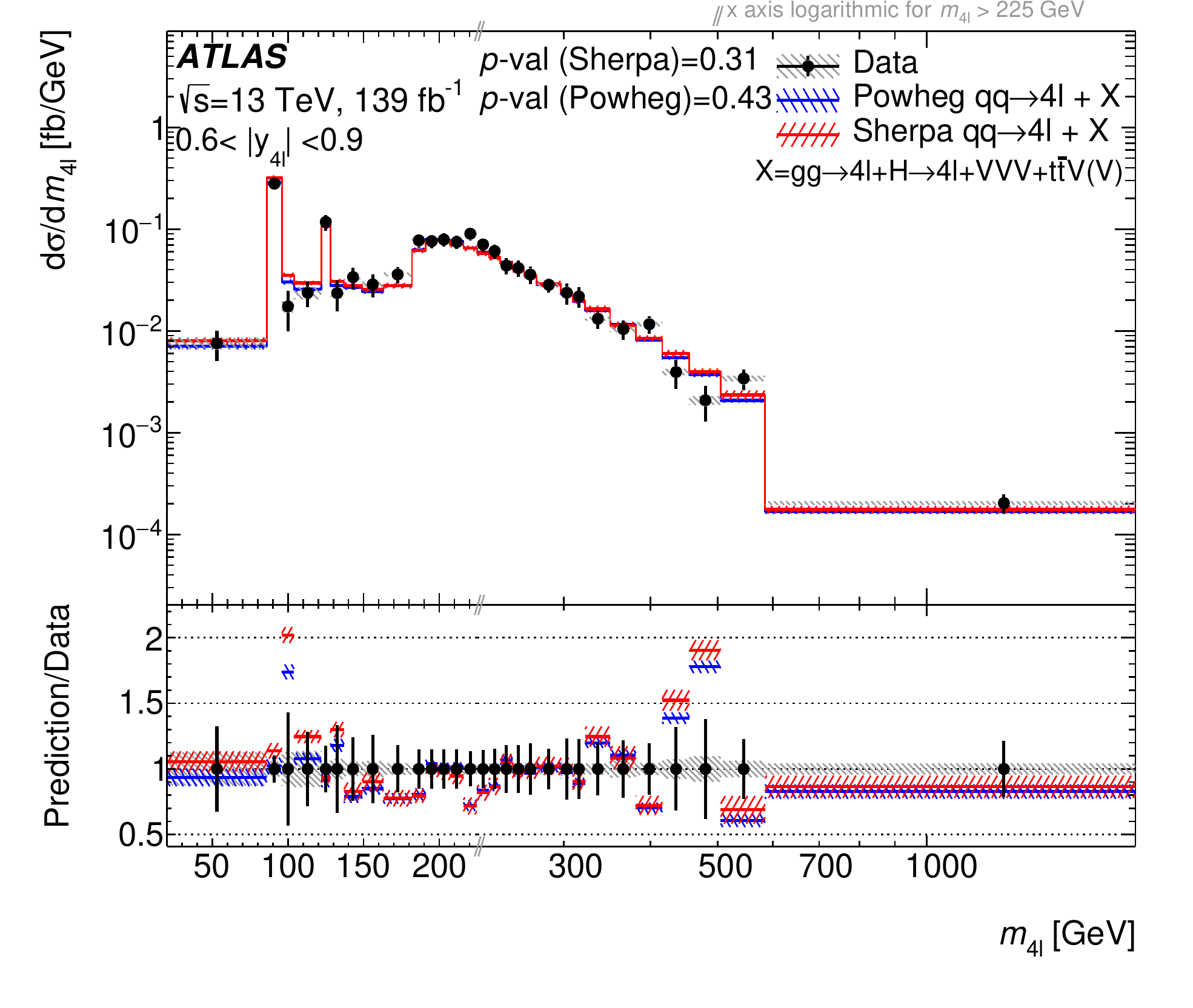}}
\subfigure[$0.9 < |\yFourL| < 1.2$
]{\includegraphics[width = 0.42\textwidth]{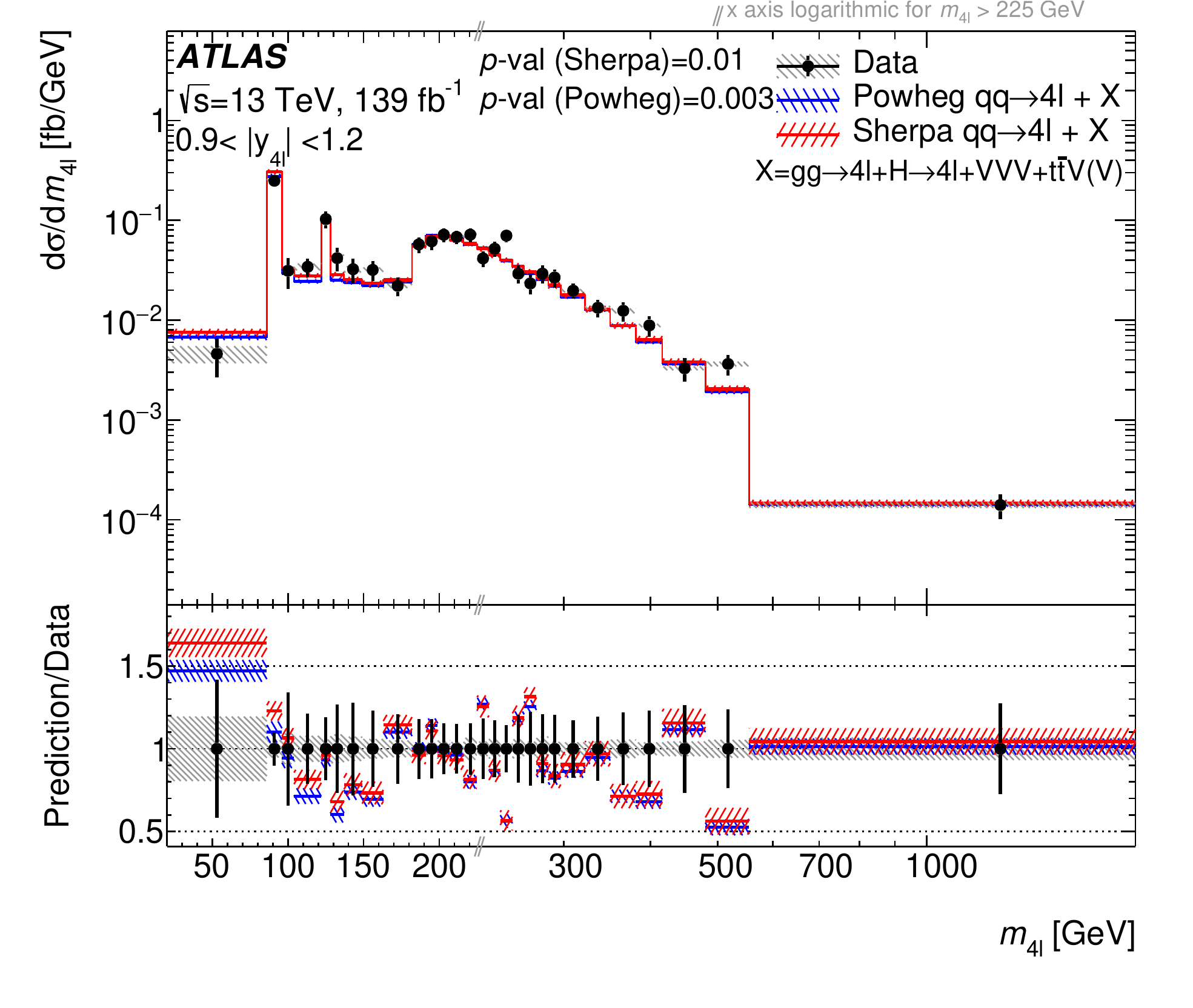}}\\
\subfigure[$1.2 < |\yFourL| < 2.5$
]{\includegraphics[width = 0.42\textwidth]{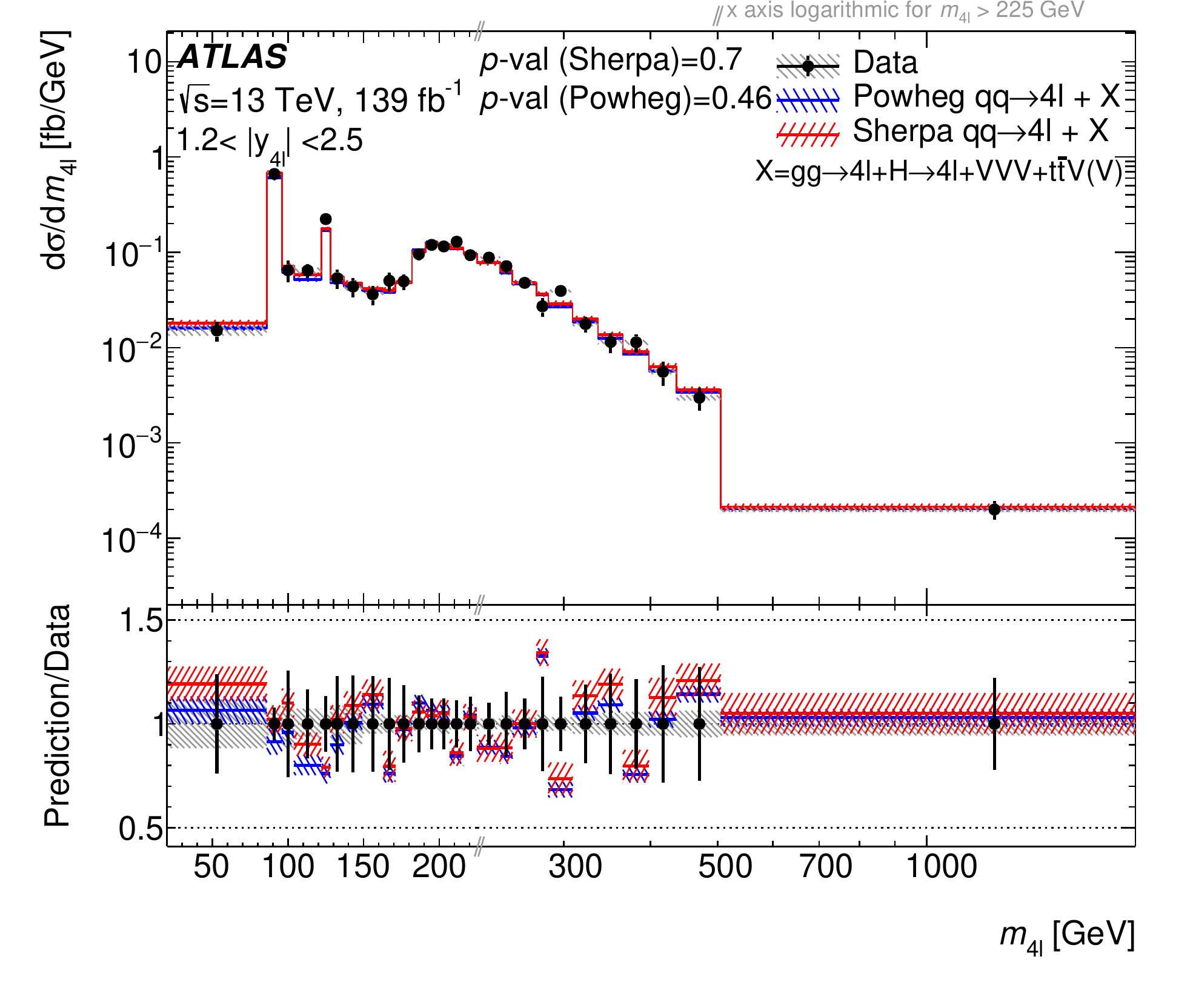}}
\caption{Differential cross-section as a function of \mFourL{} in  $ |\yFourL|$ slices. The measured data (black points) are compared with the SM
prediction using either \SHERPA{} (red, with red hashed
band for the uncertainty) or \POWHEG + \pythia{} (blue,
with blue hashed band for the uncertainty) to model the \qqFourL{} contribution. The error bars on the data points give the total uncertainty
and the grey hashed band gives the systematic uncertainty.
\Pvalue{}
The
lower panel shows the ratio of the SM predictions to the
data.  The $x$-axis is on a linear scale until $\mFourL = 225$~\GeV,
where it switches to a logarithmic scale. \label{fig:y4lres}}
\end{figure}

\begin{figure}[htb]
\centering
\subfigure[\ZFourL \ region]{\includegraphics[width = 0.48\textwidth]{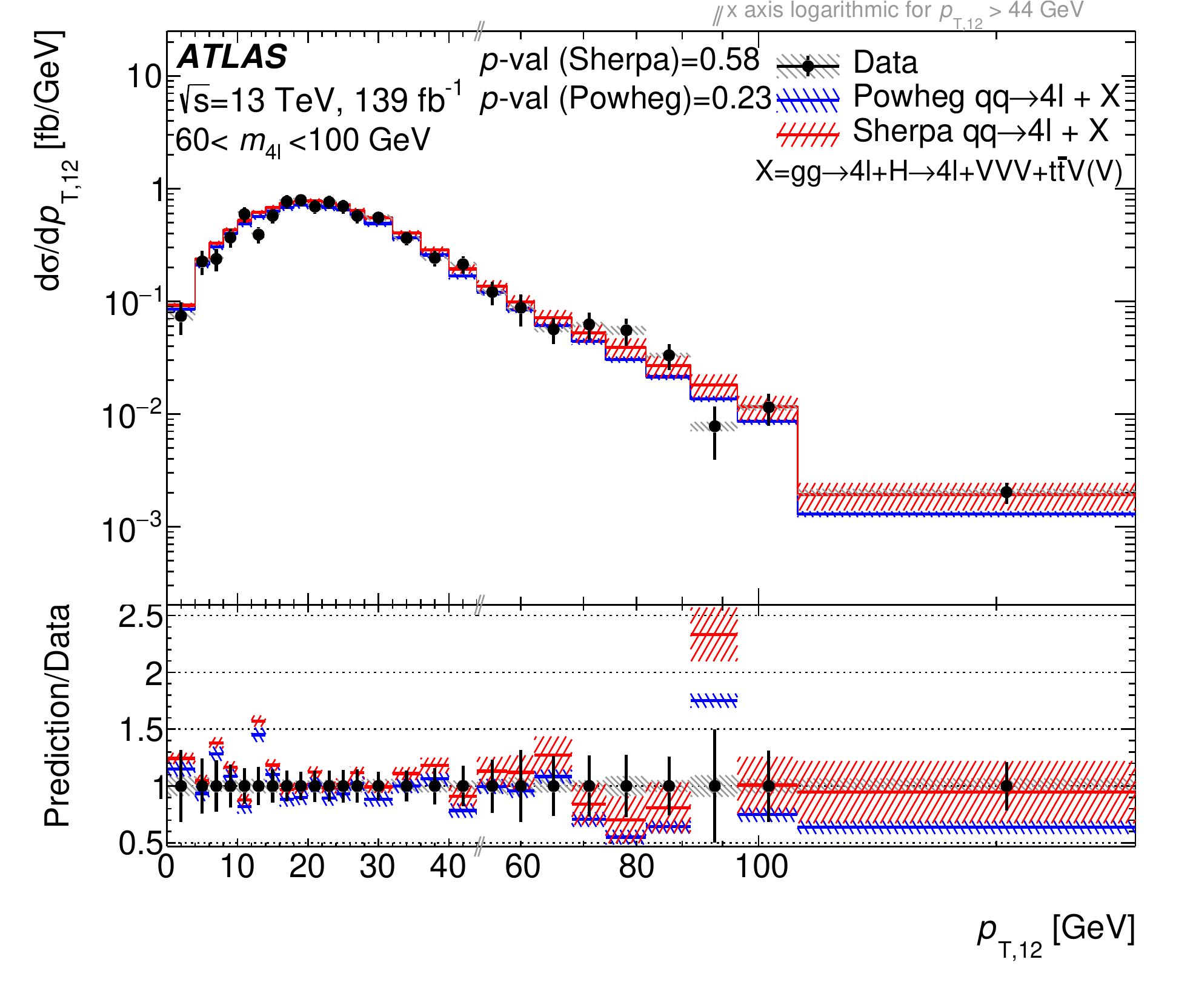}}
\subfigure[\HFourL \ region]{\includegraphics[width = 0.48\textwidth]{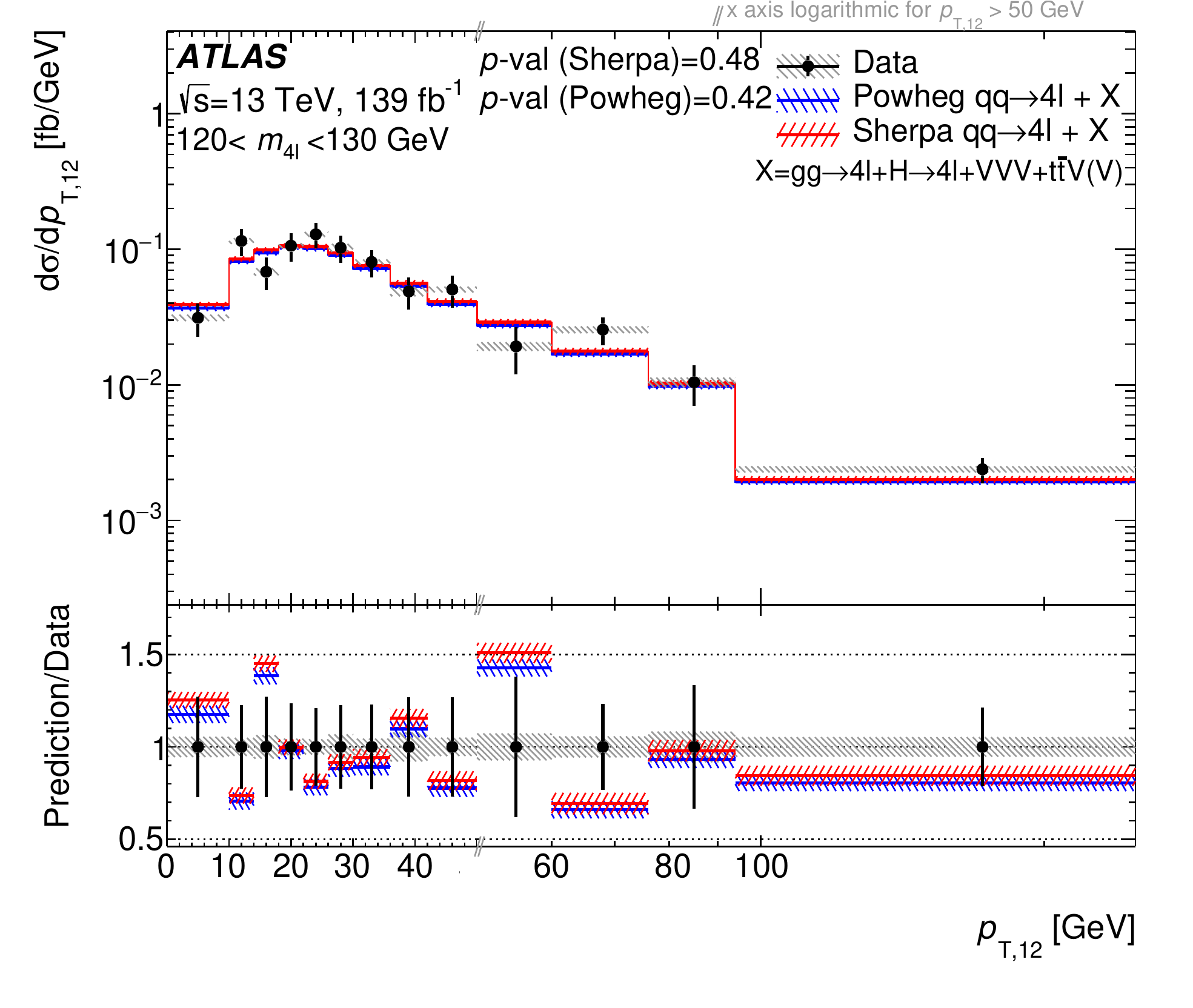}}\\
\subfigure[Off-shell $\Z\Z$ region]{\includegraphics[width = 0.48\textwidth]{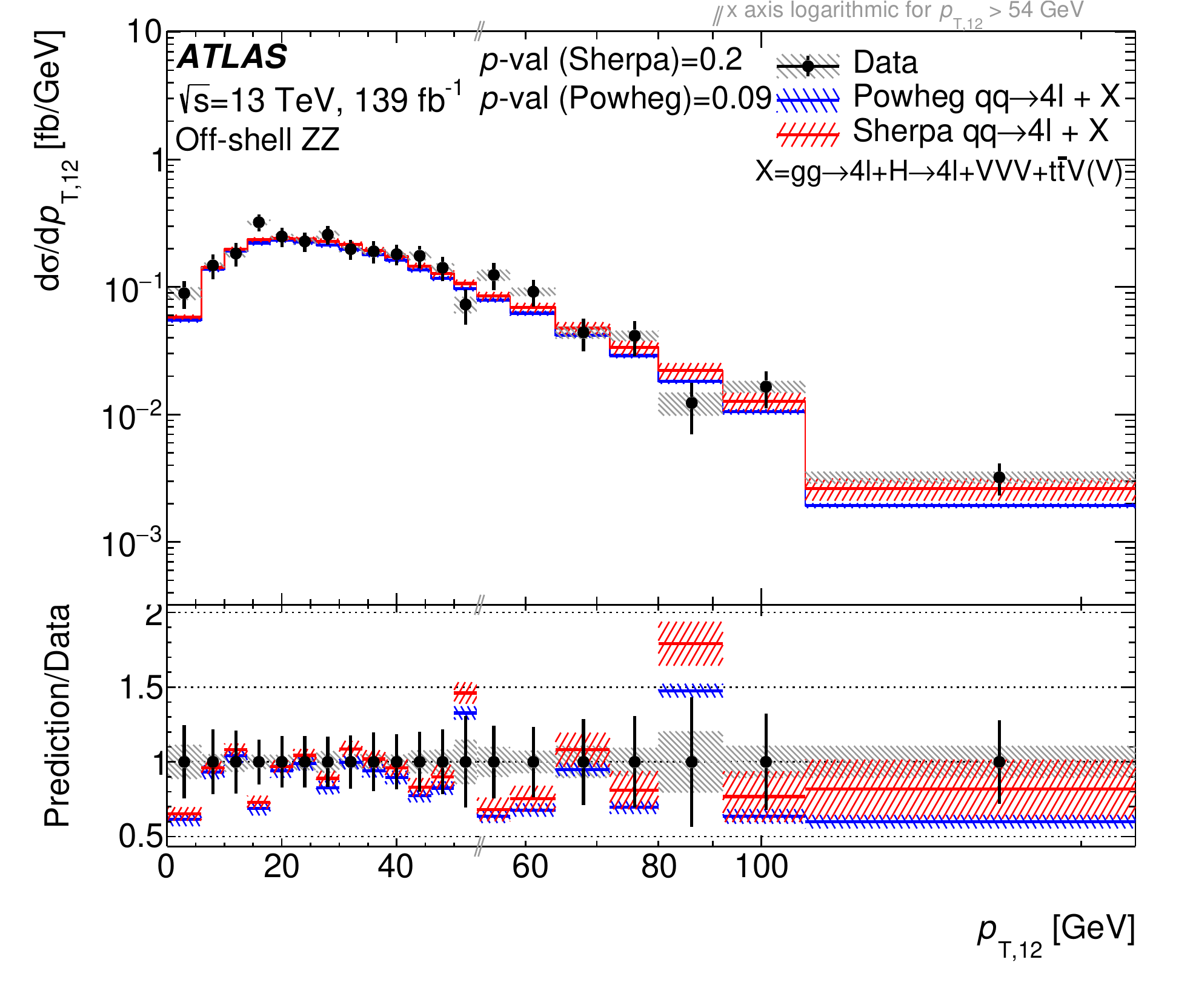}}
\caption{Differential cross-section as a function of \ptZOne{} in \mFourL{}
regions. The measured data (black points)  are compared with the SM
prediction using either \SHERPA{} (red, with red hashed
band for the uncertainty) or \POWHEG + \pythia{} (blue,
with blue hashed band for the uncertainty) to model the \qqFourL{} contribution. The error bars on the data points give the total uncertainty
and the grey hashed band gives the systematic uncertainty.
\Pvalue{}
The  lower panel shows the ratio of the SM predictions to the   data.
In (a) the $x$-axis is on a linear scale until $\ptZOne = 44$~\GeV,
where it switches to a logarithmic scale; in (b) it switches at
$\ptZOne = 50$~\GeV, and in (c) at
$\ptZOne = 54$~\GeV.
\label{fig:ptz1res}}
\end{figure}
 
\begin{figure}[htb]
\centering
\subfigure[\ZFourL \  region]{\includegraphics[width = 0.48\textwidth]{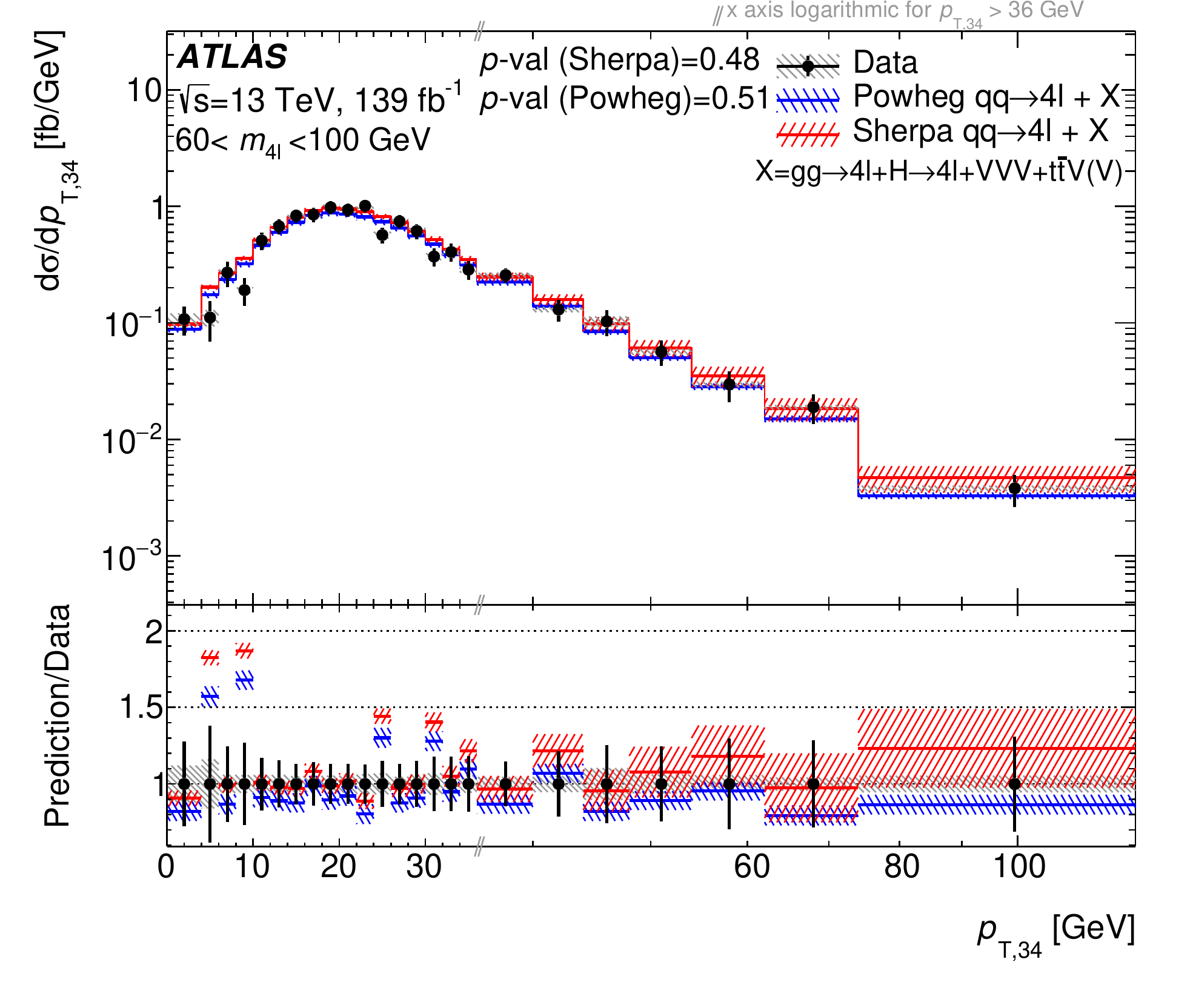}}
\subfigure[\HFourL \ region]{\includegraphics[width = 0.48\textwidth]{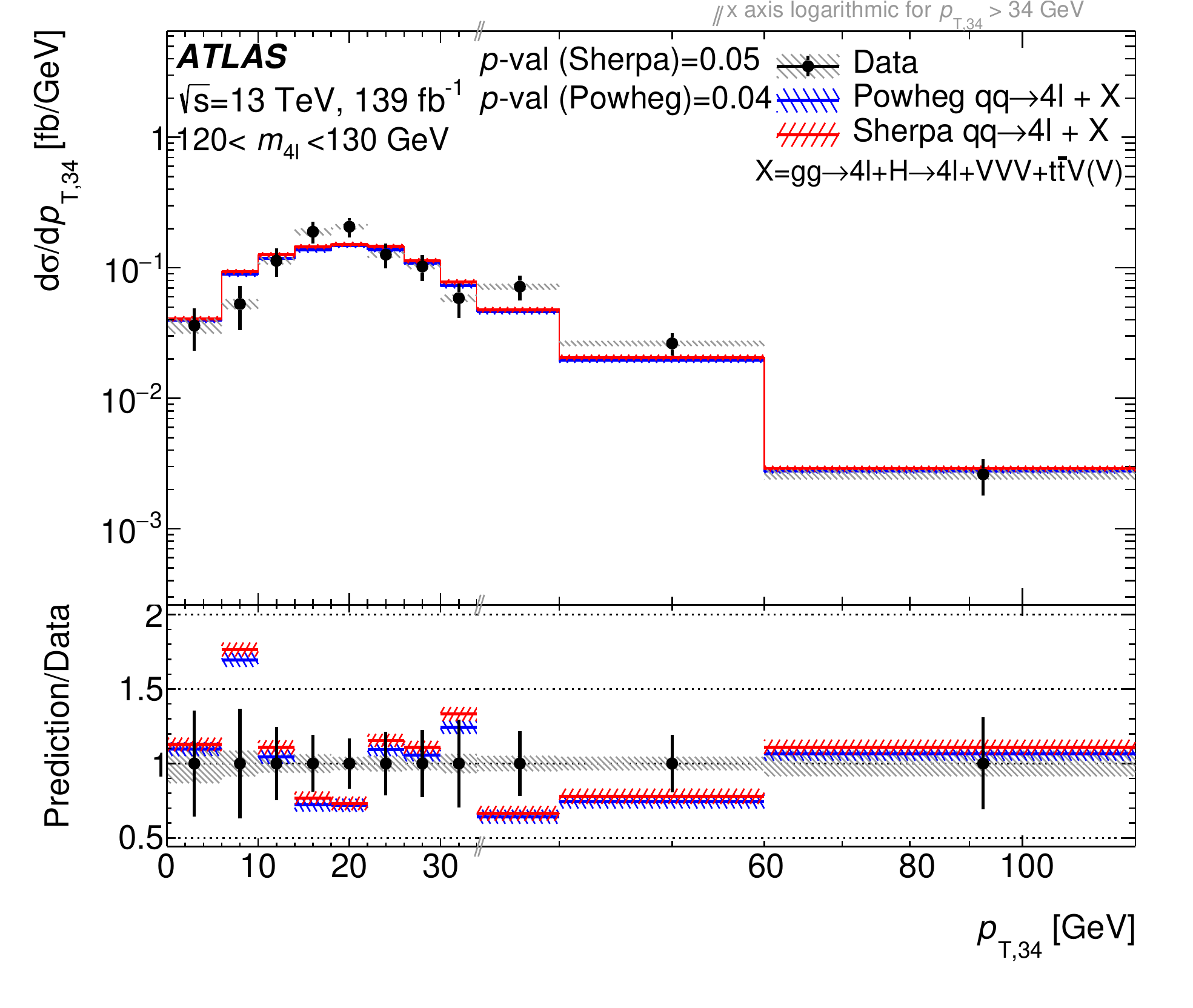}}\\
\subfigure[Off-shell $\Z\Z$ region]{\includegraphics[width = 0.48\textwidth]{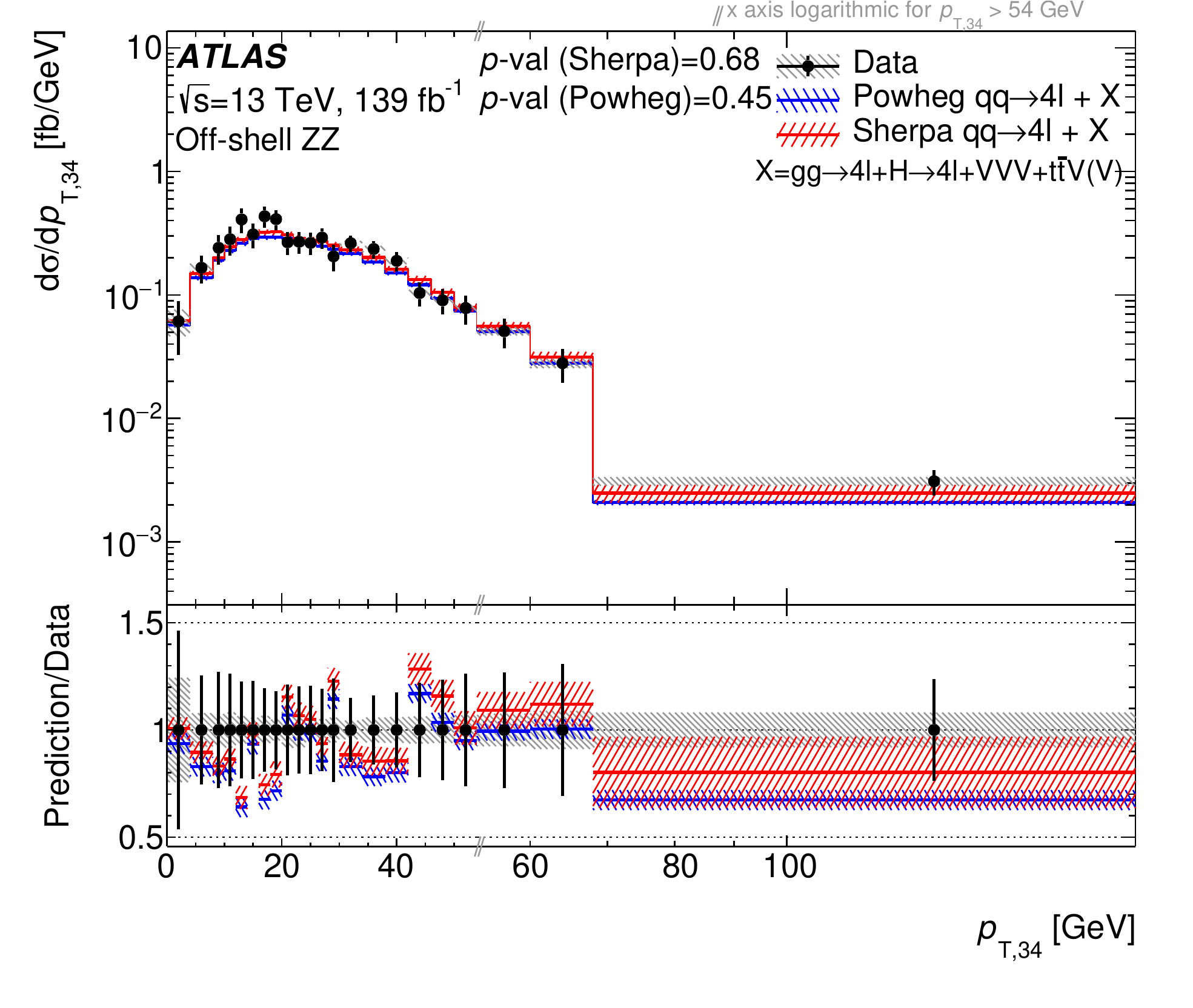}}
\caption{Differential cross-section as a function of \ptZTwo{} in \mFourL{} regions. The measured data (black points) are compared with the SM
prediction using either \SHERPA{} (red, with red hashed
band for the uncertainty) or \POWHEG + \pythia{} (blue,
with blue hashed band for the uncertainty) to model the \qqFourL{} contribution. The error bars on the data points give the total uncertainty
and the grey hashed band gives the systematic uncertainty.
\Pvalue{}
The  lower panel shows the ratio of the SM predictions to the   data.
In (a) the $x$-axis is on a linear scale until $\ptZTwo = 36$~\GeV,
where it switches to a logarithmic scale; in (b) it switches at
$\ptZTwo = 34$~\GeV, and in (c) at
$\ptZTwo = 54$~\GeV.
\label{fig:ptz2res} }
\end{figure}

\begin{figure}[htb]
\centering
\subfigure[\ZFourL \  region]{\includegraphics[width = 0.48\textwidth]{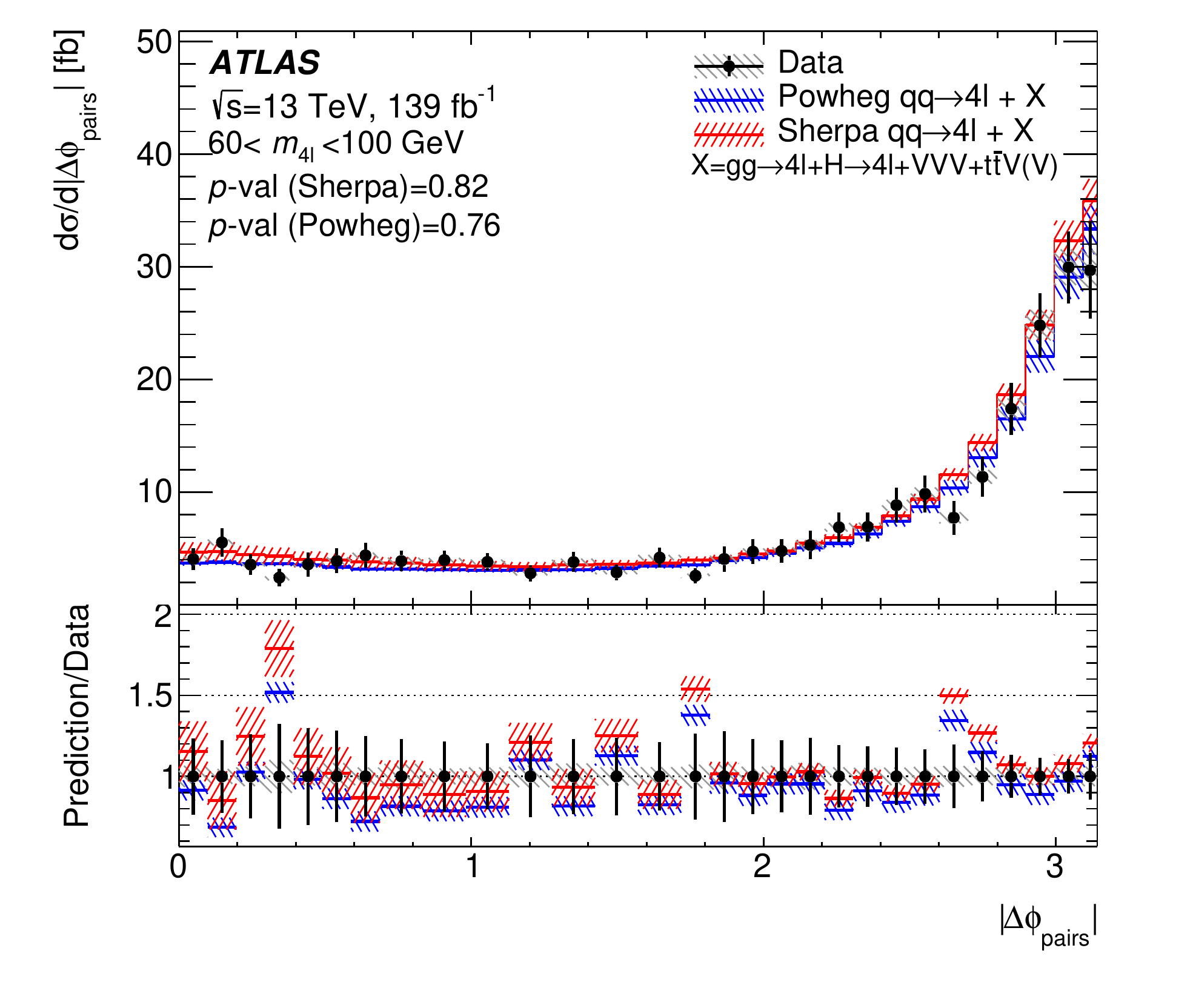}}
\subfigure[\HFourL \  region]{\includegraphics[width = 0.48\textwidth]{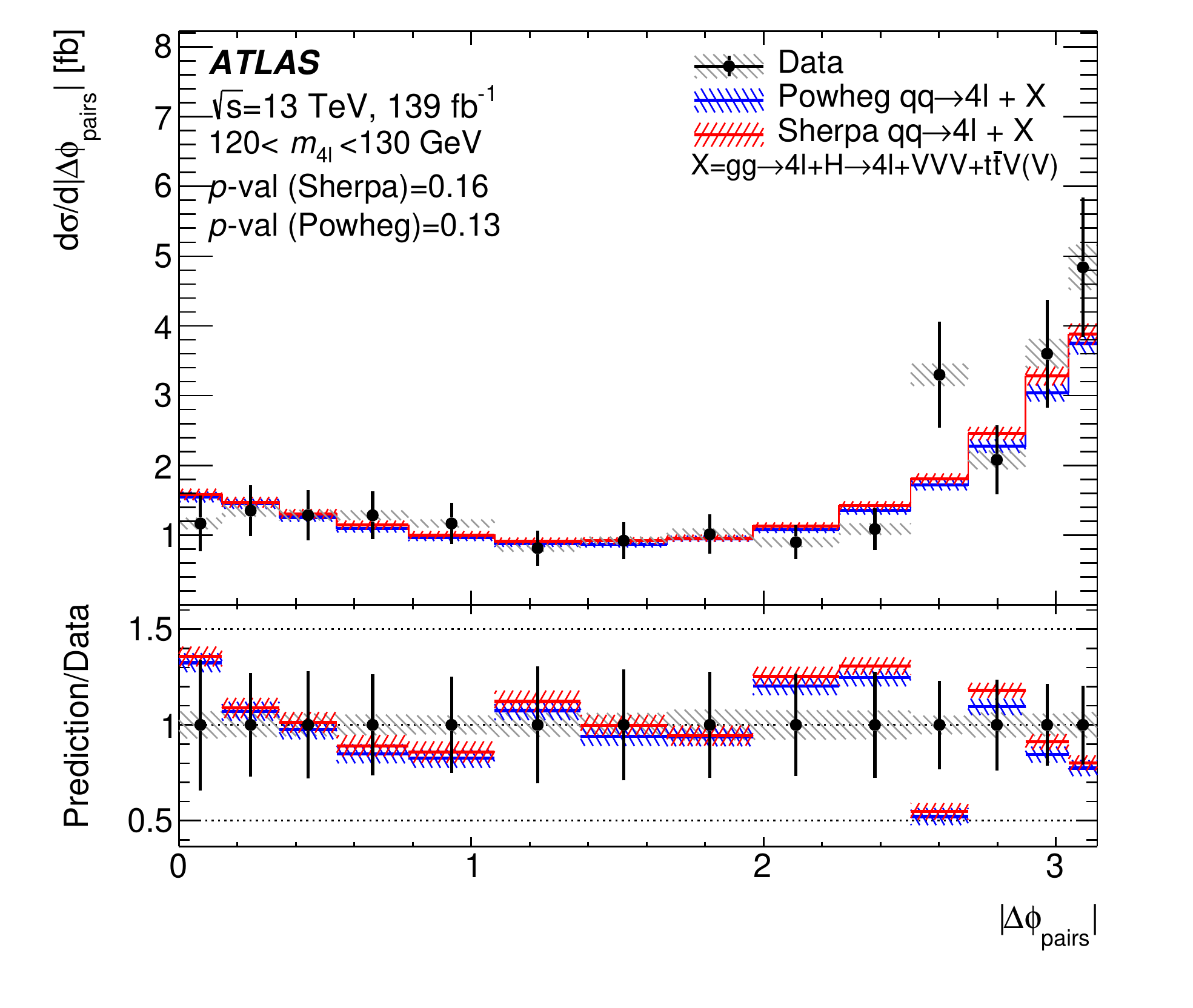}}\\
\subfigure[Off-shell $\Z\Z$ region]{\includegraphics[width = 0.48\textwidth]{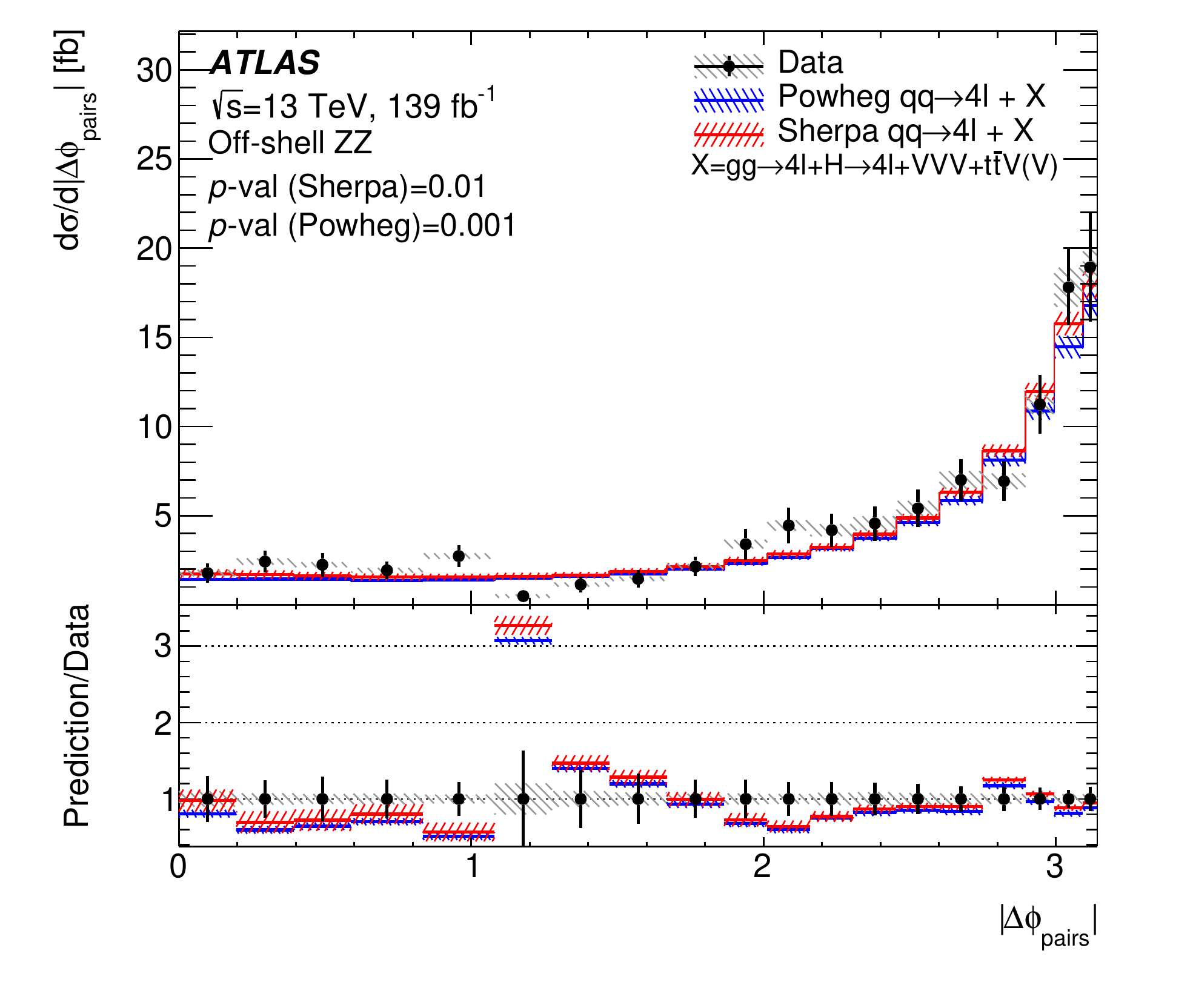}}
\caption{Differential cross-section as a function of \dPhiPairs{} in \mFourL{}
regions. The measured data (black points)  are compared with the SM
prediction using either \SHERPA{} (red, with red hashed
band for the uncertainty) or \POWHEG + \pythia{} (blue,
with blue hashed band for the uncertainty) to model the \qqFourL{} contribution. The error bars on the data points give the total uncertainty
and the grey hashed band gives the systematic uncertainty.
\Pvalue{}
The  lower panel shows the ratio of the SM predictions to the   data. \label{fig:dphipairs-rest} }
\end{figure}
 
\begin{figure}[htb]
\centering
\subfigure[\ZFourL \ region]{\includegraphics[width = 0.48\textwidth]{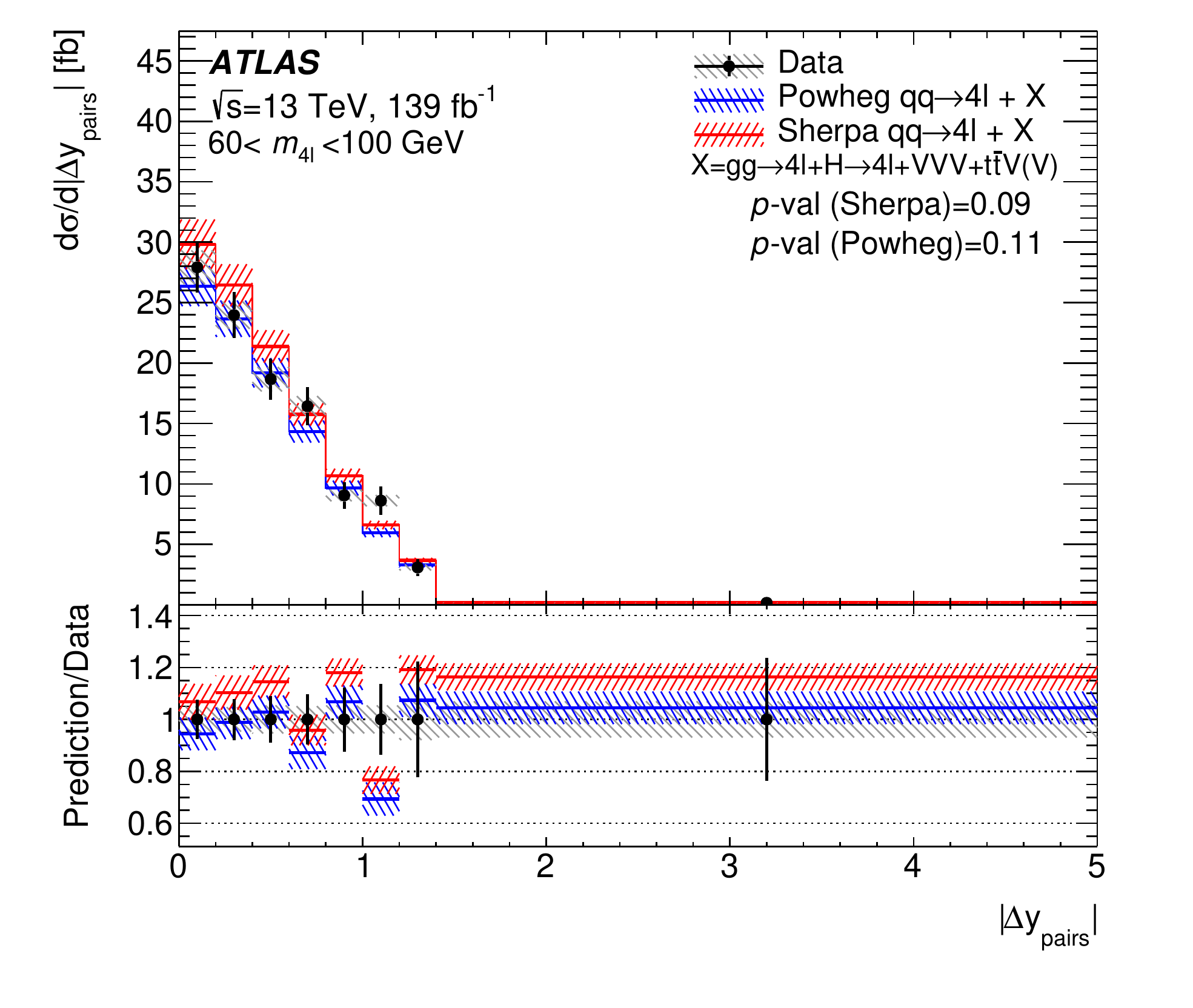}}
\subfigure[\HFourL \ region]{\includegraphics[width = 0.48\textwidth]{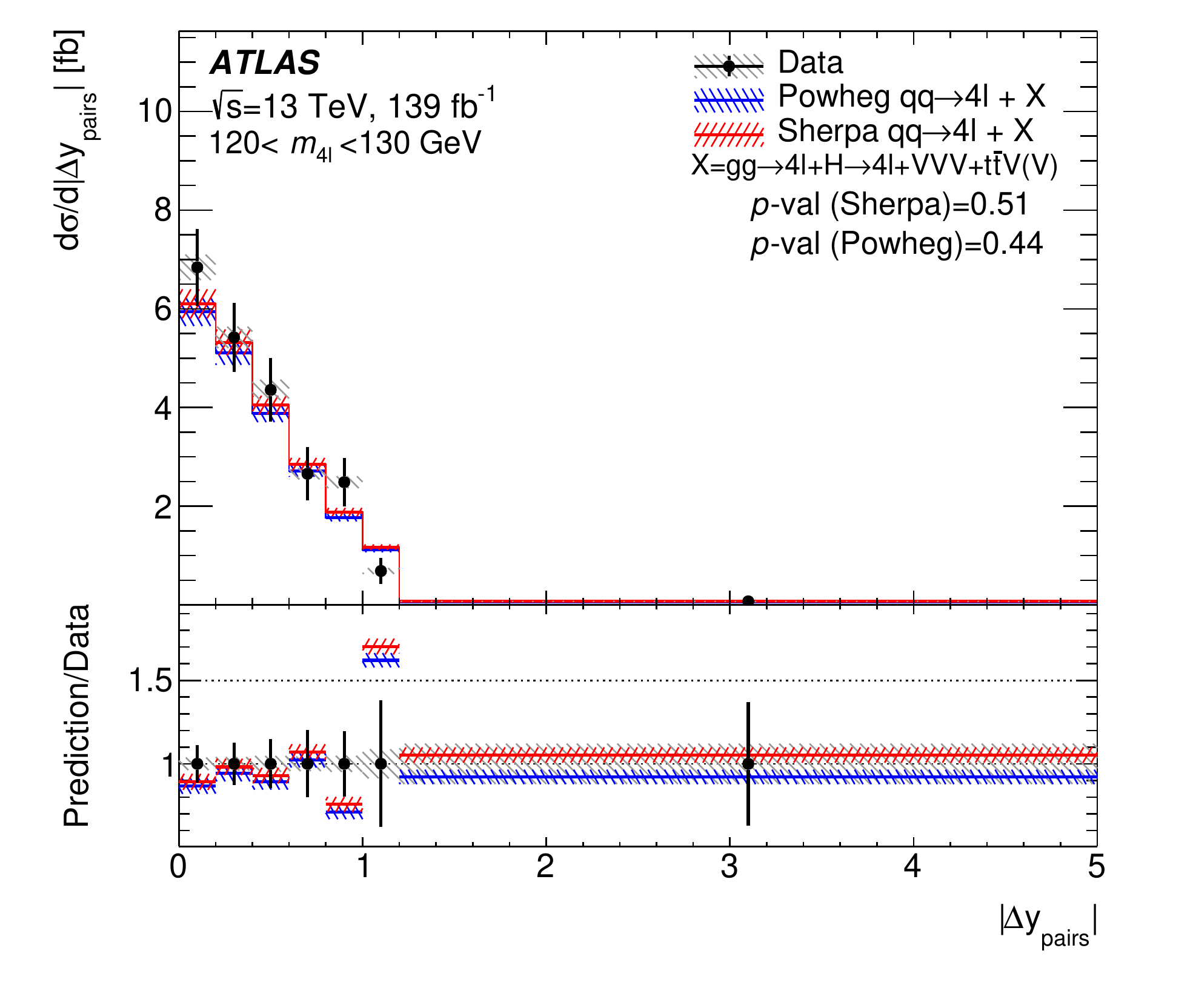}}\\
\subfigure[Off-shell $\Z\Z$ region]{\includegraphics[width = 0.48\textwidth]{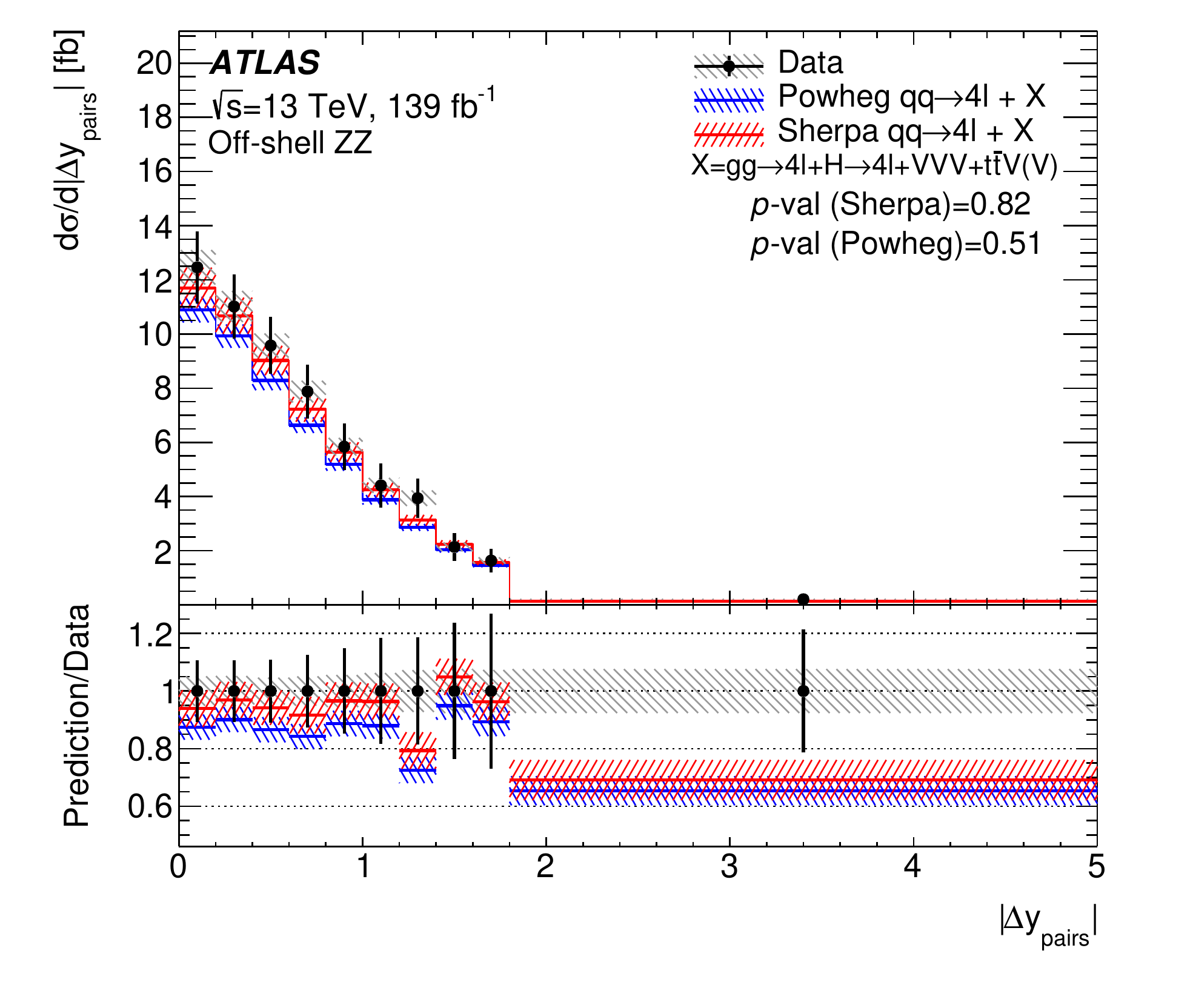}}
\caption{Differential cross-section as a function of \dYPairs{} in \mFourL{}
regions. The measured data (black points)  are compared with the SM
prediction using either \SHERPA{} (red, with red hashed
band for the uncertainty) or \POWHEG + \pythia{} (blue,
with blue hashed band for the uncertainty) to model the \qqFourL{} contribution. The error bars on the data points give the total uncertainty
and the grey hashed band gives the systematic uncertainty.
\Pvalue{}
The  lower panel shows the ratio of the SM predictions to the   data. \label{fig:dy-rest}}
\end{figure}

\clearpage
 
-------------------------------------------------------------------------------
\section*{Acknowledgements}
-------------------------------------------------------------------------------
 

We thank CERN for the very successful operation of the LHC, as well as the
support staff from our institutions without whom ATLAS could not be
operated efficiently.
 
We acknowledge the support of ANPCyT, Argentina; YerPhI, Armenia; ARC, Australia; BMWFW and FWF, Austria; ANAS, Azerbaijan; SSTC, Belarus; CNPq and FAPESP, Brazil; NSERC, NRC and CFI, Canada; CERN; ANID, Chile; CAS, MOST and NSFC, China; COLCIENCIAS, Colombia; MSMT CR, MPO CR and VSC CR, Czech Republic; DNRF and DNSRC, Denmark; IN2P3-CNRS and CEA-DRF/IRFU, France; SRNSFG, Georgia; BMBF, HGF and MPG, Germany; GSRT, Greece; RGC and Hong Kong SAR, China; ISF and Benoziyo Center, Israel; INFN, Italy; MEXT and JSPS, Japan; CNRST, Morocco; NWO, Netherlands; RCN, Norway; MNiSW and NCN, Poland; FCT, Portugal; MNE/IFA, Romania; JINR; MES of Russia and NRC KI, Russian Federation; MESTD, Serbia; MSSR, Slovakia; ARRS and MIZ\v{S}, Slovenia; DST/NRF, South Africa; MICINN, Spain; SRC and Wallenberg Foundation, Sweden; SERI, SNSF and Cantons of Bern and Geneva, Switzerland; MOST, Taiwan; TAEK, Turkey; STFC, United Kingdom; DOE and NSF, United States of America. In addition, individual groups and members have received support from BCKDF, CANARIE, Compute Canada, CRC and IVADO, Canada; Beijing Municipal Science \& Technology Commission, China; COST, ERC, ERDF, Horizon 2020 and Marie Sk{\l}odowska-Curie Actions, European Union; Investissements d'Avenir Labex, Investissements d'Avenir Idex and ANR, France; DFG and AvH Foundation, Germany; Herakleitos, Thales and Aristeia programmes co-financed by EU-ESF and the Greek NSRF, Greece; BSF-NSF and GIF, Israel; La Caixa Banking Foundation, CERCA Programme Generalitat de Catalunya and PROMETEO and GenT Programmes Generalitat Valenciana, Spain; G\"{o}ran Gustafssons Stiftelse, Sweden; The Royal Society and Leverhulme Trust, United Kingdom.
 
The crucial computing support from all WLCG partners is acknowledged gratefully, in particular from CERN, the ATLAS Tier-1 facilities at TRIUMF (Canada), NDGF (Denmark, Norway, Sweden), CC-IN2P3 (France), KIT/GridKA (Germany), INFN-CNAF (Italy), NL-T1 (Netherlands), PIC (Spain), ASGC (Taiwan), RAL (UK) and BNL (USA), the Tier-2 facilities worldwide and large non-WLCG resource providers. Major contributors of computing resources are listed in Ref.~\cite{ATL-SOFT-PUB-2020-001}.
 

\clearpage
\printbibliography
 
\clearpage 
 
\begin{flushleft}
\hypersetup{urlcolor=black}
{\Large The ATLAS Collaboration}

\bigskip

\AtlasOrcid[0000-0002-6665-4934]{G.~Aad}$^\textrm{\scriptsize 102}$,    
\AtlasOrcid[0000-0002-5888-2734]{B.~Abbott}$^\textrm{\scriptsize 128}$,    
\AtlasOrcid[0000-0002-7248-3203]{D.C.~Abbott}$^\textrm{\scriptsize 103}$,    
\AtlasOrcid[0000-0002-2788-3822]{A.~Abed~Abud}$^\textrm{\scriptsize 36}$,    
\AtlasOrcid[0000-0002-1002-1652]{K.~Abeling}$^\textrm{\scriptsize 53}$,    
\AtlasOrcid[0000-0002-2987-4006]{D.K.~Abhayasinghe}$^\textrm{\scriptsize 94}$,    
\AtlasOrcid[0000-0002-8496-9294]{S.H.~Abidi}$^\textrm{\scriptsize 167}$,    
\AtlasOrcid[0000-0002-8279-9324]{O.S.~AbouZeid}$^\textrm{\scriptsize 40}$,    
\AtlasOrcid{N.L.~Abraham}$^\textrm{\scriptsize 156}$,    
\AtlasOrcid[0000-0001-5329-6640]{H.~Abramowicz}$^\textrm{\scriptsize 161}$
% ...
\bigskip
\\

% ...
$^{36}$CERN, Geneva; Switzerland.\\
% ...
$^{40}$Niels Bohr Institute, University of Copenhagen, Copenhagen; Denmark.\\
% ...
$^{53}$II. Physikalisches Institut, Georg-August-Universit\"{a}t G\"ottingen, G\"ottingen; Germany.\\
% ...
$^{94}$Department of Physics, Royal Holloway University of London, Egham; United Kingdom.\\
% ...
$^{102}$CPPM, Aix-Marseille Universit\'e, CNRS/IN2P3, Marseille; France.\\
$^{103}$Department of Physics, University of Massachusetts, Amherst MA; United States of America.\\
% ...
$^{128}$Homer L. Dodge Department of Physics and Astronomy, University of Oklahoma, Norman OK; United States of America.\\
% ...
$^{156}$Department of Physics and Astronomy, University of Sussex, Brighton; United Kingdom.\\
% ...
$^{161}$Raymond and Beverly Sackler School of Physics and Astronomy, Tel Aviv University, Tel Aviv; Israel.\\
% ...
$^{167}$Department of Physics, University of Toronto, Toronto ON; Canada.\\
% ...

\end{flushleft}


\end{document}